\newcommand*{\ATLASLATEXPATH}{}
\documentclass[UKenglish,PAPER,cernpreprint,paper=a4,maketitle=true,texlive=2011]{\ATLASLATEXPATH atlasdoc}
\usepackage[siunitx=false,subfigure=true,biblatex=false]{\ATLASLATEXPATH atlaspackage}
\usepackage{\ATLASLATEXPATH atlasphysics}
\graphicspath{{logos/}}
\hypersetup{pdftitle={ATLAS draft},pdfauthor={The ATLAS Collaboration}}

\usepackage{tikz}
\usepackage{cite}
\usepackage{units}
\usepackage{xspace}
\usepackage{mathrsfs}
\usepackage{units}
\usepackage{amssymb}
\usepackage{placeins}
\usepackage{subfigure}
\usepackage{chngcntr}
\usepackage{booktabs}
\usepackage{rotating}
\usepackage{multirow}
\usepackage{url}
\usepackage{enumitem}
\counterwithout{equation}{section}

\newcommand{\Alpgen}{{\textsc{Alpgen}}\xspace}

\newcommand{\Mcatnlo}{{\textsc{MC@NLO}}\xspace}
\newcommand{\aMcAtNlo}{a{\textsc{MC@NLO}}\xspace}
\newcommand{\AcerMC}{{\textsc{AcerMC}}\xspace}
\newcommand{\Sherpa}{{\textsc{Sherpa}}\xspace}
\newcommand{\Herwig}{{\textsc{Herwig}}\xspace}
\newcommand{\VBFnlo}{{\textsc{VBFNLO}}\xspace}
\newcommand{\Herwigpp}{{\textsc{Herwig{\tt++}}}\xspace}

\newcommand{\Prospino}{{\textsc{Prospino2}}\xspace}
\newcommand{\Madgraph}{{\textsc{MadGraph}}\xspace}

\newcommand{\CTEQSixL}{{\textsc{CTEQ6L1}}\xspace}

\newcommand{\CTTen}{{\textsc{CT10}}\xspace}

\newcommand{\POWHEGBOX}{{\textsc{Powheg Box}}\xspace}
\newcommand{\Perugia}{{\textsc{Perugia2011C}}\xspace}
\newcommand{\AUET}{{\textsc{AUET2B}}\xspace}
\newcommand{\AU}{{\textsc{AU2}}\xspace}
\newcommand{\UEEE}{{\textsc{UE-EE-3}}\xspace}
\newcommand{\ggVV}{{\text{gg2VV}}\xspace}

\newcommand{\HistFitter}{{\textsc{HistFitter}}\xspace}
\newcommand{\ninoone}{{\ensuremath{\mathchoice%
      {\displaystyle\raise.4ex\hbox{$\displaystyle\tilde\chi^0_1$}}%
         {\textstyle\raise.4ex\hbox{$\textstyle\tilde\chi^0_1$}}%
       {\scriptstyle\raise.3ex\hbox{$\scriptstyle\tilde\chi^0_1$}}%
 {\scriptscriptstyle\raise.3ex\hbox{$\scriptscriptstyle\tilde\chi^0_1$}}}}}
\newcommand{\ninotwo}{{\ensuremath{\mathchoice%
      {\displaystyle\raise.4ex\hbox{$\displaystyle\tilde\chi^0_2$}}%
         {\textstyle\raise.4ex\hbox{$\textstyle\tilde\chi^0_2$}}%
       {\scriptstyle\raise.3ex\hbox{$\scriptstyle\tilde\chi^0_2$}}%
 {\scriptscriptstyle\raise.3ex\hbox{$\scriptscriptstyle\tilde\chi^0_2$}}}}}
\newcommand{\ninothree}{{\ensuremath{\mathchoice%
      {\displaystyle\raise.4ex\hbox{$\displaystyle\tilde\chi^0_3$}}%
         {\textstyle\raise.4ex\hbox{$\textstyle\tilde\chi^0_3$}}%
       {\scriptstyle\raise.3ex\hbox{$\scriptstyle\tilde\chi^0_3$}}%
 {\scriptscriptstyle\raise.3ex\hbox{$\scriptscriptstyle\tilde\chi^0_3$}}}}}
\newcommand{\ninoonetwothreefour}{{\ensuremath{\mathchoice%
      {\displaystyle\raise.4ex\hbox{$\displaystyle\tilde\chi^0_{1,2,3,4}$}}%
         {\textstyle\raise.4ex\hbox{$\textstyle\tilde\chi^0_{1,2,3,4}$}}%
       {\scriptstyle\raise.3ex\hbox{$\scriptstyle\tilde\chi^0_{1,2,3,4}$}}%
 {\scriptscriptstyle\raise.3ex\hbox{$\scriptscriptstyle\tilde\chi^0_{1,2,3,4}$}}}}}
\newcommand{\ninotwothreefour}{{\ensuremath{\mathchoice%
      {\displaystyle\raise.4ex\hbox{$\displaystyle\tilde\chi^0_{2,3,4}$}}%
         {\textstyle\raise.4ex\hbox{$\textstyle\tilde\chi^0_{2,3,4}$}}%
       {\scriptstyle\raise.3ex\hbox{$\scriptstyle\tilde\chi^0_{2,3,4}$}}%
 {\scriptscriptstyle\raise.3ex\hbox{$\scriptscriptstyle\tilde\chi^0_{2,3,4}$}}}}}
\newcommand{\nino}[1]{{\ensuremath{\mathchoice%
      {\displaystyle\raise.4ex\hbox{$\displaystyle\tilde\chi^0_{#1}$}}%
         {\textstyle\raise.4ex\hbox{$\textstyle\tilde\chi^0_{#1}$}}%
       {\scriptstyle\raise.3ex\hbox{$\scriptstyle\tilde\chi^0_{#1}$}}%
 {\scriptscriptstyle\raise.3ex\hbox{$\scriptscriptstyle\tilde\chi^0_{#1}$}}}}}
\newcommand{\chinoonep}{{\ensuremath{\mathchoice%
      {\displaystyle\raise.4ex\hbox{$\displaystyle\tilde\chi^+_1$}}%
         {\textstyle\raise.4ex\hbox{$\textstyle\tilde\chi^+_1$}}%
       {\scriptstyle\raise.3ex\hbox{$\scriptstyle\tilde\chi^+_1$}}%
 {\scriptscriptstyle\raise.3ex\hbox{$\scriptscriptstyle\tilde\chi^+_1$}}}}}
\newcommand{\chinoonem}{{\ensuremath{\mathchoice%
      {\displaystyle\raise.4ex\hbox{$\displaystyle\tilde\chi^-_1$}}%
         {\textstyle\raise.4ex\hbox{$\textstyle\tilde\chi^-_1$}}%
       {\scriptstyle\raise.3ex\hbox{$\scriptstyle\tilde\chi^-_1$}}%
 {\scriptscriptstyle\raise.3ex\hbox{$\scriptscriptstyle\tilde\chi^-_1$}}}}}
\newcommand{\chinoonepm}{{\ensuremath{\mathchoice%
      {\displaystyle\raise.4ex\hbox{$\displaystyle\tilde\chi^\pm_1$}}%
         {\textstyle\raise.4ex\hbox{$\textstyle\tilde\chi^\pm_1$}}%
       {\scriptstyle\raise.3ex\hbox{$\scriptstyle\tilde\chi^\pm_1$}}%
 {\scriptscriptstyle\raise.3ex\hbox{$\scriptscriptstyle\tilde\chi^\pm_1$}}}}}
\newcommand{\chinopmonetwo}{{\ensuremath{\mathchoice%
      {\displaystyle\raise.4ex\hbox{$\displaystyle\tilde\chi^\pm_{1,2}$}}%
         {\textstyle\raise.4ex\hbox{$\textstyle\tilde\chi^\pm_{1,2}$}}%
       {\scriptstyle\raise.3ex\hbox{$\scriptstyle\tilde\chi^\pm_{1,2}$}}%
 {\scriptscriptstyle\raise.3ex\hbox{$\scriptscriptstyle\tilde\chi^\pm_{1,2}$}}}}}
\newcommand{\chinomponetwo}{{\ensuremath{\mathchoice%
      {\displaystyle\raise.4ex\hbox{$\displaystyle\tilde\chi^\mp_{1,2}$}}%
         {\textstyle\raise.4ex\hbox{$\textstyle\tilde\chi^\mp_{1,2}$}}%
       {\scriptstyle\raise.3ex\hbox{$\scriptstyle\tilde\chi^\mp_{1,2}$}}%
 {\scriptscriptstyle\raise.3ex\hbox{$\scriptscriptstyle\tilde\chi^\mp_{1,2}$}}}}}
\newcommand{\chinopm}[1]{{\ensuremath{\mathchoice%
      {\displaystyle\raise.4ex\hbox{$\displaystyle\tilde\chi^\pm_{#1}$}}%
         {\textstyle\raise.4ex\hbox{$\textstyle\tilde\chi^\pm_{#1}$}}%
       {\scriptstyle\raise.3ex\hbox{$\scriptstyle\tilde\chi^\pm_{#1}$}}%
 {\scriptscriptstyle\raise.3ex\hbox{$\scriptscriptstyle\tilde\chi^\pm_{#1}$}}}}}
\newcommand{\chinomp}[1]{{\ensuremath{\mathchoice%
      {\displaystyle\raise.4ex\hbox{$\displaystyle\tilde\chi^\mp_{#1}$}}%
         {\textstyle\raise.4ex\hbox{$\textstyle\tilde\chi^\mp_{#1}$}}%
       {\scriptstyle\raise.3ex\hbox{$\scriptstyle\tilde\chi^\mp_{#1}$}}%
 {\scriptscriptstyle\raise.3ex\hbox{$\scriptscriptstyle\tilde\chi^\mp_{#1}$}}}}}
\newcommand{\stau}{\ensuremath{\tilde{\tau}}\xspace}
\newcommand{\slep}{\ensuremath{\tilde{\ell}}\xspace}

\newcommand{\stauL}{\ensuremath{\tilde{\tau}_{L}}\xspace}
\newcommand{\slepL}{\ensuremath{\tilde{\ell}_{L}}\xspace}

\newcommand{\stauR}{\ensuremath{\tilde{\tau}_{R}}\xspace}
\newcommand{\slepR}{\ensuremath{\tilde{\ell}_{R}}\xspace}
\newcommand{\snu}{\ensuremath{\tilde{\nu}}\xspace}

\def\slepton{\ensuremath{\tilde{\ell}}}

\def\gravitino{\ensuremath{\tilde{G}}}

\newcommand{\mt}{\ensuremath{m_\mathrm{T}}}
\newcommand{\mtlone}{\ensuremath{\mt^\mathrm{lep 1}}}
\newcommand{\mtltwo}{\ensuremath{\mt^\mathrm{lep 2}}}
\newcommand{\Ht}{\ensuremath{H_\mathrm{T}}}
\newcommand{\msfos}{\ensuremath{m_\mathrm{SFOS}}}
\newcommand{\minmsfos}{\ensuremath{m_\mathrm{SFOS}^\mathrm{min}}}
\newcommand{\meff}{\ensuremath{m_\mathrm{eff}}}
\newcommand{\mttwo}{\ensuremath{m_\mathrm{T2}}}

\newcommand{\mtt}{\ensuremath{m_{\tau\tau}}}
\newcommand{\mlll}{\ensuremath{m_{\ell\ell\ell}}}

\newcommand{\pTvec}{\vec{p}_\mathrm{T}}
\newcommand{\qTvec}{\vec{q}_\mathrm{T}}

\newcommand{\metrel}{{\ensuremath{E_{\mathrm{T}}^{\mathrm{miss, rel}}}}}
\newcommand{\lumi}{\unit[20.3]{\ifb}\xspace}
\newcommand{\mDeltaR}{\ensuremath{M_{\Delta}^{R}}}
\newcommand{\dPhiBr}{\ensuremath{\Delta\phi_{R}^{\beta}}}
\newcommand{\rTwo}{\ensuremath{R_{2}}}
\newcommand{\pTll}{\ensuremath{p_{\mathrm{T}}^{\ell\ell}}}

\newcommand{\mll}{\ensuremath{m_{\ell\ell}}}
\newcommand{\dEtajj}{\ensuremath{|\Delta\eta_{jj}|}}
\newcommand{\mjj}{\ensuremath{m_{jj}}}
\newcommand{\pTobj}[2]{\ensuremath{\pt^{\mathrm{#1\,#2}}}}
\newcommand{\ptjone}{\ensuremath{\pTobj{jet}{1}}}
\newcommand{\ptjtwo}{\ensuremath{\pTobj{jet}{2}}}
\newcommand{\ptlone}{\ensuremath{\pTobj{\ell}{1}}}
\newcommand{\ptltwo}{\ensuremath{\pTobj{\ell}{2}}}

\def\TeV{\ifmmode {\mathrm{\ Te\kern -0.1em V}}\else
                   \textrm{Te\kern -0.1em V}\fi}%
\def\GeV{\ifmmode {\mathrm{\ Ge\kern -0.1em V}}\else
                   \textrm{Ge\kern -0.1em V}\fi}%
\def\MeV{\ifmmode {\mathrm{\ Me\kern -0.1em V}}\else
                   \textrm{Me\kern -0.1em V}\fi}%
\def\keV{\ifmmode {\mathrm{\ ke\kern -0.1em V}}\else
                   \textrm{ke\kern -0.1em V}\fi}%
\def\eV{\ifmmode  {\mathrm{\ e\kern -0.1em V}}\else
                   \textrm{e\kern -0.1em V}\fi}%

\def\checkmark{\tikz\fill[scale=0.4](0,.35) -- (.25,0) -- (1,.7) -- (.25,.15) -- cycle;}

\skipbeforetitle{-100pt}

\AtlasTitle{\boldmath Search for the electroweak production of supersymmetric particles in $\sqrt{s}\,=\,$8$\TeV$ $pp$ collisions with the ATLAS detector}
\author{The ATLAS Collaboration}
\PreprintIdNumber{CERN-PH-EP-2015-218}
\AtlasJournal{Phys.\ Rev.\ D.}
\AtlasAbstract{
The ATLAS experiment has performed extensive searches for the electroweak production of charginos, neutralinos and staus. 
This article summarizes and extends the search for electroweak supersymmetry with new analyses targeting scenarios not covered by previously published searches. 
New searches use vector-boson fusion production, initial-state radiation jets, and low-momentum lepton final states, as well as multivariate analysis techniques to improve the sensitivity to scenarios with small mass splittings and low-production cross-sections.
Results are based on 20 \ifb\ of proton--proton collision data at $\sqrt{s}=8\TeV$ recorded with the ATLAS experiment at the Large Hadron Collider.
No significant excess beyond Standard Model expectations is observed.
The new and existing searches are combined and interpreted in terms of 95\% confidence-level exclusion limits in simplified models, where a single production process and decay mode is assumed, as well as within phenomenological supersymmetric models.
}

\begin{document}
\tableofcontents
\clearpage
\FloatBarrier

\section{Introduction  \label{sec:intro}}

Supersymmetry (SUSY)~\cite{Miyazawa:1966,Ramond:1971gb,Golfand:1971iw,Neveu:1971rx,Neveu:1971iv,Gervais:1971ji,Volkov:1973ix,Wess:1973kz,Wess:1974tw} 
is a space-time symmetry that postulates for each Standard Model (SM) particle the existence of a partner state whose spin differs by one-half unit.
The introduction of these new SUSY particles (sparticles) provides a potential solution to the hierarchy problem~\cite{Weinberg:1975gm,Gildener:1976ai,Weinberg:1979bn,Susskind:1978ms}.
If $R$-parity is conserved~\cite{Fayet:1976et,Fayet:1977yc,Farrar:1978xj,Fayet:1979sa,Dimopoulos:1981zb}, as assumed in this article, 
sparticles are always produced in pairs and the lightest supersymmetric particle (LSP) emerges as a stable dark-matter candidate.

The charginos and neutralinos are mixtures of the bino, winos and higgsinos, collectively referred to as the electroweakinos, that are superpartners of the U(1), SU(2) gauge bosons and the Higgs bosons, respectively.
Their mass eigenstates are referred to as $\tilde{\chi}_{i}^{\pm}$ $(i=1,2)$ and $\tilde{\chi}_{j}^{0}$ $(j=1,2,3,4)$ in order of increasing mass.
The direct production of charginos, neutralinos and sleptons ($\slep$) through electroweak (EW) interactions may dominate the SUSY production at the Large Hadron Collider (LHC) if the masses of the gluinos and squarks are large. 
Previous searches for electroweak SUSY production at ATLAS
targeted the production of $\slepton^{+}\slepton^{-}$, $\stau^{+}\stau^{-}$, $\chinoonep\chinoonem$ (decaying through $\slepton$ or $W$ bosons), $\chinoonepm\ninotwo$ (decaying through $\slepton$ or $W$ and $Z/h$ bosons), and $\ninotwo\ninothree$ (decaying through $\slepton$ or $Z$ bosons)~\cite{Aad:2014vma,Aad:2014nua,Aad:2014iza,Aad:2014yka,Aad:2015jqa}, and found no significant excess beyond SM expectations.
These searches are typically sensitive to scenarios where there is a relatively large $O(m_{W,Z})$ splitting between the produced sparticles and the LSP, leaving uncovered territory for smaller mass splittings. 

This article addresses EW SUSY production based on the 20.3 fb$^{-1}$ of $\sqrt{s}=8\TeV$ proton--proton collisions collected by the ATLAS experiment in 2012. 
A series of new analyses targeting regions in parameter space not covered by previous ATLAS analyses~\cite{Aad:2014vma,Aad:2014nua,Aad:2014iza,Aad:2014yka,Aad:2015jqa} are presented. 
The results from new and published searches are combined and reinterpreted to provide the final 8$\TeV$ ATLAS limits on the production of EW SUSY particles in a variety of models. 
The dependence of the limits on the mass of the intermediate slepton in models of electroweakino production with $\slepton$-mediated decays is also studied, thus generalizing the results of Refs.~\cite{Aad:2014nua,Aad:2014vma,Aad:2014iza}.

In cases where the LSP is wino- or higgsino-dominated,  the lighter electroweakino states $\chinoonepm$, $\ninotwo$ can have mass differences with the $\ninoone$ ranging from a few $\MeV$ to a few tens of $\GeV$, depending on the values of the other parameters in the mixing matrix~\cite{Han:2014kaa}. 
In particular, in naturalness-inspired models~\cite{Barbieri:1987fn,deCarlos:1993yy} the higgsino must be light, so the $\ninoone$, $\ninotwo$ and $\chinoonepm$ are usually higgsino-dominated and have a small mass splitting. 
Therefore, a situation with a light $\ninoone$ approximately mass degenerate with the $\chinoonepm$ and $\ninotwo$ has a strong theoretical motivation. 
A relatively low mass splitting between the produced sparticles and the LSP (referred to as compressed scenarios) results in low-momentum decay products that are difficult to reconstruct efficiently, and probing these signatures is experimentally challenging. 
The new analyses introduced in this article improve the sensitivity to the compressed spectra. 
The two- and three-lepton searches for $\chinoonep\chinoonem$ and $\chinoonepm\ninotwo$ production in Refs.~\cite{Aad:2014vma,Aad:2014nua} are extended by lowering the transverse momentum threshold on reconstructed leptons, and by boosting the electroweak SUSY system through the requirement of QCD initial state radiation (ISR). 
The search for the vector-boson fusion (VBF) production of $\chinoonepm\chinoonepm$ 
uses the signature of a same-sign light lepton ($e,\mu$) pair with two jets to probe compressed spectra. 

In many SUSY scenarios with large $\tan\beta$, the stau ($\stau$) is lighter than the selectron and smuon~\cite{Delannoy:2013dla}, resulting in tau-rich final states. 
Co-annihilation processes~\cite{PhysRevD.43.3191} favor a light $\stau$ that has a small mass splitting with a bino LSP, as it can set the relic density to the observed value~\cite{Hinshaw:2012aka}. 
An additional new search is presented here, which uses a final state with two hadronically decaying $\tau$ leptons and multivariate techniques to improve the sensitivity to direct $\stau$ production compared to the search presented in Ref.~\cite{Aad:2014yka}. 

Searches for the electroweak production of SUSY particles have been conducted at the Tevatron~\cite{D0-2009,CDF-2008} and by the CMS Collaboration~\cite{Khachatryan:2014mma,Khachatryan:2014qwa,Khachatryan:2015kxa}. 
At LEP~\cite{LEPSUSYWG:01-03.1,Heister:2003zk,Abdallah:2003xe,Acciarri:1999km,Abbiendi:2003sc}, 
searches set lower limits of $103.5 \GeV$, $99.9 \GeV$, $94.6 \GeV$, and $86.6 \GeV$ at 95\% confidence level (CL) 
on the mass of promptly decaying charginos, selectrons, smuons, and staus respectively. 
For the interval $0.1\,$$\lesssim\Delta m(\chinoonepm,\ninoone)\,$$\lesssim\,$$3 \GeV$, the chargino mass limit set by LEP degrades to $91.9 \GeV$.
The slepton mass limits from LEP assume gaugino mass unification, which is not assumed in the results presented here.

The article is organized as follows: 
Section~\ref{sec:susysignals} describes the signal models studied in this article;
Section~\ref{sec:atlasdet} provides a brief description of the ATLAS detector;
Sections~\ref{sec:mcgen} and~\ref{sec:evtreco} outline the Monte Carlo (MC) simulation and event selection, respectively;
Section~\ref{sec:GenAna} discusses the analysis strategy common to all analyses studied in this article;
Section~\ref{sec:taumvachannel} presents the direct stau production search;
Section~\ref{sec:compressedchannels} presents the compressed spectra searches in direct production;
Section~\ref {sec:samesignvbf} presents the search for same-sign chargino-pair production via VBF;
Section~\ref{sec:interpretation} provides a global overview of the results of the ATLAS searches for electroweakino production at 8$\TeV$, integrating the results of the new analyses with published analyses in the framework of several relevant signal models;
finally conclusions are drawn in Section~\ref{sec:conclusion}.

\section{SUSY scenarios  \label{sec:susysignals}}

The SUSY scenarios considered in this article can be divided into two categories: simplified models and phenomenological models.
The simplified models~\cite{Alwall:2008ag} target the production of charginos, neutralinos and sleptons, 
where the masses and the decay modes of the relevant particles are the only free parameters.  
In each of the simplified models, a single production process with a fixed decay chain is considered 
for optimization of the event selection and interpretation of the results.
To illustrate the range of applicability of the searches, several classes of phenomenological models that consider all relevant SUSY production and decay processes are also used to interpret the results.
These models include the five-dimensional EW phenomenological Minimal Supersymmetric Standard Model (pMSSM)~\cite{Djouadi:1998di},
the Non Universal Higgs Masses (NUHM) model~\cite{Ellis:2002iu,Ellis:2002wv},
and a Gauge-Mediated SUSY Breaking (GMSB) model~\cite{Dine:1981gu,AlvarezGaume:1981wy,Nappi:1982hm,Dine:1993yw,Dine:1994vc,Dine:1995ag}. 

$R$-parity is assumed to be conserved in all SUSY scenarios considered in this article. 
The LSP is assumed to be the lightest neutralino \ninoone\ except in the GMSB scenarios, 
where it is the gravitino \gravitino. 
The next-to-LSP (NLSP) is usually one or more of the charginos, neutralinos or sleptons.
All SUSY particles are assumed to decay promptly, with the exception of the LSP, which is stable. 
Finally, SUSY particles that are not considered in a given model are decoupled by setting their masses to values inaccessible at the LHC.

\begin{figure}\centering
	\includegraphics[width=0.5\textwidth]{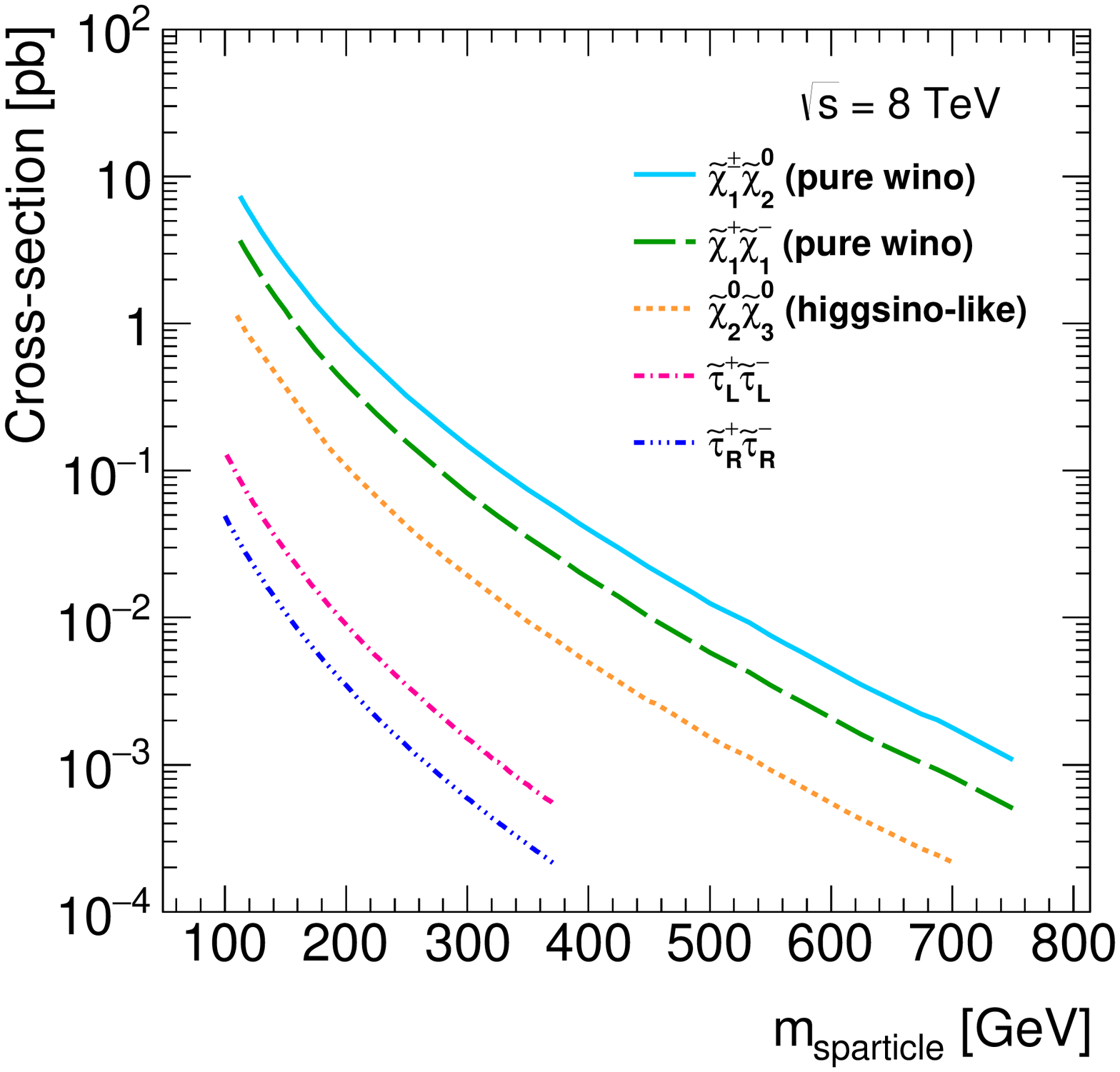}
	\caption{\label{crossSections}
		The production cross-sections for the simplified models of the direct production of \chinoonep\chinoonem, 
		\chinoonepm\ninotwo\ (where m$(\chinoonepm)$=m$(\ninotwo)$), 
		\ninotwo\ninothree\ (where m$(\ninotwo)$=m$(\ninothree)$), and 
		$\stau^{+}\stau^{-}$ studied in this article. 
		The left-handed and right-handed stau-pair production cross-sections are shown separately.}
\end{figure}

Unless stated otherwise, signal cross-sections are calculated to next-to-leading order (NLO) 
in the strong coupling constant using \Prospino~\cite{Beenakker:1996ch},
and are shown in Figure~\ref{crossSections} for a number of selected simplified-model production modes.
The cross-sections for the production of charginos and neutralinos are in agreement with the NLO calculations matched 
to resummation at next-to-leading logarithmic accuracy (NLO+NLL) within about two percent~\cite{Fuks:2012qx,Fuks:2013vua,Fuks:2013lya}.
The nominal cross-section and the uncertainty are taken from the center and spread, respectively, 
of the envelope of cross-section predictions using different parton distribution function (PDF) sets and factorization 
and renormalization scales, as described in Ref.~\cite{Kramer:2012bx}.

\subsection{Direct stau-pair production simplified model}

Two simplified models describing the direct production of $\tilde{\tau}^{+}\tilde{\tau}^{-}$ are used in this article: 
one considers stau partners of the left-handed $\tau$ lepton ($\stauL$), and a second considers stau partners of the right-handed $\tau$ lepton ($\stauR$).
In both models, the stau decays with a branching fraction of 100\% to the SM tau-lepton and the LSP. 
The diagram for this model can be seen in Figure~\ref{diagrams1}(a). 

\subsection{Direct chargino-pair, chargino--neutralino, and neutralino-pair production simplified models}

In the simplified models describing the direct production of \chinoonep\chinoonem\ and \chinoonepm\ninotwo\,, 
both the \chinoonepm\ and \ninotwo\ are assumed to be pure wino and mass-degenerate, while the \ninoone\ is assumed to be pure bino.
However, it is possible to reinterpret the results from these simplified models by assuming different compositions of the $\ninoone$, $\ninotwo$ and $\chinoonepm$ for the same masses of the states.
Two different scenarios for the decays of the \chinoonepm\ and \ninotwo\ are considered, as shown in Figures~\ref{diagrams1}(b) and \ref{diagrams1}(c):

\begin{itemize}[nolistsep]
\item {\bf $\chinoonep\chinoonem$/$\chinoonepm\ninotwo$ production with $\tilde{\ell}_{L}$-mediated decays:} The \chinoonepm\ and \ninotwo\ decay with a branching fraction of 1/6 via
$\tilde{e}_{L}$, $\tilde{\mu}_{L}$, $\tilde{\tau}_{L}$, $\tilde{\nu}_{e}$, $\tilde{\nu}_{\mu}$,  or $\tilde{\nu}_{\tau}$ 
with masses $m_{\tilde{\nu}_{\ell}}=m_{\tilde{\ell}_{L}}= x\  (m_{\chinoonepm} -m_{\ninoone}) + m_{\ninoone}$ with $x$ = 0.05, 0.25, 0.5, 0.75 or 0.95,
\item {\bf $\chinoonepm\ninotwo$ production with $\tilde{\tau}_{L}$-mediated decay:} The first- and second-generation sleptons and sneutrinos are assumed to be very heavy, so that the \chinoonepm\ and \ninotwo\ decay with a branching fraction of 1/2 via $\tilde{\tau}_{L}$ or $\tilde{\nu}_{\tau}$ with masses $m_{\tilde{\nu}_{\tau}}=m_{\tilde{\tau}_{L}}= 0.5  (m_{\chinoonepm} + m_{\ninoone})$.
\end{itemize}
In the simplified models considered here, the slepton mass is assumed to lie between the $\ninoone$ and $\chinoonepm/\ninotwo$ masses, 
which increases the branching fraction to leptonic final states compared to scenarios without sleptons. 

The compressed spectra searches in this article are less sensitive to scenarios where the $\chinoonepm$/$\ninotwo$ decay through SM $W$, $Z$ or Higgs bosons, as the branching fraction to leptonic final states is significantly suppressed. 
The results of the ATLAS searches for $\chinoonep\chinoonem$ production with $WW$-mediated decays~\cite{Aad:2014vma}, $\chinoonepm\ninotwo$ production with $WZ$-mediated decays~\cite{Aad:2014nua} and $\chinoonepm\ninotwo$ production with $Wh$-mediated decays~\cite{Aad:2015jqa} are summarized in Section~\ref{sec:summaryplots}. 
In these scenarios with decays mediated by SM bosons, the $W$, $Z$ and $h$ bosons are assumed to decay with SM branching fractions.

In the simplified models of the direct production of \ninotwo\ninothree\,, the \ninotwo\  and \ninothree\ 
are assumed to be pure higgsino and mass-degenerate, while the \ninoone\ is assumed to be pure bino. 
The \ninotwo\ and \ninothree\ are assumed to decay with a branching fraction of one half via $\tilde{e}_{R}$, $\tilde{\mu}_{R}$ 
with mass $m_{\tilde{\ell}_{R}} = x\  (m_{\ninotwo} -m_{\ninoone}) + m_{\ninoone}$ with $x$ = 0.05, 0.25, 0.5, 0.75 or 0.95 ($\ninotwo\ninothree$ production with $\tilde{\ell}_{R}$-mediated decay).
The associated diagram is shown in Figure~\ref{diagrams1}(d). 
In this \ninotwo\ninothree\, simplified model, the choice of right-handed sleptons in the decay chain ensures high lepton multiplicities in the final state while suppressing the leptonic branching fraction of any associated chargino, thus enhancing the rate of four-lepton events with respect to events with lower lepton multiplicities. 

\subsection{Simplified model of same-sign chargino-pair production via vector-boson fusion}

\begin{figure}[h!]
\centering
  \subfigure[]{\includegraphics[width=0.35\textwidth]{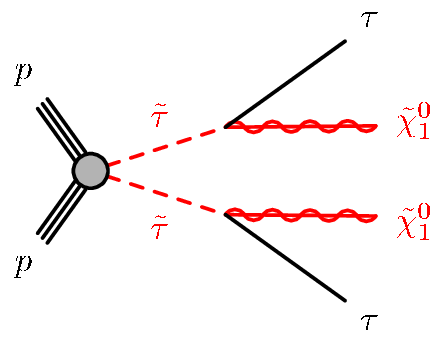}} \hfil
  \subfigure[]{\includegraphics[width=0.35\textwidth]{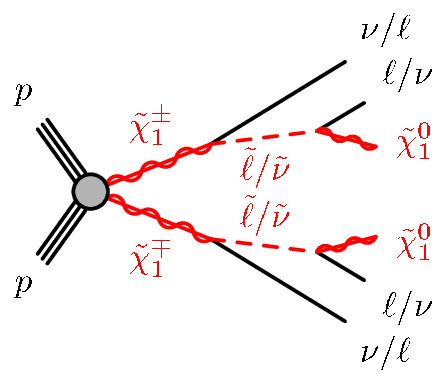}}
  \subfigure[]{\includegraphics[width=0.35\textwidth]{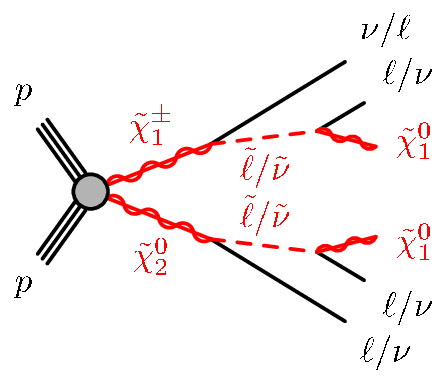}}\hfil
  \subfigure[]{\includegraphics[width=0.35\textwidth]{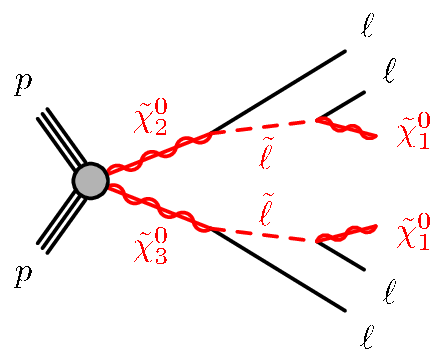}}
  \subfigure[]{\includegraphics[width=0.35\textwidth]{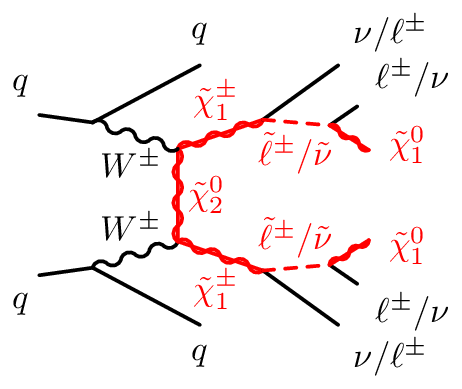}}\hfil
	\caption{\label{diagrams1}
		The diagrams for the simplified models of the direct pair production of staus and 
		the direct production of \chinoonep\chinoonem, \chinoonepm\ninotwo\ and \ninotwo\ninothree, 
		and the VBF production of \chinoonepm\chinoonepm\ studied in this article. 
		All three generations are included in the definition of $\tilde{\ell}/\tilde{\nu}$, 
		except for the direct production of \ninotwo\ninothree\ where only the first two generations are assumed.
		The different decay modes are discussed in the text.}
\end{figure}

A simplified model for \chinoonepm\chinoonepm\ production via VBF~\cite{Cho:2006sx,Dutta:2012xe} is also considered. 
As in the case of direct production, the \chinoonepm\ is assumed to be pure wino, and mass-degenerate with the \ninotwo, 
and the \ninoone\ is assumed to be pure bino. 
The \chinoonepm\ decays with a branching fraction of 1/6 via 
$\tilde{e}_{L}$, $\tilde{\mu}_{L}$, $\tilde{\tau}_{L}$, $\tilde{\nu}_{e}$, $\tilde{\nu}_{\mu}$,  or $\tilde{\nu}_{\tau}$ with masses 
$m_{\tilde{\nu}_{\ell}}=m_{\tilde{\ell}_{L}}= 0.5 (m_{\chinoonepm} + m_{\ninoone})$.
The diagram for \chinoonepm\chinoonepm\ production via VBF, 
where the sparticles are produced along with two jets, is shown in Figure~\ref{diagrams1}(e).  
The jets are widely separated in pseudorapidity\footnote{ATLAS uses a right-handed coordinate system with its origin at the nominal interaction point (IP) 
in the center of the detector and the $z$-axis along the beam pipe. 
The $x$-axis points from the IP to the center of the LHC ring, and the $y$-axis points upward. 
Cylindrical coordinates $(r,\phi)$ are used in the transverse plane, $\phi$ being the azimuthal angle around the $z$-axis. 
The pseudorapidity is defined in terms of the polar angle $\theta$ as $\eta=-\ln\tan(\theta/2)$.}
 $\eta$ and have a relatively high dijet invariant mass \mjj.
Due to the VBF topology, the charginos are often boosted in the transverse plane, forcing the decay products to be more collinear and energetic, even in highly compressed spectra.
This feature of VBF production makes it a good candidate to probe compressed SUSY scenarios 
that are experimentally difficult to explore via the direct production modes.
The  signal cross-sections are calculated to leading order (LO) 
in the strong coupling constant using \Madgraph5-1.3.33 ~\cite{Alwall:2007st} (more details on the cross-section calculation are given in Appendix~\ref{sec:app_vbfxsec}). 
The uncertainties on the signal cross-sections are calculated by using different PDF sets (2\%) and by 
varying the renormalization and factorization scales between 0.5 and 2 times the nominal values (6\%)~\cite{Alwall:2014hca}. 
For a \chinoonepm\ with mass of $120\GeV$, the cross-section for \chinoonepm\chinoonepm\ production in association with two jets satisfying the criteria \mjj$\,>350\GeV$ and \dEtajj$\,>1.6$ is 1.1\,fb. 
For the assumed mixings in the chargino--neutralino sector, and the mass values considered in the analysis,
the cross-section for $\chinoonepm\chinoonepm$ VBF production is found to be independent of the $\ninoone$ mass.

\subsection{Phenomenological Minimal Supersymmetric Standard Model \label{sec:ModelpMSSM}}

The analysis results are interpreted in a pMSSM scenario. 
The masses of the sfermions, the gluino, and of the CP-odd Higgs boson are set to high values ($2\TeV$, $2\TeV$ and $500\GeV$ respectively), thus decoupling the production of these particles and allowing only the direct production of charginos and neutralinos decaying via SM gauge bosons and the lightest Higgs boson. 
The remaining four parameters, the ratio of the expectation values of the two Higgs doublets ($\tan\beta$), the gaugino mass parameters $M_{1}$ and $M_{2}$, and the higgsino mass parameter $\mu$, determine the phenomenology of direct electroweak SUSY production. 
For the analysis presented here, $\mu$ and $M_{2}$ are treated as free parameters. 
The remaining parameters are fixed to tan $\beta=10$ and $M_{1}=50\GeV$, so that the relic dark-matter density is below the 
cosmological bound~\cite{Hinshaw:2012aka} across most of the $\mu$--$M_{2}$ grid. 
The lightest Higgs boson has a mass close to $125\GeV$, which is set by tuning the mixing in the top squark sector, 
and decays to SUSY as well as SM particles where kinematically allowed. 

\subsection{Two-parameter Non Universal Higgs Masses model}

Radiatively-driven natural SUSY~\cite{Baer:2013xua} 
allows the $Z$ and Higgs boson masses to be close to
$100\GeV$, with gluino and squark masses beyond the \TeV\ scale.
In the two-parameter NUHM model (NUHM2) that is considered in this article, the direct production of charginos and neutralinos is dominant in a large area of the parameter space considered. 
The mass hierarchy, composition and production cross-section of the SUSY particles
are governed by the universal soft SUSY-breaking scalar mass $m_0$, the soft SUSY-breaking gaugino mass $m_{1/2}$, 
the trilinear SUSY-breaking parameter $A_0$, the pseudoscalar Higgs boson mass $m_A$, $\tan\beta$ and $\mu$.
Both $\mu$ and $m_{1/2}$ are treated as free parameters and the other parameters are fixed to $m_0\,=\,5\TeV$, $A_0\,=\,-1.6\,m_0$, $\tan\beta\,=\,15$, $m_A = 1 \TeV$, and sign($\mu$) $>$ 0.
These conditions ensure a low level of electroweak fine tuning, while keeping the lightest Higgs boson mass close to $125\GeV$ 
and the squark masses to a few \TeV. The gluino mass typically satisfies $m_{\tilde{g}}\simeq 2.5m_{1/2}$.
For low gluino masses, the production of strongly interacting SUSY particles dominates;
as the gluino mass increases the production of electroweakinos becomes more important.
The charginos and neutralinos decay via $W$, $Z$ and Higgs bosons.

\subsection{Gauge-Mediated SUSY Breaking model}

Minimal GMSB models are described by six parameters: the SUSY-breaking mass scale in the low-energy sector ($\Lambda$),
the messenger mass ($M_{\mathrm{mess}}$), 
the number of SU(5) messenger fields ($N_{5}$), 
the scale factor for the gravitino mass ($C_{\mathrm{grav}}$), 
$\tan\beta$, and $\mu$. 
In the model presented here, $\Lambda$ and $\tan\beta$ are treated as free parameters, and the remaining parameters are fixed to $M_{\mathrm{mess}}=250\TeV$, $N_{5}=3$, $C_{\mathrm{grav}}=1$ and sign($\mu$) $>0$. 
For high $\Lambda$ values, the EW production of SUSY particles dominates over other SUSY processes. 
In most of the relevant parameter space, the NLSP is the $\stau$ for large values of $\tan\beta$ ($\tan\beta$>20), and the final states contain two, three or four tau-leptons. 
In the region where the mass difference between the stau and selectron/smuon is smaller than the sum of the tau and the electron/muon masses, the stau, selectron and smuon decay directly into the LSP and a lepton, defining the phenomenology.
The charginos and neutralinos decay as $\chinoonepm$$\rightarrow$$W^{\pm} \ninoone$ and $\ninotwo$$\rightarrow$$ Z \ninoone$, where the $\ninoone$ decays as $\ninoone$$\rightarrow$$\ell^{\pm}\slepton^{\mp}$$\rightarrow$$\ell^{+}\ell^{-}\gravitino$ and the LSP is the gravitino \gravitino.

\section{The ATLAS detector  \label{sec:atlasdet}}
The ATLAS detector~\cite{atlas-det} is a multipurpose particle physics detector with forward-backward symmetric cylindrical geometry.
 The inner tracking detector (ID) covers $|\eta|\,$$<\,$2.5 and consists of a silicon pixel detector, a semiconductor microstrip detector, and a transition radiation tracker. The ID is surrounded by a thin superconducting solenoid providing a 2$\,$T axial magnetic field. A high-granularity lead/liquid-argon sampling calorimeter measures the energy and the position of electromagnetic showers within $|\eta|\,$$<\,$3.2. 
Sampling calorimeters with liquid argon as the active medium are also used to measure hadronic showers in the endcap (1.5$\,$$<\,$$|\eta|\,$$<\,$3.2) and forward (3.1$\,$$<\,$$|\eta|\,$$<\,$4.9) regions, while a steel/scintillator tile calorimeter measures hadronic showers in the central region ($|\eta|\,$$<\,$1.7). 
The muon spectrometer (MS) surrounds the calorimeters and consists of three large superconducting air-core toroid magnets, each with eight coils, a system of precision tracking chambers ($|\eta|\,$$<\,$2.7), and fast trigger chambers ($|\eta|\,$$<\,$2.4). 
A three-level trigger system~\cite{atlas} selects events to be recorded for offline analysis.

\section{Monte Carlo simulation \label{sec:mcgen} }
Monte Carlo (MC) generators are used to simulate SM processes and new physics signals. 
The SM processes considered are those that can lead to leptonic signatures. 
Details of the signal and background MC simulation samples used in this article, as well as the order of cross-section calculations in perturbative QCD used for yield normalization are shown in Table~\ref{tab:MCsamples}.

\begin{sidewaystable}[!h]
  \centering
  \caption{The MC simulation samples used in this article for background and signal estimates. Shown are the generator type, the order of cross-section calculations used for yield normalization, the names of the sets of tunable parameters (tunes) used for the underlying-event generation, and the PDF sets. 
  \label{tab:MCsamples} }
  \footnotesize{
    \begin{tabular}{ ccccc }
      \toprule
Process & Generator & Cross-section  & Tune   & PDF set  \\
 & + fragmentation/hadronization & & & \\
\midrule
{\bf Diboson (VV)} & & & & \\
         $W^{+}W^{-}$, $WZ$, $ZZ$ & \POWHEGBOX-r2129~\cite{Nason:2004rx,Frixione:2007vw} & NLO QCD  & \AU~\cite{Pythia8tunes} & \CTTen~\cite{CT10pdf} \\ 
 & + \PYTHIA-8.165~\cite{Sjostrand:2006za} & with \MCFM-6.2~\cite{mcfm1,mcfm2}  & & \\
 & (or + \PYTHIA-6.426) &   & & \\
         $W^{\pm}W^{\pm}$ & \Sherpa-1.4.0~\cite{Sherpa} & NLO  & (\Sherpa\ internal) & \CTTen \\ 
         $W^{\pm}W^{\pm}$ via vector-boson fusion & \Sherpa-1.4.0 & NLO  & (\Sherpa\ internal) & \CTTen \\ 
         $ZZ$, $W^{+}W^{-}$ via gluon fusion & \ggVV~\cite{Kauer:2012hd}  & NLO & \AUET~\cite{mc11ctunes} & \CTTen \\ 
 {\it ~~~~(not incl. in \POWHEGBOX)} &  + \Herwig-6.520 & & & \\ 
         $W\gamma$, $Z\gamma$ & \Sherpa-1.4.1 & NLO & (\Sherpa\ internal) & \CTTen \\  
\midrule
{\bf Triboson (VVV)} & & & & \\
        $WWW$, $ZWW$ & \Madgraph5-1.3.33 + \PYTHIA-6.426 & NLO~\cite{Campanario:2008yg} & \AUET & \CTEQSixL~\cite{Pumplin:2002vw} \\ 
\midrule
{\bf Higgs} & & & & \\
      via gluon fusion & \POWHEGBOX-r2092 + \PYTHIA-8.165  & NNLO+NNLL QCD, NLO EW~\cite{Dittmaier:2012vm} & \AU & \CTTen  \\ 
      via vector-boson fusion & \POWHEGBOX-r2092 + \PYTHIA-8.165 & NNLO QCD, NLO EW~\cite{Dittmaier:2012vm} & \AU & \CTTen  \\ 
      associated $W$/$Z$ production & \PYTHIA-8.165 & NNLO QCD, NLO EW~\cite{Dittmaier:2012vm} & \AU & \CTEQSixL \\
      associated $\ttbar$-production & \PYTHIA-8.165 & NNLO QCD~\cite{Dittmaier:2012vm} & \AU & \CTEQSixL \\
\midrule
{\bf Top+Boson $\ttbar V$} & & & & \\
          $\ttbar W$, $\ttbar Z$ & \Alpgen-2.14~\cite{Mangano:2002ea} + \Herwig-6.520 & NLO~\cite{ttZ, ttW} & \AUET & \CTEQSixL \\ 
          $\ttbar WW$ & \Madgraph5-1.3.33 + \PYTHIA-6.426 &  NLO~\cite{ttW} & \AUET &  \CTEQSixL \\ 
\midrule
\boldmath $\ttbar$ & \POWHEGBOX-r2129 + \PYTHIA-6.426 & NNLO+NNLL~\cite{Cacciari:2011hy,Baernreuther:2012ws,Czakon:2012zr,Czakon:2012pz,Czakon:2013goa,Czakon:2011xx} & \Perugia~\cite{Skands:2010ak} & \CTTen  \\ 
\midrule
{\bf Single top} & & & & \\
        $t$-channel & \AcerMC-38~\cite{Kersevan:2004yg} + \PYTHIA-6.426 & NNLO+NNLL~\cite{Kidonakis:2011wy}  & \AUET & \CTEQSixL \\  
        $s$-channel, $Wt$ &  \Mcatnlo-4.06~\cite{Frixione:2005vw,Frixione:2008yi} + \Herwig-6.520 & NNLO+NNLL~\cite{Kidonakis:2010tc, Kidonakis:2010ux} & \AUET & \CTTen \\
        $tZ$ & \Madgraph5-1.5.11 + \PYTHIA-6.426 &  NLO~\cite{tZ} & \AUET &  \CTEQSixL \\   
\midrule
{\bf \boldmath $W$+jets, $Z$+jets} & \Alpgen-2.14 + \PYTHIA-6.426 & NNLO QCD using DYNNLO-1.1~\cite{Catani:2009sm} & \Perugia & \CTEQSixL \\ 
 & (or + \Herwig-6.520) & with MSTW2008 NNLO~\cite{Martin:2009iq} & & \\
 & or \Sherpa-1.4.0 & NNLO QCD using DYNNLO-1.1 & & \CTTen \\
 &&with MSTW2008 NNLO&& \\
\midrule
{\bf Low-mass resonances } & & & & \\
      $J/\Psi$, $\Upsilon$ & \PYTHIA-8.165  & NLO & \AU & \CTEQSixL  \\ 
\midrule
 {\bf SUSY signal} & & & & \\
 $\stau\stau$, $\chinoonep\chinoonem$, $\chinoonepm\ninotwo$ simplified models & \Herwigpp-2.5.2~\cite{herwigplusplus} & NLO using \Prospino~\cite{Beenakker:1996ch} & \UEEE~\cite{ueee3} & \CTEQSixL \\
 $\ninotwo\ninothree$ simplified models & \Madgraph5-1.5.12 + \PYTHIA-6.426 & NLO using \Prospino & \AUET & \CTEQSixL \\
 VBF $\chinoonep\chinoonem$ simplified models & \Madgraph5\_\aMcAtNlo-2.1.1 + \PYTHIA-6.426 & LO using \Madgraph5-1.3.33~\cite{Alwall:2007st} & \AUET & \CTEQSixL \\
 NUHM2, GMSB & \Herwigpp-2.5.2 & NLO using \Prospino & \UEEE & \CTEQSixL \\
      \bottomrule
    \end{tabular} 
  }       
\end{sidewaystable}

For all MC simulation samples, the propagation of particles through the ATLAS detector is modeled with \GEANT4~\cite{Agostinelli:2002hh} using the full ATLAS detector simulation~\cite{:2010wqa}, or a fast simulation using a parametric response of the electromagnetic and hadronic calorimeters~\cite{atlfastII} and \GEANT4 elsewhere. 
The effect of multiple proton--proton collisions in the same or nearby beam bunch crossings (in-time and out-of-time pileup) is incorporated into the simulation by overlaying additional minimum-bias events generated with \PYTHIA-8 onto hard-scatter events.  
Simulated events are weighted to match the distribution of the mean number of interactions per bunch crossing in data, and are reconstructed in the same manner as data. 
The simulated MC samples are corrected to account for differences with respect to the data in the heavy-flavor quark jet selection efficiencies and misidentification probabilities, lepton efficiencies, tau misidentification probabilities, as well as the energy and momentum measurements of leptons and jets. 
The $\chinoonep\chinoonem$ ($\chinoonepm\ninotwo$) signal samples simulated with \Herwigpp\  are reweighted to match the 
$\chinoonep\chinoonem$ ($\chinoonepm\ninotwo$) system transverse momentum distribution obtained from the \Madgraph\ samples that 
are generated with an additional parton in the matrix element to give a better description of the ISR.

\FloatBarrier 

\section{Event reconstruction\label{sec:evtreco}}
Events recorded during stable data-taking conditions are analyzed if the reconstructed primary vertex has five or more tracks with transverse momentum $\pt\,$$>\,$400$\MeV$ associated with it.
The primary vertex of an event is identified as the vertex with the highest $\Sigma \pt^2$ of associated tracks.
After the application of beam, detector and data-quality requirements, the total luminosity considered in these analyses corresponds to \lumi (20.1$\,$fb$^{-1}$ for the direct stau production analysis due to a different trigger requirement).

Electron candidates are required to have $|\eta|\,$$<\,$2.47 and $\pt\,$$>\,$7$ \GeV$, where the $\pt$ and $\eta$ are determined from the calibrated clustered energy deposits in the electromagnetic calorimeter and the matched ID track, respectively.
Electrons must satisfy ``medium'' identification criteria, following Ref.~\cite{Aad:2014nim}. 
Muon candidates are reconstructed by combining tracks in the ID and tracks in the MS~\cite{Aad:2014rra}, and are required to have $|\eta|\,$$<\,$2.5 and $\pt\,$$>\,$5$ \GeV$. 
Events containing one or more muons that have transverse impact parameter with respect to the primary vertex $|d_0|\,$$>\,$0.2$\,$mm or longitudinal impact parameter with respect to the primary vertex $|z_0|\,$$>\,$1$\,$mm are rejected to suppress cosmic-ray muon background. 
In the direct stau production analysis, and the two-lepton compressed spectra analyses, electrons and muons are required to have $\pt\,$$>\,$10$ \GeV$.

Jets are reconstructed with the anti-$k_t$ algorithm~\cite{Cacciari:2008gp} with a radius parameter of $R\,$$=\,$0.4. 
Three-dimen\-sional calorimeter energy clusters are used as input to the jet reconstruction. 
The clusters are calibrated using the local hadronic calibration~\cite{Aad:2014bia}, which gives different weights to the energy deposits from the electromagnetic and hadronic components of the showers. 
The final jet energy calibration corrects the calorimeter response to the particle-level jet energy~\cite{Aad:2014bia,Aad:2012vm}, where correction factors are obtained from simulation and then refined and validated using data. 
Corrections for in-time and out-of-time pileup are also applied based on the jet area method~\cite{Aad:2014bia}. 
Central jets must have $|\eta|\,$$<\,$2.4 and $\pt\,$$>\,$20$ \GeV$, and a ``jet vertex fraction''~\cite{Aad:2014bia} (JVF) larger than 0.5 if $\pt\,$$<\,$50$ \GeV$. 
The JVF is the $\pt$-weighted fraction of the tracks in the jet that are associated with the primary vertex. 
Requiring large JVF values suppresses jets from pileup. 
Forward jets are those with $2.4<|\eta|<4.5$ and $\pT>30\GeV$. 
Events containing jets failing to satisfy the quality criteria described in Ref.~\cite{Aad:2014bia} are rejected to suppress events with large calorimeter noise and noncollision backgrounds. 

Central jets are identified as containing $b$-hadrons (referred to as $b$-tagged) using a multivariate technique based on quantities related to reconstructed secondary vertices. 
The chosen working point of the $b$-tagging algorithm~\cite{btag} correctly identifies $b$-hadrons in simulated $\ttbar$ samples with an efficiency of 80\%, with a light-flavor jet misidentification probability of about 4\% and a $c$-jet misidentification probability of about 30\%.

Hadronically decaying $\tau$ leptons ($\tau_{\rm had}$) are reconstructed using jets described above with $|\eta|\,$$<\,$2.47 and a lower $\pt$ threshold of 10$ \GeV$. 
The $\tau_{\rm had}$ reconstruction algorithm uses information about the tracks within $\Delta R \equiv \sqrt{(\Delta\phi)^2+(\Delta\eta)^2}\,$$=\,$0.2 of the seed jet, in addition to the electromagnetic and hadronic shower shapes in the calorimeters. 
The $\tau_{\rm had}$ candidates are required to have one or three associated tracks (prongs), as $\tau$ leptons predominantly decay to either one or three charged pions together with a neutrino and often additional neutral pions. 
The $\tau_{\rm had}$ candidates are required to have $\pt\,$$>\,$20$ \GeV$ and unit total charge of their constituent tracks.
A boosted decision tree algorithm (BDT) uses discriminating track and cluster variables to optimize $\tau_{\rm had}$ identification, where ``loose'', ``medium'' and ``tight'' working points are defined~\cite{Aad:2014rga}. 
Electrons misidentified as $\tau_{\rm had}$ candidates are vetoed using transition radiation and calorimeter information. 
The $\tau_{\rm had}$ candidates are corrected to the $\tau$ energy scale~\cite{Aad:2014rga} using an $\eta$- and $\pt$-dependent calibration. 
Kinematic variables built using taus in this article use only the visible decay products from the hadronically decaying tau. 

The missing transverse momentum is the negative vector sum of the transverse momenta of all muons with $\pt\,$$>\,$10$ \GeV$, electrons with $\pt\,$$>\,$10$ \GeV$, photons with $\pt\,$$>\,$10$ \GeV$~\cite{Aad:2014nim}, jets with $\pt\,$$>\,$20$ \GeV$, and calibrated calorimeter energy clusters with $|\eta|\,$$<\,$4.9 not associated with these objects. 
Hadronically decaying $\tau$ leptons are included in the $\met$ calculation as jets. 
Clusters associated with electrons, photons and jets are calibrated to the scale of the corresponding objects. 
Calorimeter energy clusters not associated with these objects are calibrated using both calorimeter and tracker information~\cite{Aad:2012re}.
For jets, the calibration includes the pileup correction described above, whilst the JVF requirement is not considered when selecting jet candidates. 

To avoid potential ambiguities among objects, ``tagged'' leptons are candidate leptons separated from each other and from jets in the following order:
\vspace{-5pt}
\begin{enumerate}[nolistsep]
\item If two electron candidates are reconstructed with $\Delta R\,$$<\,$0.1, the lower energy candidate is discarded.
\item Jets within $\Delta R\,$$=\,$0.2 of an electron candidate,
and $\tau_{\rm had}$ candidates within $\Delta R\,$$=\,$0.2 of an electron or muon, are discarded.
\item Electron and muon candidates are discarded if found within $\Delta R\,$$=\,$0.4 of a remaining jet to suppress leptons from semileptonic decays of $c$- and $b$-hadrons. 
\item To reject bremsstrahlung from muons, $e\mu$ ($\mu\mu$) pairs are discarded if the two leptons are within $\Delta R\,$$=\,$0.01 (0.05) of one another.
\item jets found within $\Delta R\,$$=\,$0.2 of a ``signal'' $\tau$ lepton (see below) are discarded.
\end{enumerate}
Finally, to suppress low-mass decays, if tagged electrons and muons form a same-flavor opposite-sign (SFOS) pair with $\msfos\,$$<\,$2$ \GeV$, both leptons in the pair are discarded. 

Tagged leptons satisfying additional identification criteria are called ``signal'' leptons. 
To maximize the search sensitivity, some analyses presented in this article require different additional criteria for signal leptons and these are highlighted where necessary. 
Signal $\tau$ leptons must satisfy ``medium'' identification criteria~\cite{Aad:2014rga}, while for the final signal-region selections, both the ``medium'' and ``tight'' criteria are used. 
Unless stated otherwise, signal electrons (muons) are tagged electrons (muons) for which the scalar sum of the transverse momenta of tracks within a cone of $\Delta R\,$$=\,$0.3 around the lepton candidate is less than 16\% (12\%) of the lepton $\pt$. 
Tracks used for the electron (muon) isolation requirement defined above are those that have $\pt\,$$>\,$0.4 (1.0)$ \GeV$ and $|z_0| \,$$<\,$2$\,$mm with respect to the primary vertex of the event. 
Tracks of the leptons themselves as well as tracks closer in $z_0$ to another vertex (that is not the primary vertex) are not included. 
The isolation requirements are imposed to reduce the contributions from semileptonic decays of hadrons and jets misidentified as leptons.
Signal electrons must also satisfy ``tight'' identification criteria~\cite{Aad:2014nim} and the sum of the extra transverse energy deposits in the calorimeter (corrected for pileup effects) within a cone of $\Delta R = 0.3$ around the electron candidate must be less than 18\% of the electron $\pt$. 
To further suppress electrons and muons originating from secondary vertices, the $d_0$ normalized to its uncertainty is required to be small, with $|d_{0}|/\sigma(d_{0})\,<\,5\,(3)$, and $|z_{0}\sin{\theta}|\,$$<\,$0.4$\,$mm (1$\,$mm) for electrons (muons).

Events must satisfy the relevant trigger for the analysis, and satisfy the corresponding $\pt$-threshold requirements shown in Table~\ref{tab:triggers}. 

\begin{table}[h]
  \begin{center}
  \caption{The triggers used in the analyses and the offline $\pt$ threshold used, ensuring that the lepton(s) or $\met$ triggering the event are in the plateau region of the trigger efficiency. 
Where multiple triggers are listed for an analysis, events are used if any of the triggers is passed. 
Muons are triggered within a restricted range of $|\eta|<2.4$.
  \label{tab:triggers} }
  \small{
    \begin{tabular}{ cc c }
      \toprule
      Trigger   & $\pt$ threshold [$\GeV$] & Analysis \\
      \midrule
      Single $\tau$ & 150 & \multirow{2}{*}{Direct stau production} \\
      Double $\tau$ & 40,25 & \\
      \midrule
      Single Isolated $e$ & 25 & \multirow{2}{*}{Compressed spectra $\ell^{+}\ell^{-}$, 3$\ell$} \\
      Single Isolated $\mu$ & 25 & \\
      \midrule
      \multirow{2}{*}{Double $e$} & 14,14 & \multirow{2}{*}{Compressed spectra $\ell^{+}\ell^{-}$, $\ell^{\pm}\ell^{\pm}$, 3$\ell$} \\
                                  & 25,10 &  \\
      \midrule
      \multirow{2}{*}{Double $\mu$} &  14,14  & \multirow{2}{*}{Compressed spectra $\ell^{+}\ell^{-}$, $\ell^{\pm}\ell^{\pm}$, 3$\ell$} \\
                                    & 18,10 &  \\
      \midrule
      \multirow{1}{*}{Triple $e$} & 20,9,9 & \multirow{1}{*}{Compressed spectra 3$\ell$} \\
      \midrule
      \multirow{2}{*}{Triple $\mu$} &  7,7,7  & \multirow{2}{*}{Compressed spectra 3$\ell$} \\
                                    & 19,5,5 &  \\
      \midrule
      \multirow{4}{*}{Combined $e\mu$ } & 14($e$),10($\mu$) & \multirow{4}{*}{Compressed spectra 3$\ell$} \\
                                        & 18($\mu$),10($e$)  & \\
                                        & 9($e$),9($e$),7($\mu$)  & \\
                                        & 9($e$),7($\mu$),7($\mu$)  & \\
      \midrule
      $\met$ & 120 & Chargino production via VBF \\
      \bottomrule
    \end{tabular} 
  }       
  \end{center}
\end{table}

\section{General analysis strategy \label{sec:GenAna}}

The broad range of EW SUSY scenarios considered by the ATLAS experiment is accompanied by a large number of experimental signatures: from the two-tau signature from direct stau production, to three-lepton signatures from $\chinoonepm\ninotwo$ production. 
As much as possible the individual analyses follow a common approach. 
Signal regions (SR) are defined to target one or more EW SUSY scenarios, using kinematic variables with good signal--background separation, as described in Section~\ref{sec:evtvarbles}. 
The optimization of key selection variables is performed by maximizing the expected sensitivity to the signal model. 
A common background estimation strategy is used for the analyses in this article: 
 the main SM backgrounds are estimated by normalizing MC simulation samples to data in dedicated control regions (CRs);  
 backgrounds due to non-prompt and fake leptons are derived from data as outlined in Section~\ref{sec:matrixmethod}, while small backgrounds are estimated purely using MC simulation samples. 
The \HistFitter~\cite{Baak:2014wma} software framework is used in all analyses for constraining the background normalizations and the statistical interpretation of the results.

The CRs are defined with kinematic properties similar to the SRs, yet are disjoint from the SR, and have high purity for the background process under consideration.
The CRs are designed in a way that minimizes the contamination from the signal model and cross-contamination between multiple CRs is taken into account in the normalization to data.
To validate the modeling of the SM backgrounds, the yields and shapes of key kinematic variables are compared to data in validation regions (VR). 
The VRs are defined to be close to, yet disjoint from the SR and CR, and be dominated by the background process under consideration. 
The VRs are designed such that the contamination from the signal model is low.
Three different fit configurations are used. 
The ``background-only fit'' is used for estimating the expected background in the SRs and VRs using observations in the CRs, with no assumptions made on any signal model.
In the absence of an observed excess of events in one or more signal regions, the ``model-dependent signal fit'' is used to set exclusion limits in a particular model, where the signal contribution from the particular model that is being tested is taken into account in all CR and SR.
Finally, in the ``model-independent signal fit'', both the CRs and SRs are used in the same manner as for the model-dependent signal fit, but signal contamination is not accounted for in the CRs. 
A likelihood function is built as the product of Poisson probability functions, 
describing the observed and expected number of events in the CRs and SRs. 
The observed number of events in various CRs and SRs are used in a combined 
profile likelihood fit to determine the expected SM background yields in each of the SRs. 
The systematic uncertainties on the expected background yields described in Section~\ref{sec:commsyste} are included as nuisance parameters, constrained to be Gaussian with a width determined by the size of the uncertainty. 
Correlations between control and signal regions, and background processes, are taken into account with common nuisance parameters. 
The free parameters and the nuisance parameters are determined by maximizing the product of the Poisson probability functions and the Gaussian constraints on the nuisance parameters. 

After the background modeling is understood and validated, the predicted background in the SR is compared to the observed data. 
In order to quantify the probability for the background-only hypothesis to fluctuate to the observed number of events or higher, the one-sided $p_0$-value is calculated. 
For this calculation, the profile likelihood ratio is used as a test statistic to exclude the signal-plus-background hypothesis if no significant excess is observed.
A signal model can be excluded at 95\% confidence level (CL) if the CL$_s$~\cite{Read:2002hq} of the signal plus background hypothesis is $<$0.05. 
For each signal region, the expected and observed upper limits at 95\% CL on the number of beyond-the-SM events ($S^{95}_{\rm exp}$ and $S^{95}_{\rm obs}$) are calculated using the model-independent signal fit.
The 95\% CL upper limits on the signal cross-section times efficiency ($\langle\epsilon{\rm \sigma}\rangle_{\rm obs}^{95}$) and the CL$_{b}$ value for the background-only hypothesis are also calculated for each analysis in this article. 

\subsection{Event variables \label{sec:evtvarbles}}

A large set of discriminating variables is used in the analysis strategies presented here. The following kinematic variables are defined and their use in the various analyses is detailed in Sections~\ref{sec:taumvachannel}--\ref{sec:samesignvbf}:
\begin{description}
  \item[$\pT^{X}$] The transverse momentum of a reconstructed object $X$.
  \item[$\Delta\phi(X,Y)$, $\Delta\eta(X,Y)$] The separation in $\phi$ or $\eta$ between two reconstructed objects $X$ and $Y$, e.g. $\Delta\phi(\met,\,\ell)$.
  \item[$\dEtajj$] The separation in $\eta$ between the leading two jets. 
  \item[$\met$] The magnitude of the missing transverse momentum in the event.
  \item[$\metrel$] The quantity $\metrel$ is defined as 
\begin{equation}
	\metrel =  \left\{   
	\begin{array}{ll}  
	   \met & \quad \text{if $\Delta \phi(\met,\ell/j) \geq \pi$/2} \\
	   \met\times\sin{\Delta \phi(\met,\ell/j)} & \quad \text{if  $\Delta \phi(\met,\ell/j) < \pi$/2} \\
    \end{array} \right., \nonumber
\end{equation}
where $\Delta \phi(\met,\ell/j)$ is the azimuthal angle between the direction of $\met$ and that of the nearest  electron,  muon, or central jet. 
  \item[$\pTll$] The transverse momentum of the two-lepton system.
  \item[$\Ht$] The scalar sum of the transverse momenta of the leptons and jets in the event.
  \item[$\mt$] The transverse mass formed using the $\met$ and the leading lepton or tau in the event
\begin{equation}
 m_{\rm T}(\pTvec^{\, \ell/\tau},\met)=\sqrt{2 \pt^{\, \ell/\tau} \met - 2 \pTvec^{\, \ell/\tau} \cdot \met}.  \nonumber
\end{equation}
In the three-lepton analysis, the lepton not forming the SFOS lepton pair with mass closest to the $Z$ boson mass is used. 
In cases where the second lepton or tau is used, the variable is labeled as $\mt^{X}$, where $X$ is the object used with the $\met$ to form the transverse mass.
  \item[$\msfos$] The invariant mass of the SFOS lepton pair in the event. 
In the three-lepton analysis, the SFOS pair with mass closest to the $Z$ boson mass is used.
  \item[$\minmsfos$] The lowest $\msfos$ value among the possible SFOS combinations. 
  \item[$\mlll$] The three-lepton invariant mass.
  \item[$\mtt$] The two-tau invariant mass.
  \item[$\mttwo$] The ``stransverse mass'' is calculated as 
\begin{equation}
   \mttwo = \min_{\qTvec}\left[\max\left(\mt(\vec{\pT}^{\ell1/\tau1}, \,\qTvec),\mt(\vec{\pT}^{\ell2/\tau2}, \,\met-\qTvec)\right)\right],  \nonumber
\end{equation}
where $\ell 1/\tau 1$ and $\ell 2 / \tau 2$ denote the highest- and second-highest-\pt\ leptons or taus in the event, respectively, 
and $\qTvec$ is a test transverse vector that minimizes the larger of the two transverse masses $\mt$. 
The $\mttwo$ distribution has a kinematic endpoint for events where two massive pair-produced particles each decay to two particles, one of which is detected and the other escapes undetected~\cite{Lester:1999tx,Barr:2003rg}. 
  \item[$\meff$] The scalar sum of the transverse momenta of the signal leptons, taus, jets and $\met$ in the event:
\begin{equation}
   \meff = \met + \Sigma \pt^{\rm leptons} + \Sigma \pt^{\rm taus}  + \Sigma \pt^{\rm jets}.  \nonumber
\end{equation}
In the case of the two-tau analysis, only the sum of the $\met$ and two taus is used.
\item[$\rTwo$] The quantity $\rTwo$ is defined as 
\begin{equation}
	\rTwo = \frac{\met}{\met+\ptlone+\ptltwo}.  \nonumber
\end{equation}
The \rTwo\ distribution is shifted towards unity for signal events compared to the background, due to the existence of the LSPs 
that results in a larger \met. 
\item[$\mDeltaR$, $\dPhiBr$] The super-razor quantities $\mDeltaR$ and $\dPhiBr$ are defined in Ref.~\cite{Buckley:2013kua}.
These variables are motivated by the generic process of the pair production of two massive particles, 
each decaying into a set of visible and invisible particles (i.e. \chinoonepm$\rightarrow\ell\nu_{\ell}$\ninoone). 
Similar to \mttwo, \mDeltaR\ is sensitive to the squared mass difference of the pair-produced massive particle and the invisible particle, via 
a kinematic endpoint. These two variables are expected to provide a similar performance for discriminating the signal from the background.
For systems where the invisible particle has a mass that is comparable to the pair-produced massive particle (i.e. compressed spectra), 
the variable \dPhiBr\ has a pronounced peak near $\pi$. The effect is magnified as the spectrum becomes more and 
more compressed, making this variable a good discriminator for compressed spectra searches.

\end{description}

\subsection{Common reducible background estimation~\label{sec:matrixmethod}}

Electron and muon candidates can be classified into three main types, depending on their origin: ``real'' leptons are prompt and isolated leptons from a $W$ or $Z$ boson, a prompt tau or a SUSY particle decay; 
``fake'' leptons can originate from a misidentified light-flavor quark or gluon jet (referred to as ``light flavor''); 
``non-prompt'' leptons can originate from a semileptonic decay of a heavy-flavor quark, from the decay of a meson, or an electron from a photon conversion. 
The background due to non-prompt and fake electrons and muons, collectively referred to as ``reducible'', is commonly estimated using the matrix method described in Ref.~\cite{Aad:2010ey}. 
The matrix method extracts the number of events with one or two fake or non-prompt leptons from a system of linear equations relating the number of events with two signal or tagged leptons (before signal lepton identification requirements are applied) to the number of events with two candidates that are either real, fake or non-prompt. 
The coefficients of the linear equations are functions of the real-lepton identification efficiencies and of the fake and non-prompt lepton misidentification probabilities, both defined as a fraction of the corresponding tagged leptons satisfying the signal lepton requirements. 

The real-lepton identification efficiencies are obtained from MC simulation samples in the region under consideration to account for detailed kinematic dependencies and are multiplied by correction factors to account for residual differences with respect to the data. 
The correction factors are obtained from a control region rich in $Z\rightarrow e^{+}e^{-}$ and  $Z\rightarrow \mu^{+}\mu^{-}$ decays.
The fake and non-prompt lepton misidentification probabilities are calculated as the weighted averages of the corrected type- and process-dependent misidentification probabilities defined below according to their relative contributions in a given signal or validation region.
The type- and process-dependent misidentification probabilities for each relevant fake and non-prompt lepton type (heavy-flavor, light-flavor or conversion) and for each reducible background process are corrected using the ratio (``correction factor'') of the misidentification probability in data to that in simulation obtained from dedicated control samples. 
The correction factors are assumed to be independent of the selected regions and of any potential composition or kinematic differences. 
For non-prompt electrons and muons from heavy-flavor quark decays, the correction factor is measured in a $b\bar{b}$-dominated control sample. 
The correction factor for the conversion candidates is determined in events with a converted photon radiated from a muon in $Z\rightarrow\mu\mu$ decays.

\subsection{Common systematic uncertainties \label{sec:commsyste}}
Several sources of systematic uncertainty are considered for the SM background estimates and signal yield predictions. 
When the MC simulation samples are normalized to data yields in the CR, there is a partial cancellation of both the experimental and theoretical modeling systematic uncertainties. 

The experimental systematic uncertainties affecting the simulation-based estimates include: 
the uncertainties due to the jet energy scale and resolution~\cite{Aad:2014bia,Aad:2014rra};
the uncertainties due to the lepton energy scale, energy resolution and identification efficiency~\cite{Aad:2014nim,Aad:2014rra,Aad:2014rga}; 
the uncertainty due to the hadronic tau misidentification probability~\cite{Aad:2014rga}; 
the uncertainty on the $\met$ from energy deposits not associated with reconstructed objects ($\met$ soft-term resolution)~\cite{Aad:2012re};
and the uncertainties due to $b$-tagging efficiency and mistag probability~\cite{btag}.  
The uncertainty on the integrated luminosity is $\pm$2.8\% and is derived following the same methodology as that detailed in Ref.~\cite{Aad:2011dr}. 
The uncertainty due to the modeling of the pileup in the MC simulation samples is estimated by varying the distribution of the number of interactions per bunch crossing overlaid in the MC samples by $\pm$10\%.
An uncertainty is applied to MC samples to cover differences in efficiency observed between the trigger in data and the MC trigger simulation. 

The systematic uncertainties due to the limitations in theoretical models or calculations affecting the simulation-based background estimates include:
the cross-section uncertainties that are estimated by varying the renormalization and factorization scales and the PDFs,
and the acceptance uncertainties due to PDFs and the choice of MC generator and parton shower. 
The cross-section uncertainties for the irreducible backgrounds used here are 
30\% for $\ttbar V$~\cite{ttW,ttZ}, 
50\% for $tZ$, 
5\% for $ZZ$, 7\% for $WZ$ 
and 100\% for the triboson samples. 
For the Higgs boson samples, a 20\% uncertainty is used for $VH$ and VBF production, 
while a 100\% uncertainty is assigned to $\ttbar H$ and Higgs boson production via gluon fusion~\cite{Dittmaier:2012vm}. 
For the $\chinoonep\chinoonem$ and $\chinoonepm\ninotwo$ signal simulations that are sensitive to ISR, the impact of the choice of renormalization scales, factorization scales, the scale for the first emission in the so-called MLM matching scheme~\cite{Alwall:2007fs}, and MLM matching scale are evaluated by varying these individually between 0.5 and 2 times the nominal values in \Madgraph.

\FloatBarrier

\section{Direct stau production  \label{sec:taumvachannel}}

This section presents a search for direct stau-pair production with subsequent decay into final states with two taus and $\met$. 
The search for direct stau production is very challenging, as the final state is difficult to trigger on and to separate from the SM background. 
In Ref.~\cite{Aad:2014yka}, the best observed upper limit on the direct stau production cross-section was found for a stau mass of 80$\GeV$ and a massless $\ninoone$, where the theoretical cross-section at NLO is 0.07 (0.17) pb for right-handed (left-handed) stau-pair production and the excluded cross-section is 0.22 (0.28) pb. 
This analysis is an update of Ref.~\cite{Aad:2014yka}, using a multivariate analysis technique instead of a simple cut-based method to improve the sensitivity to direct stau-pair production. 

\subsection{Event selection}

Events are selected using the basic reconstruction, object and event selection criteria described in Section~\ref{sec:evtreco}. 
In addition, if taus form an SFOS pair with $\msfos\,$$<\,$12$ \GeV$, the event is rejected. 
Events with exactly two hadronically decaying tau candidates are selected, where the two tau candidates are required to have opposite-sign (OS) charge. 
At least one tau must satisfy the ``tight'' tau identification BDT requirement and events with additional tagged light leptons are vetoed. 
Events must satisfy either the single-tau or ditau trigger criteria, as described in Section~\ref{sec:evtreco}. 

To suppress events from $Z$ boson decays, events are rejected if the invariant mass of the tau pair lies within $\pm 10\GeV$ of the peak value of 81$\GeV$ for $Z$ boson candidates.\footnote{The $Z$ boson mass in di-tau decays is
reconstructed lower than the $Z$ boson mass value due to the neutrinos from the tau decay.}
To suppress background from events containing a top quark, events with $b$-tagged jets are vetoed. 
To further select SUSY events from direct stau production and suppress $WW$ and $\ttbar$ production, $\mttwo$ is calculated using the two taus and the $\met$ in the event. 
The additional requirement of $\mttwo\,$$>\,$30$\GeV$ is applied to select events for the training and optimization of the multivariate analysis (MVA).

 After applying the preselection listed above, both the signal and background MC samples are split in two. Half is used for the BDT training and the other half for testing.
Twelve variables with good discriminatory power are considered as input for the BDT training procedure: $\met$, $\meff$, $\mttwo$, $m_{\tau\tau}$, $\Delta\phi(\tau,\tau)$, $\Delta\eta(\tau,\tau)$, $p_{\rm T}^{\tau 1}$, $p_{\rm T}^{\tau 2}$, $m_{{\rm T}\tau 1}$, $m_{{\rm T}\tau 2}$, $\Delta\phi(\met,\tau 1)$ and   $\Delta\phi(\met,\tau 2)$. 
The MC simulation samples are compared to data for these variables and their correlations to ensure that they are modeled well.

A direct stau production scenario with $m$($\stauR$,$\ninoone$)\,=\,(109,0)$\GeV$ is used for the training and optimization of the BDT, and the BDT response requirement ($t_{\rm cut}$) is chosen based on the best expected sensitivity for discovery. 
The two-tau MVA SR definition is shown in Table~\ref{tab:tauSR-def}.

\begin{table}[h]
\centering
\caption{ Two-tau MVA signal region and validation region definitions for the direct stau-pair production analysis, where $t_{\rm cut}$ is the BDT response requirement.  \label{tab:tauSR-def}}
\small{
\begin{tabular}{c c| cc | cc}
\toprule
      & \multicolumn{5}{c}{Common}   \\
\midrule
& \multicolumn{5}{c}{exactly 2 medium OS taus} \\
& \multicolumn{5}{c}{$\geq$ 1 tight tau} \\
& \multicolumn{5}{c}{tagged $\ell$ veto} \\  
& \multicolumn{5}{c}{$b$-jet veto} \\  
& \multicolumn{5}{c}{$Z$-veto} \\ 
\midrule
& Signal region SR & Multi-jet VR1 & Multi-jet VR2 & $W$-VR1 & $W$-VR2 \\
\midrule
$\mttwo$ & $>\,$30$\GeV$  &  30--50$\GeV$ &  50--80$\GeV$ & $>\,$30$\GeV$  & $>\,$30$\GeV$ \\
$\met$ & --  &  -- &  --&                                   $>\,$100$\GeV$  & $>\,$90$\GeV$ \\
$t_{\rm cut}$ & $>\,$0.07 & $<\,$0.07 & $<\,$0.07  & $-$0.2--0.07 & $-$0.2--0.07 \\
\bottomrule
\end{tabular}}
\end{table}

\subsection{Background determination \label{sec:TauMVA_BgEst}}

The main SM backgrounds in the two-tau MVA SR are $W$+jets and diboson production. Contributions from diboson, $\ttbar$, and $Z$+jets processes are estimated using MC simulation samples and validated using data in $WW$-rich, $\ttbar$-rich or $Z$-rich validation regions, as defined in Ref.~\cite{Aad:2014yka}. 

The $W$+jets contribution in the signal region is dominated by events where the $W$ decays to a tau-lepton and a jet is misidentified as another tau.
The contribution is estimated by normalizing the yields from MC simulation samples to data in a dedicated control region. 
The $W$+jets control region selects events with the $W$ boson decaying to a muon and neutrino to suppress the multi-jet background, which is larger for the electron channel. 
Events containing exactly one isolated muon and one tau satisfying the tight identification requirement are selected, where the muon and tau must have opposite electrical charge. 
To reduce the contribution from $Z$+jets production, $\mt^{\tau} + \mt^{\mu}\,$$>\,$80$\GeV$ is required, and the reconstructed invariant mass of the muon and tau must be outside the $Z$ mass window (12$\GeV$$<\,$$m_{\tau \mu}\,$$<\,$40$\GeV$ or $m_{\tau \mu}\,$$>\,$100$\GeV$). 
To further suppress multi-jet and $Z$+jets processes, $\met\,$$>\,$40$\GeV$ is required, and the muon and tau must not be back-to-back ($\Delta\phi(\tau,\mu)$$\,<$2.7 and $\Delta\eta(\tau,\mu)$$\,<$2.0). 
The contribution from events with top quarks is suppressed by rejecting events containing $b$-tagged jets. 
The multi-jet background in the $W$+jets control region is estimated using a region with the same requirements, but with a same-sign muon and tau. 
The contribution from other SM processes is subtracted using MC simulation samples, and the ratio of opposite-sign muon and tau events to same-sign events is assumed to be unity for the multi-jet background. 

The contribution from multi-jet events in the signal region, where both selected taus are misidentified jets, is small and is estimated using the so-called ABCD method. 
Four exclusive regions (A, B, C, D) are defined in a two-dimensional plane as a function of the two uncorrelated discriminating variables $\mttwo$ and the tau identification criterion.
The regions A and B are required to have two medium taus where at least one meets the tight tau identification criteria, while regions C and D are required to have two loose taus that fail to satisfy the tight tau identification criteria. 
In regions A and C (B and D) $\mttwo\,$$>\,$30$\GeV$ ($\mttwo\,$$<\,$20$\GeV$) is also required. 
The multi-jet background in signal region A can be estimated from $N_{\rm A} = N_{\rm C} \times N_{\rm B}/N_{\rm D}$, where $N_{\rm A}$, $N_{\rm B}$, $N_{\rm C}$, and $N_{\rm D}$ are the numbers of events in regions A, B, C and D respectively. 
The assumption that the ratios $N_{\rm A}/N_{\rm C}$ and $N_{\rm B}/N_{\rm D}$ are the same is confirmed using MC simulation samples and in validation regions using data.
 
A simultaneous likelihood fit to the multi-jet estimation and $W$+jets CR is performed to normalize the corresponding background estimates and obtain the expected yields in the SR (as described in Section~\ref{sec:GenAna}). 
After the simultaneous fit, the multi-jet and $W$+jets normalization factors are found to be $1.4^{+2.5}_{-1.4}$ and 0.98$\pm$0.30 respectively. 
Due to the small number of events in some of the ABCD regions, the uncertainty on the multi-jet normalization factor is large; however, the multi-jet contribution to the total background is very small and the effect on the total signal region background uncertainty is small.

\begin{figure}[h]
\centering
\subfigure[]{\includegraphics[width=0.49\textwidth]{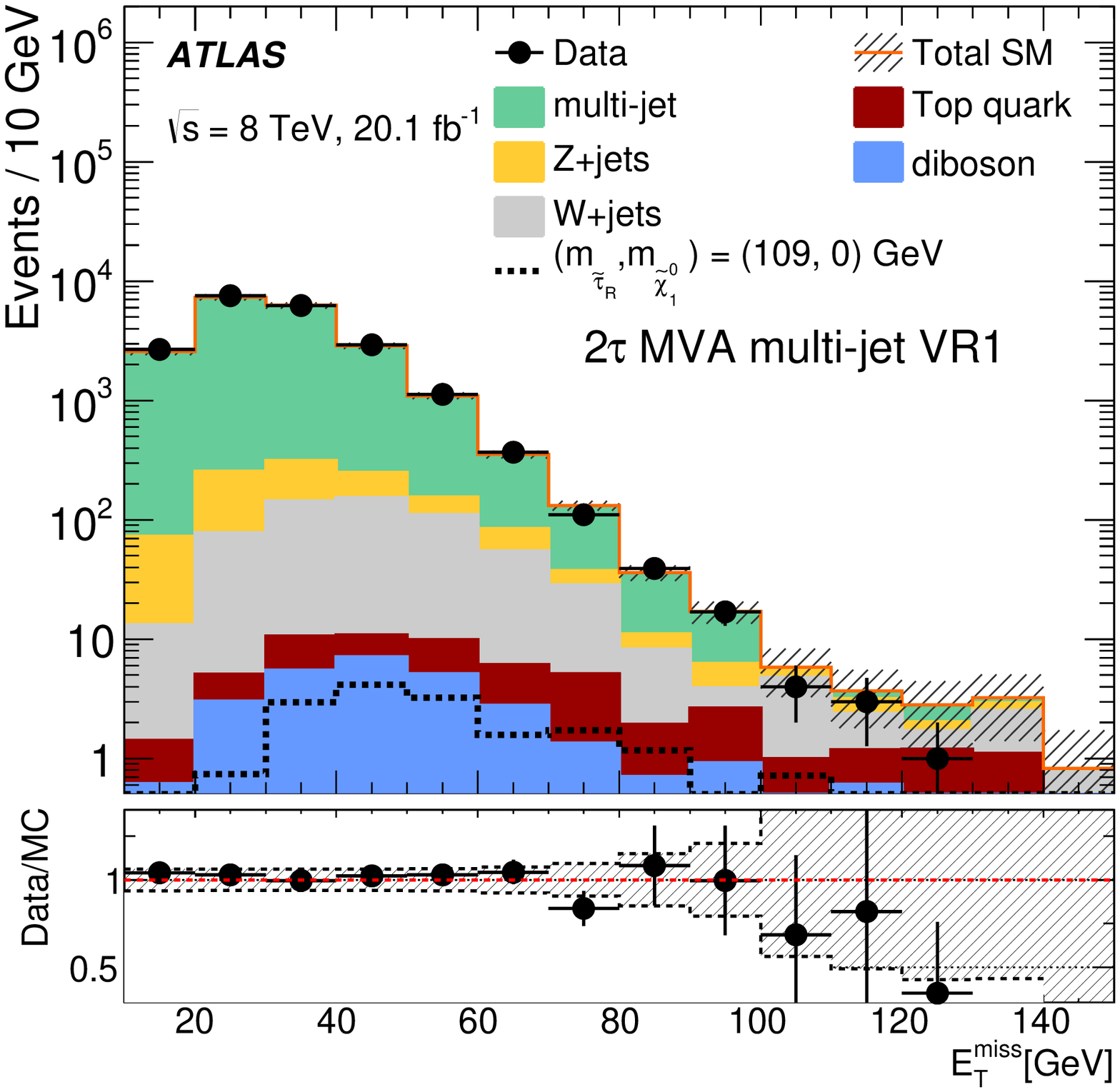}} 
\subfigure[]{\includegraphics[width=0.49\textwidth]{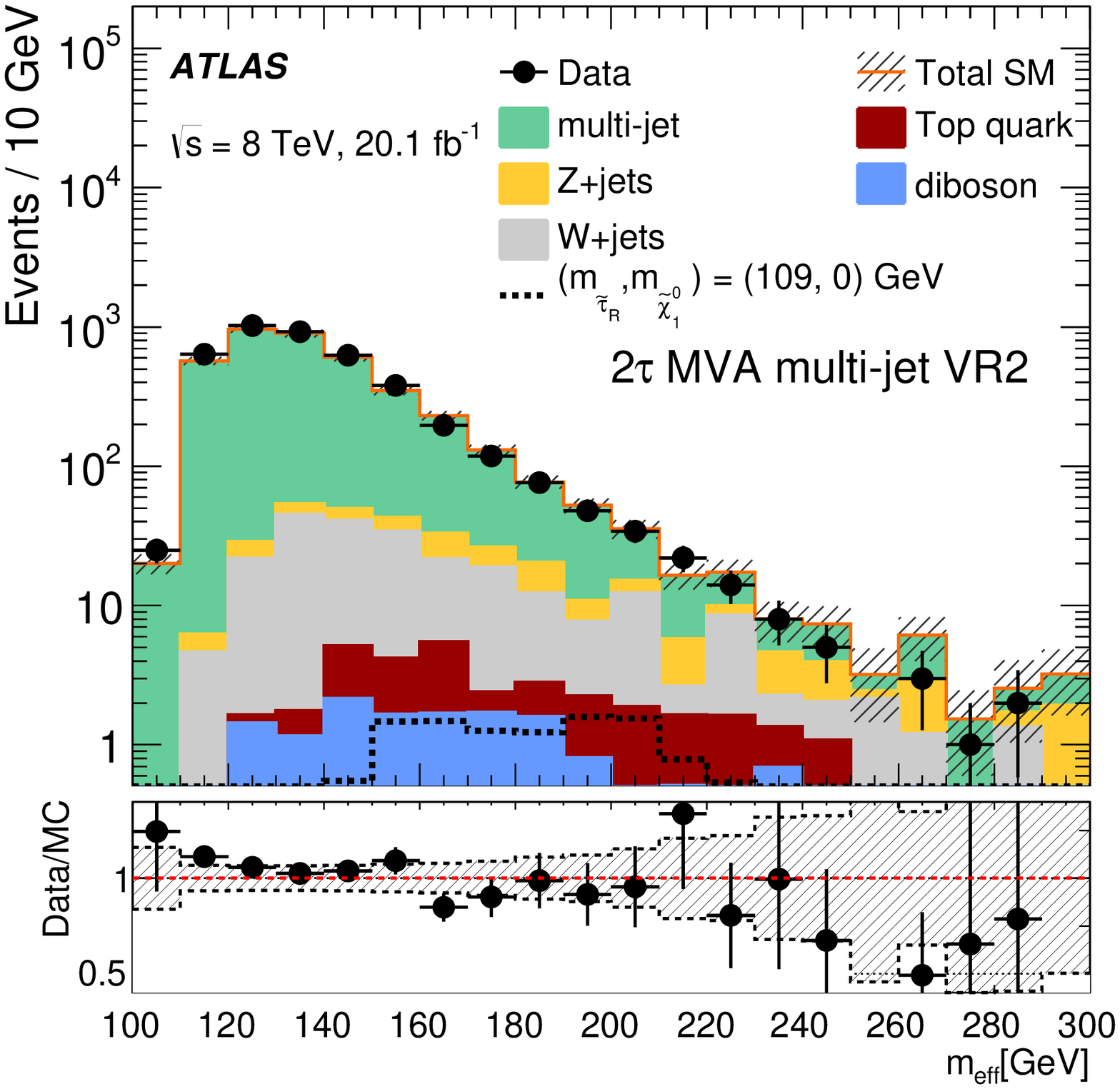}}
\subfigure[]{\includegraphics[width=0.49\textwidth]{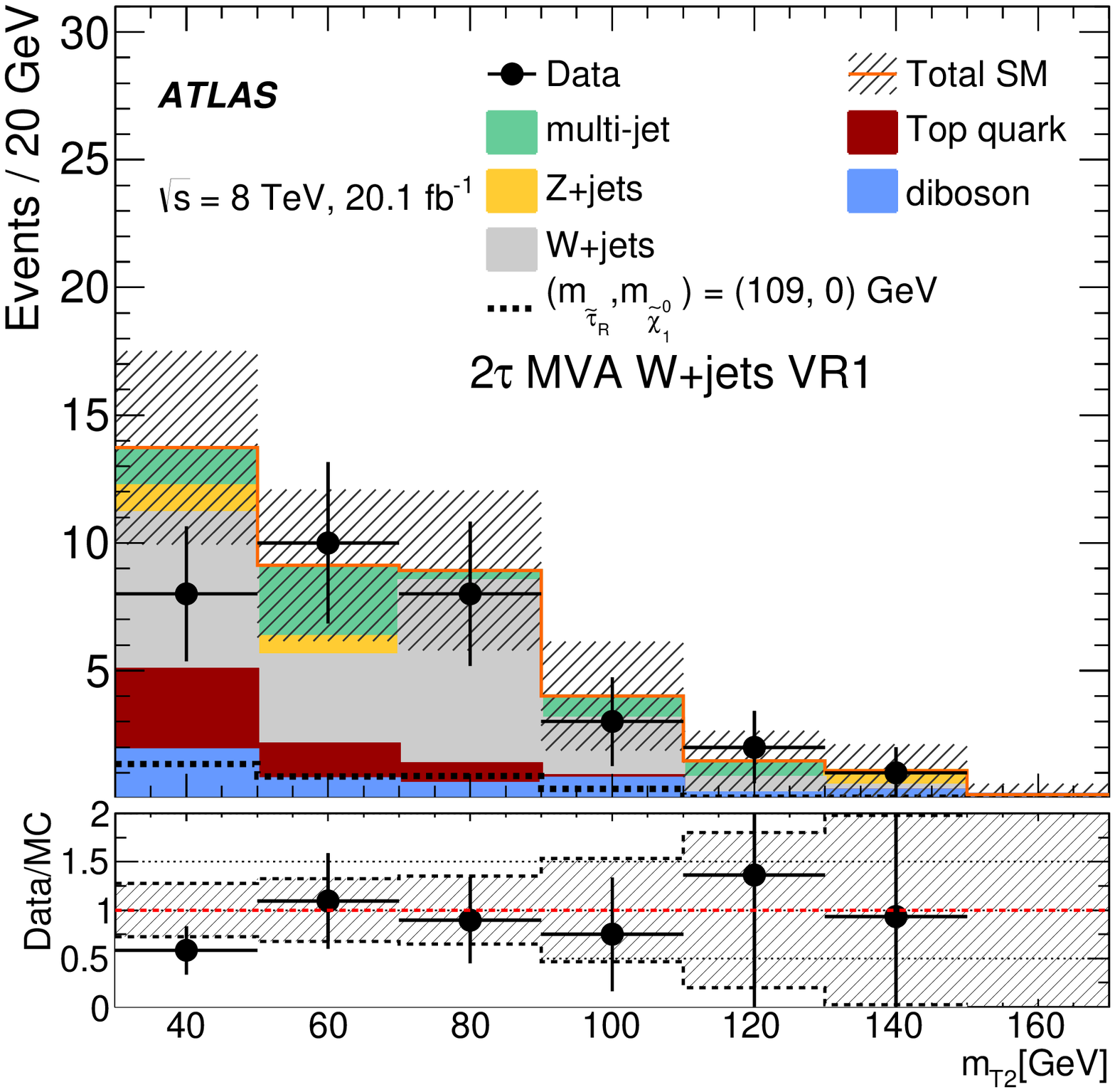}}
\subfigure[]{\includegraphics[width=0.49\textwidth]{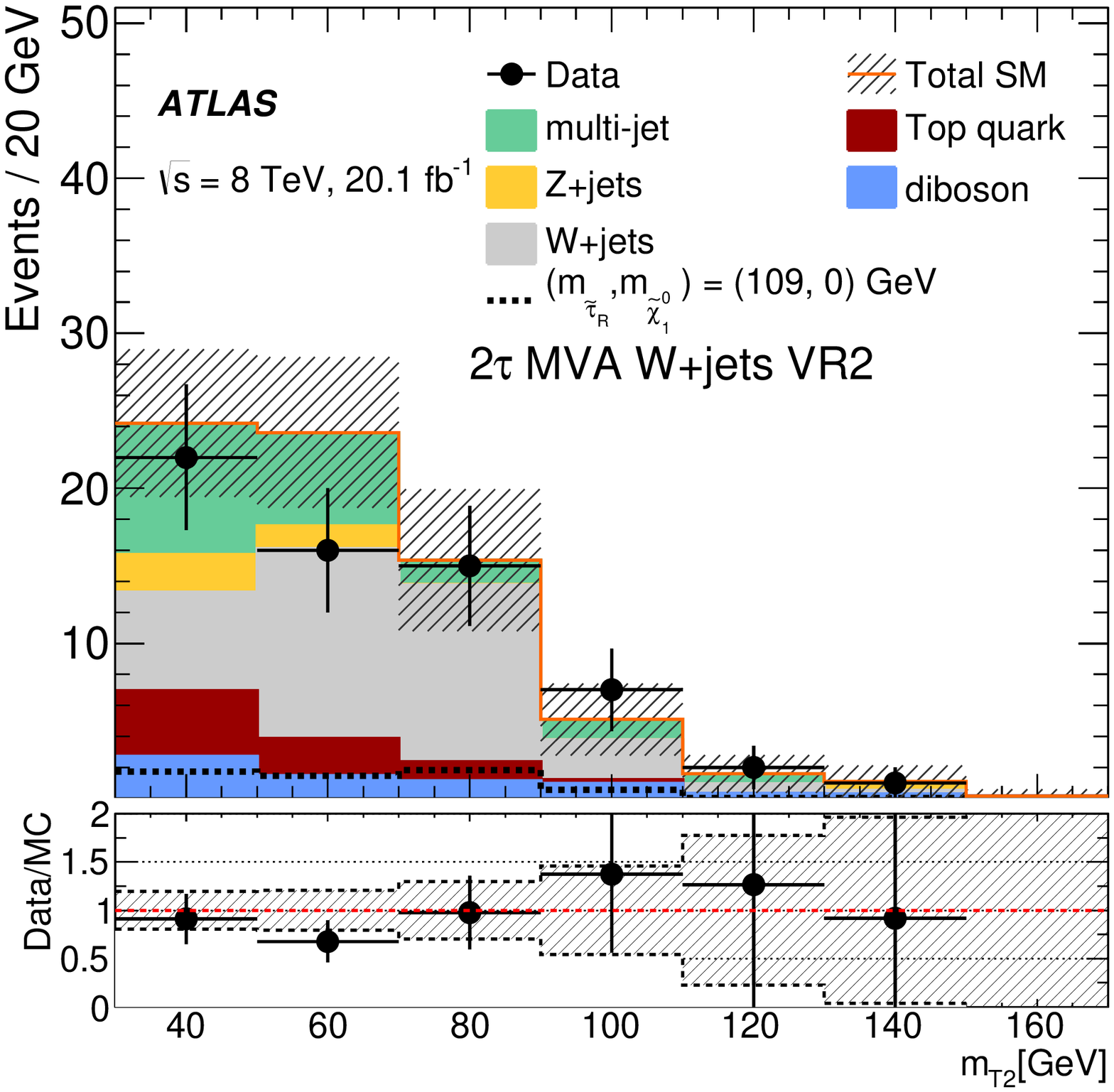}}
\caption{
Distributions in the two-tau MVA validation regions: (a) missing transverse momentum $\met$ in multi-jet VR1, (b) effective mass $\meff$ in multi-jet VR2, (c) stransverse mass $\mttwo$ in $W$-VR1, and (d) $\mttwo$ in $W$-VR2. 
The lower panel of each plot shows the ratio of data to the SM background prediction. 
The last bin in each distribution includes the overflow.
The uncertainty band includes both the statistical and systematic uncertainties on the SM prediction. 
\label{fig:tauVR-summary}}
\end{figure}

Two multi-jet validation regions are defined with the same selection as for the signal region, but with $t_{\rm cut} < 0.07$ and intermediate $\mttwo$. 
These multi-jet validation regions are enriched in events with jets misidentified as hadronic tau decays and good agreement is seen between the data and expectation across the BDT input kinematic variables. 
A further two validation regions are defined to check the modeling of the $W+$jets background. 
The intermediate BDT region $-0.2\,$$<\,$$t_{\rm cut}\,$$<\,$$0.07$ is used, with a high $\met$ selection, where the $W+$jets background is seen to be modeled well. 
The validation region definitions are shown in Table~\ref{tab:tauSR-def}. 
Table~\ref{tab:tauVRSR-results} and Figures~\ref{fig:tauVR-summary}(a), \ref{fig:tauVR-summary}(b), \ref{fig:tauVR-summary}(c), and \ref{fig:tauVR-summary}(d) show the agreement between data and expectation in the validation regions. 
The purity of the multi-jet and $W$+jets validation regions is $\sim$90\% and $\sim$50\% respectively, while the signal contamination from the $m$($\stauR$,$\ninoone$)\,=\,(109,0)$\GeV$ scenario is $<$1\% and $<$10\% respectively.

\subsection{Results}

\begin{figure}[h]
\centering
\subfigure[]{\includegraphics[width=0.49\textwidth]{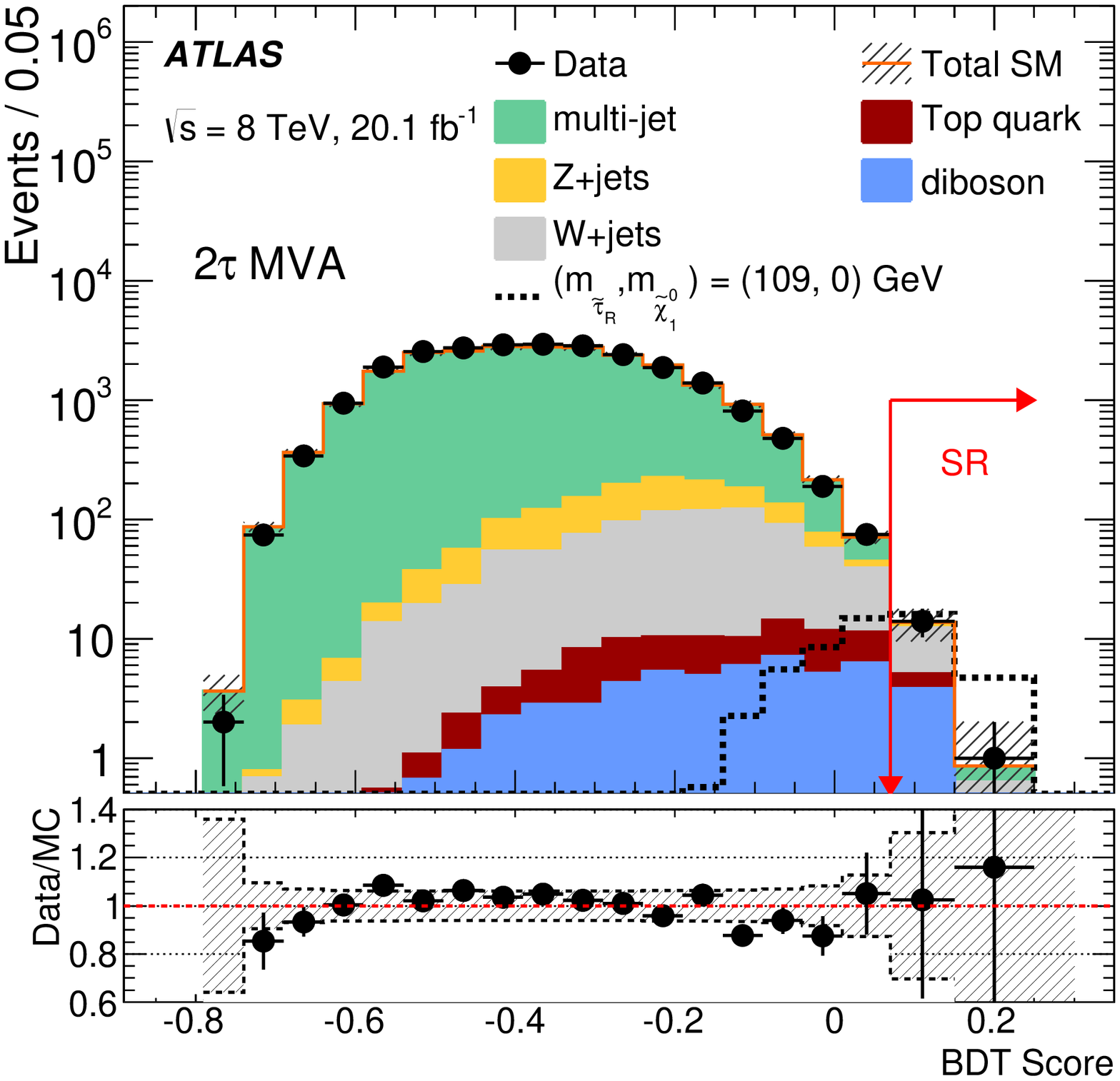}}
\subfigure[]{\includegraphics[width=0.49\textwidth]{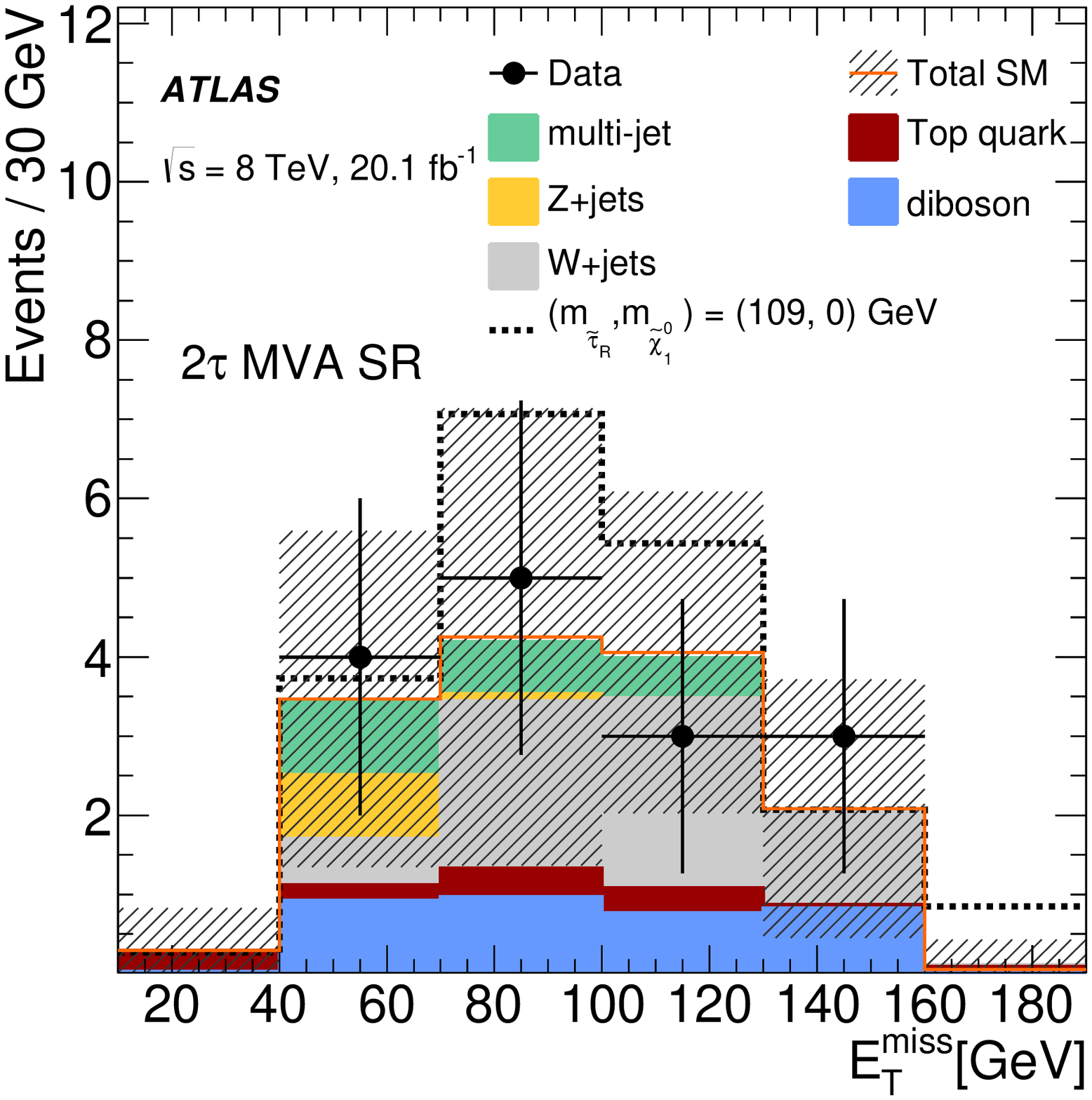}}
\subfigure[]{\includegraphics[width=0.49\textwidth]{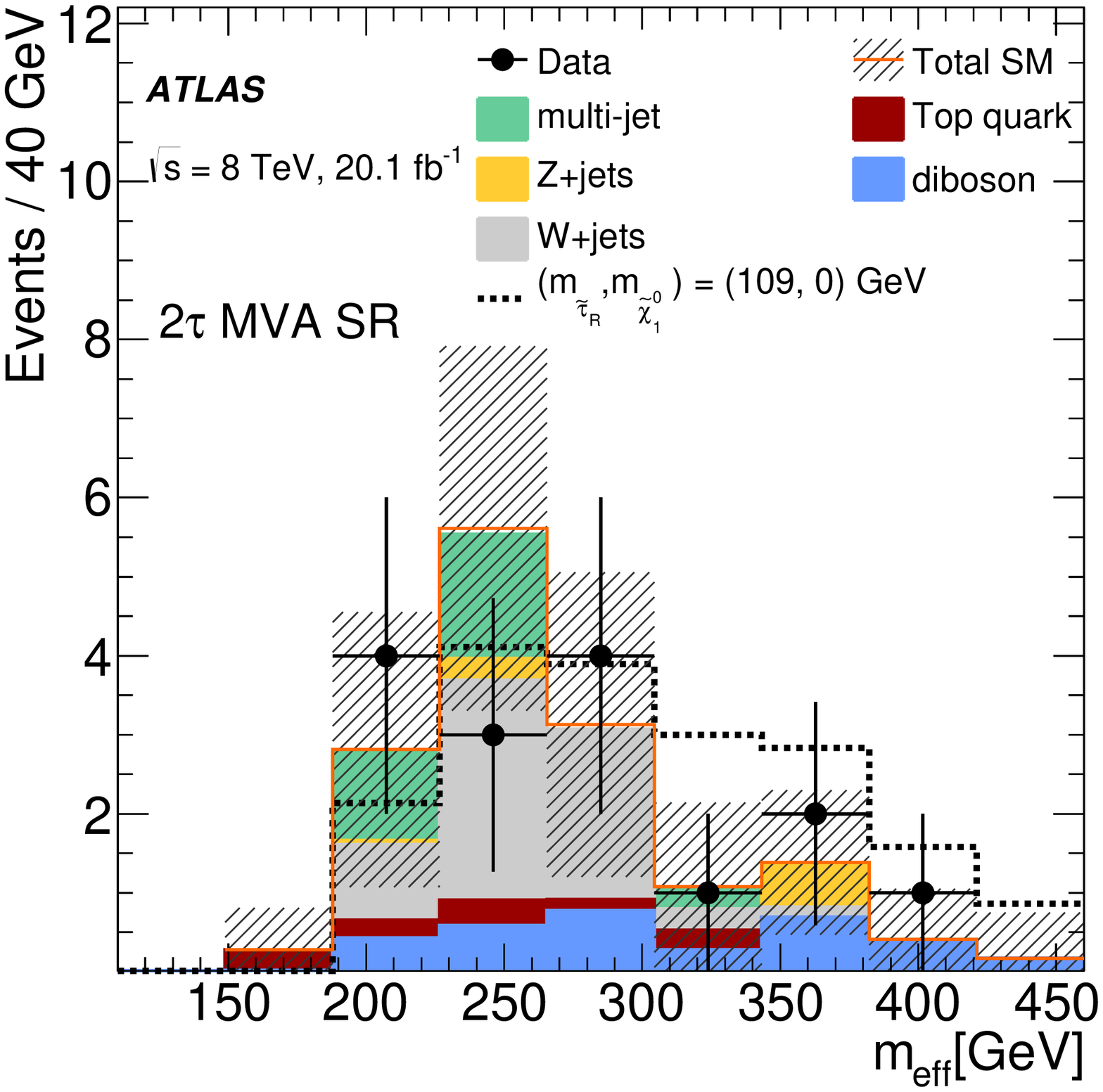}}
\subfigure[]{\includegraphics[width=0.49\textwidth]{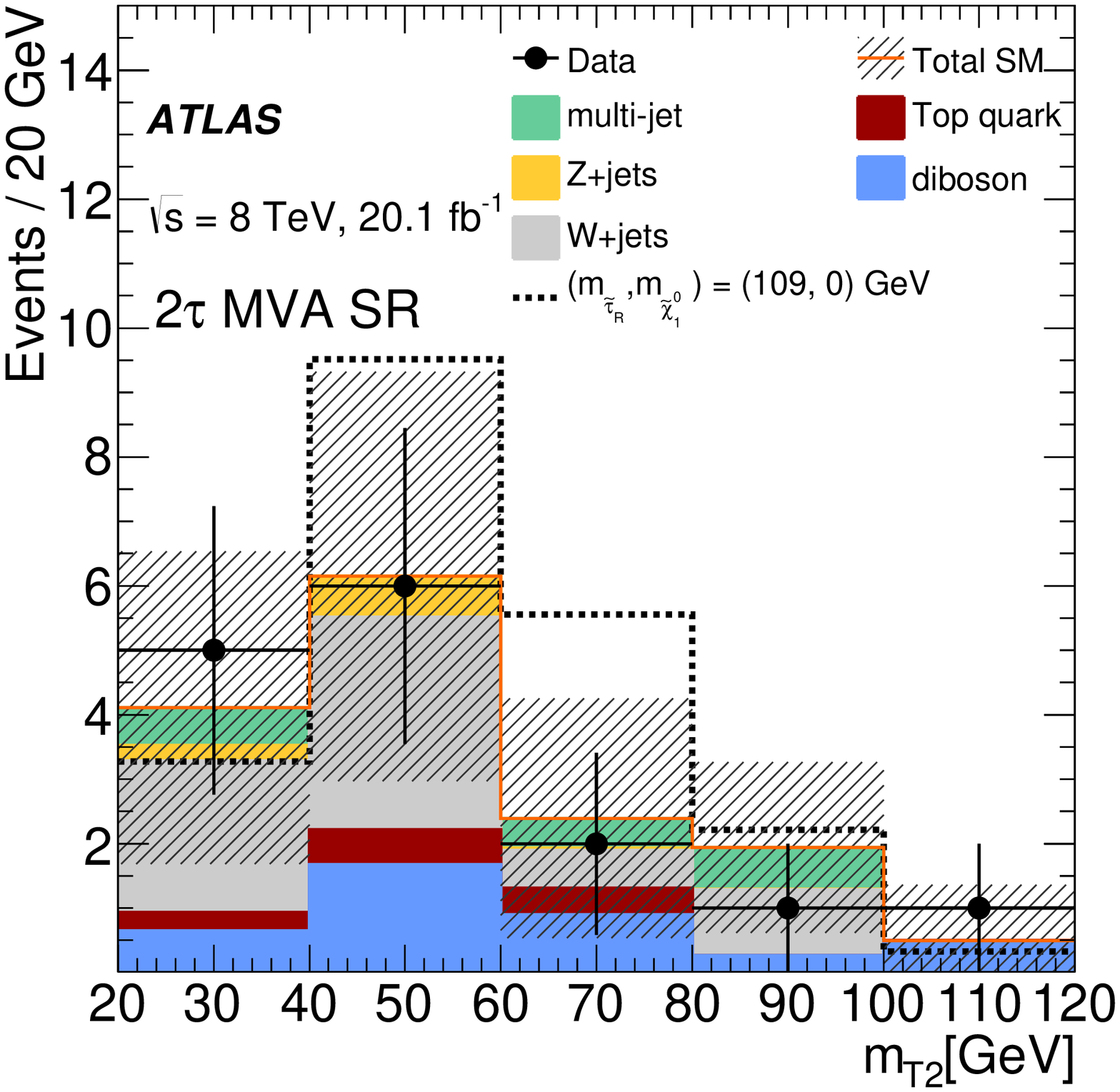}}
\caption{ The BDT response is shown in (a) prior to applying the SR $t_{\rm cut}$ requirement. 
Also shown are distributions in the two-tau MVA SR: (b) $\met$, (c) $\meff$ and (d) $\mttwo$. 
The lower panel in (a) shows the ratio of data to the SM background prediction. 
The uncertainty band includes both the statistical and systematic uncertainties on the SM prediction. 
The multi-jet and $W$+jets normalization factors from the background fits are applied in (b)--(d); only the $W$+jets normalization factor is applied in (a). 
\label{fig:tauSR-summary} }
\end{figure}

The observed number of events in the signal region is shown in Table~\ref{tab:tauVRSR-results} along with the background expectations, uncertainties, $p_0$-value, $S^{95}_{\rm exp}$, $S^{95}_{\rm obs}$, $\langle\epsilon{\rm \sigma}\rangle_{\rm obs}^{95}$, and the CL$_{b}$ value. 
The individual sources of uncertainty on the background estimation in the SR are shown in Table~\ref{tab:tauSR_syst}, where the dominant sources are the statistical uncertainty on the MC simulation samples, the uncertainty on the $\met$ from energy deposits not associated with reconstructed objects and the statistical uncertainty on the normalization factor applied to the $W$+jets background. 
Generator modeling uncertainties for the $W$+jets background are estimated by varying the renormalization and factorization scales individually between 0.5 and 2 times the nominal values in \Alpgen. 
Additionally, the impact of the jet $\pt$ threshold used for parton–jet matching in \Alpgen\ $W$+jets simulation is assessed by changing the jet $\pt$ threshold from 15$\GeV$ to 25$\GeV$. 
Figures~\ref{fig:tauSR-summary}(a), \ref{fig:tauSR-summary}(b), \ref{fig:tauSR-summary}(c), and \ref{fig:tauSR-summary}(d) show the distributions of the BDT response prior to the $t_{\rm cut}$ selection, and the $\met$, $\meff$ and $\mttwo$ quantities in the SR, where good agreement between the expected background and the observed data is seen.

\begin{table}[h]
\centering
\caption{Numbers of events observed in data and expected from SM processes and the SUSY reference point $m$($\stauR$,$\ninoone$)\,=\,(109,0)$\GeV$ in the two-tau MVA validation and signal regions. 
The uncertainties shown include both statistical and systematic components.
The ``top'' contribution includes the single top, $\ttbar$, and $\ttbar V$ processes. 
The multi-jet background estimation is taken from data, as described in the text.
In the VR, the multi-jet scale factor from fitting the background is not applied, while the $W$+jets scale factor is applied.
In the SR, both the multi-jet and the $W$+jets scale factors are applied.
Also shown are the model-independent limits calculated from the signal region observations: the one-sided $p_0$-value; the expected and observed upper limit at 95\% CL on the number of beyond-the-SM events ($S^{95}_{\rm exp}$ and $S^{95}_{\rm obs}$) for each signal region, calculated using pseudoexperiments and the CL$_s$ prescription;  the observed 95\% CL upper limit on the signal cross-section times efficiency ($\langle\epsilon{\rm \sigma}\rangle_{\rm obs}^{95}$); and the CL$_{b}$ value for the background-only hypothesis.
\label{tab:tauVRSR-results}}
\small
  \renewcommand\arraystretch{1.3}
\begin{tabular}{c | cccc| c}
\toprule
    SM process       &   Multi-jet VR1             &   Multi-jet VR2                        &   $W$-VR1        &   $W$-VR2         & SR              \\ 
\midrule
 Top                 &      30$\pm$9         &       19$\pm$6       &    5.4$\pm$2.6  &       8.1$\pm$3.4     &        1.2$\pm$0.9     \\
 $Z$+jets            &      590$\pm$100      &       86$\pm$21      &    2.3$\pm$1.7  &       4.4$\pm$2.5     &        0.9$\pm$1.2    \\
 $W$+jets            &      570$\pm$190      &       210$\pm$70     &    20$\pm$8     &      33$\pm$13        &        7.3$\pm$3.4   \\
 Diboson             &      29$\pm$8         &       16$\pm$5       &    4.7$\pm$2.4  &       7.1$\pm$3.1     &        4.4$\pm$1.6    \\
Multi-jet            &      19400$\pm$1200   &       3840$\pm$230   &    5.9$\pm$2.7  &      17$\pm$12        &        0.9$\pm$2.6   \\
\midrule
SM total             &      20700$\pm$1200   &     4170$\pm$250     &    38$\pm$9     &      70$\pm$19        &        15$\pm$5          \\
Observed             &             21107                   &               4002                 &             33                   &          65   & 15  \\
\midrule
$m$($\stauR$,$\ninoone$)\,=\,(109,0)$\GeV$      &   17$\pm$7  &  13$\pm$5  &    3.4$\pm$2.2  &      5.6$\pm$2.9   &        21$\pm$5   \\   
 \midrule
$p_0$   & --- & --- & --- & --- & 0.48\\
$S_{\rm obs}^{95}$  & --- & --- & --- & --- & 15.3 \\
$S_{\rm exp}^{95}$  & --- & --- & --- & --- & $15.1^{+5.1}_{-3.5}$\\
$\langle\epsilon{\rm \sigma}\rangle_{\rm obs}^{95}$ [fb]  & --- & --- & --- & --- & 0.76 \\
CL$_{b}$  & --- & --- & --- & --- & 0.52 \\
 \bottomrule

\end{tabular}
\end{table}

\begin{table}[!hbp]   
\caption{ The relative systematic uncertainty (\%) on the background estimate in the two-tau MVA SR from the leading sources. 
Uncertainties from different sources may be correlated, and do not necessarily add in quadrature to the total uncertainty. \label{tab:tauSR_syst} 
} 
\centering
\small
\begin{tabular}{cr}
\toprule
Systematic Source & Uncertainty \\
\midrule
Statistical uncertainty on MC samples & 20\% \\
$\met$ soft-term resolution & 20\% \\
Statistical uncertainty on the $W$+jets scale factor & 15\% \\
Tau misidentification probability & 14\% \\
$W$+jets theory and modeling  & 13\% \\
Jet energy scale & 11\% \\
$\met$ soft-term scale & 10\% \\
\midrule
Total & 35\% \\
\bottomrule
\end{tabular}
\end{table}

\FloatBarrier

\section{Compressed spectra in direct production of $\chinoonep\chinoonem$ or $\chinoonepm\ninotwo$ \label{sec:compressedchannels}}

In many SUSY scenarios, one or more of the mass differences between the charginos and neutralinos is small, resulting in final states with low-momentum leptons that require dedicated searches. 
The two-lepton analysis in Ref.~\cite{Aad:2014vma} excluded $\chinoonep\chinoonem$ scenarios with $\slepL$-mediated decays with \chinoonepm--\ninoone\ mass splittings down to approximately $100\GeV$, while the three-lepton analysis in Ref.~\cite{Aad:2014nua} excluded $\chinoonepm\ninotwo$ scenarios with $\slepL$-mediated decays down to $\ninotwo$--$\ninoone$ mass splittings of 20$\GeV$. 
The analyses presented in this section focus on event selections based on low-momentum leptons, and also on the production in association with ISR jets to provide improved sensitivity to the compressed spectra scenarios not covered by previous searches. 
As discussed in Section~\ref{sec:susysignals}, simplified models describing $\chinoonep\chinoonem$ and $\chinoonepm\ninotwo$ production are considered for these compressed spectra searches, where the $\chinoonepm$/$\ninotwo$ decay only through sleptons or sneutrinos. 
The compressed spectra searches are less sensitive to scenarios where the $\chinoonepm$/$\ninotwo$ decay through SM $W$, $Z$ or Higgs bosons, as the branching fraction to leptonic final states is significantly suppressed. 
The experimental sensitivity to these scenarios is expected to be recovered with a larger dataset.

\subsection{Searches with two opposite-sign light leptons \label{sec:superrazor}}
Previous searches for direct \chinoonep\chinoonem\ production using two opposite-sign light-lepton final states are extended here to increase the sensitivity to compressed SUSY scenarios. 
The opposite-sign, two-lepton analysis presented here probes \chinoonepm--\ninoone\ mass splittings below $100\GeV$ using an ISR-jet selection. 

\subsubsection{Event selection}

Events are reconstructed as described in Section~\ref{sec:evtreco}, with the signal light-lepton $\pt$ threshold raised to $\pt=10\GeV$. 
In addition, in events where tagged light leptons form an SFOS pair with $\msfos\,$$<\,$12$ \GeV$, both leptons in the pair are rejected. 
Events must have exactly two signal light leptons with opposite charge, and satisfy the symmetric or asymmetric dilepton trigger criteria, 
as described in Section~\ref{sec:evtreco}. 

To suppress the top-quark ($t\bar{t}$ and $Wt$) production contribution to the background, 
events containing central $b$-tagged jets or forward jets are rejected.
To suppress events from $Z$ boson decays, events with invariant mass of the reconstructed SFOS pair 
within 10$\GeV$ of the $Z$ boson mass (91.2$\GeV$) are rejected in the same-flavor channel. 

Two SRs, collectively referred to as SR2$\ell$-1, are defined. 
Both are designed to provide sensitivity to \chinoonep\chinoonem\ production with $\slepL$-mediated decays and low \chinoonepm--\ninoone\ mass splittings 
and rely on a high-\pT\ ISR jet to boost the leptons, which would otherwise have too low momentum to be reconstructed. 
The super-razor variables that are discussed in Section~\ref{sec:evtvarbles} are used to discriminate between signal and backgrounds.
Both the same-flavor (SF) and different-flavor (DF) channels are used.
The first SR, SR2$\ell$-1a, requires ${\rTwo>0.5(0.7)}$ in the SF (DF) channel, 
whereas the second SR, SR2$\ell$-1b, requires \rTwo\,$>0.65\,(0.75)$.
Both SRs require \mDeltaR\,$>20\GeV$ to reduce SM $Z$+jets background, 
and \dPhiBr\,$>2\,(2.5)$ in the SF (DF) to further increase the signal sensitivity.
Table~\ref{tab:2LOSSRCRdefs} summarizes the complete definitions of the SRs.
SR2$\ell$-1a provides sensitivity for moderate \chinoonepm--\ninoone\ mass splittings from $50\GeV$ to $100\GeV$, 
while SR2$\ell$-1b provides sensitivity for \chinoonepm--\ninoone\ mass splittings less than $50\GeV$.

\begin{table}[h]
\centering
\caption{The selection requirements for the opposite-sign, two-lepton signal and control regions, targeting $\chinoonep\chinoonem$ production with small mass splittings between the $\chinoonepm$ and LSP. \label{tab:2LOSSRCRdefs}}
\small{
\begin{tabular}{c | c  c | c  c | c | c | c }
\toprule
 & \multicolumn{7}{c}{Common} \\
\midrule
Central light-flavor jets &  \multicolumn{7}{c}{$=$1} \\
Forward jets         & \multicolumn{7}{c}{veto}  \\
$\mDeltaR$ [$\GeV$] & \multicolumn{7}{c}{$>\,$20} \\
\midrule
 & \multicolumn{2}{c}{SR2$\ell$-1a} &  \multicolumn{2}{|c|}{SR2$\ell$-1b}  & CR2$\ell$-Top &  CR2$\ell$-$WW$ &  CR2$\ell$-$ZV$  \\
\midrule
$\ell$ flavor/sign & $\ell^{\pm}\ell^{\mp}$ &  $\ell^{\pm}\ell^{\prime\mp}$ & $\ell^{\pm}\ell^{\mp}$ &  $\ell^{\pm}\ell^{\prime\mp}$ & $\ell^{\pm}\ell^{\prime\mp}$ & $\ell^{\pm}\ell^{\prime\mp}$ & $\ell^{\pm}\ell^{\mp}$  \\
Central $b$-tagged jets  & \multicolumn{4}{c|}{veto}  & $\ge\,$1  & veto & veto \\
$\msfos$ [$\GeV$]   & \multicolumn{4}{c|}{veto 81.2--101.2} & -- & -- & select 81.2--101.2 \\
$\pTll$ [$\GeV$] & -- & -- & $<\,$40 & $<\,$50 & --  & $>\,$70 & $>\,$70 \\
$\pt^{\mathrm{jet}}$ [$\GeV$]  &  $>\,$80 & $>\,$80  & $>\,$60 & $>\,$80 & -- & -- & -- \\
$\rTwo$ & $>\,$0.5 & $>\,$0.7 & $>\,$0.65 & $>\,$0.75 & -- & -- & -- \\
$\dPhiBr$ [rad] & $>\,$2 & $>\,$2.5 & $>\,$2 & $>\,$2.5 & -- & $<\,2$ & $>\,2$ \\
$p^{\mathrm{central\,light\,jet}}_{\mathrm{T}}$ [$\GeV$]  	& -- & -- & -- & -- &  $>\,$80  &  --  &  --  \\
\bottomrule
\end{tabular}}
\end{table}

\subsubsection{Background determination \label{sec:superrazor-bgdet}}

The SM background is dominated by $WW$ diboson and top-quark production.
The MC predictions for these SM sources, in addition to contributions from $ZV$ production, where $V=W$ or $Z$, 
are normalized in dedicated control regions for each background. 
The reducible background is estimated using the matrix method as described in Section~\ref{sec:matrixmethod}.
Finally, contributions from remaining sources of SM background,
which include Higgs boson production and $Z$+jets, are small and are estimated from simulation.
These are collectively referred to as ``Others''.

\begin{figure}[h!]
\centering
	\subfigure[]{\includegraphics[width=0.49\textwidth]{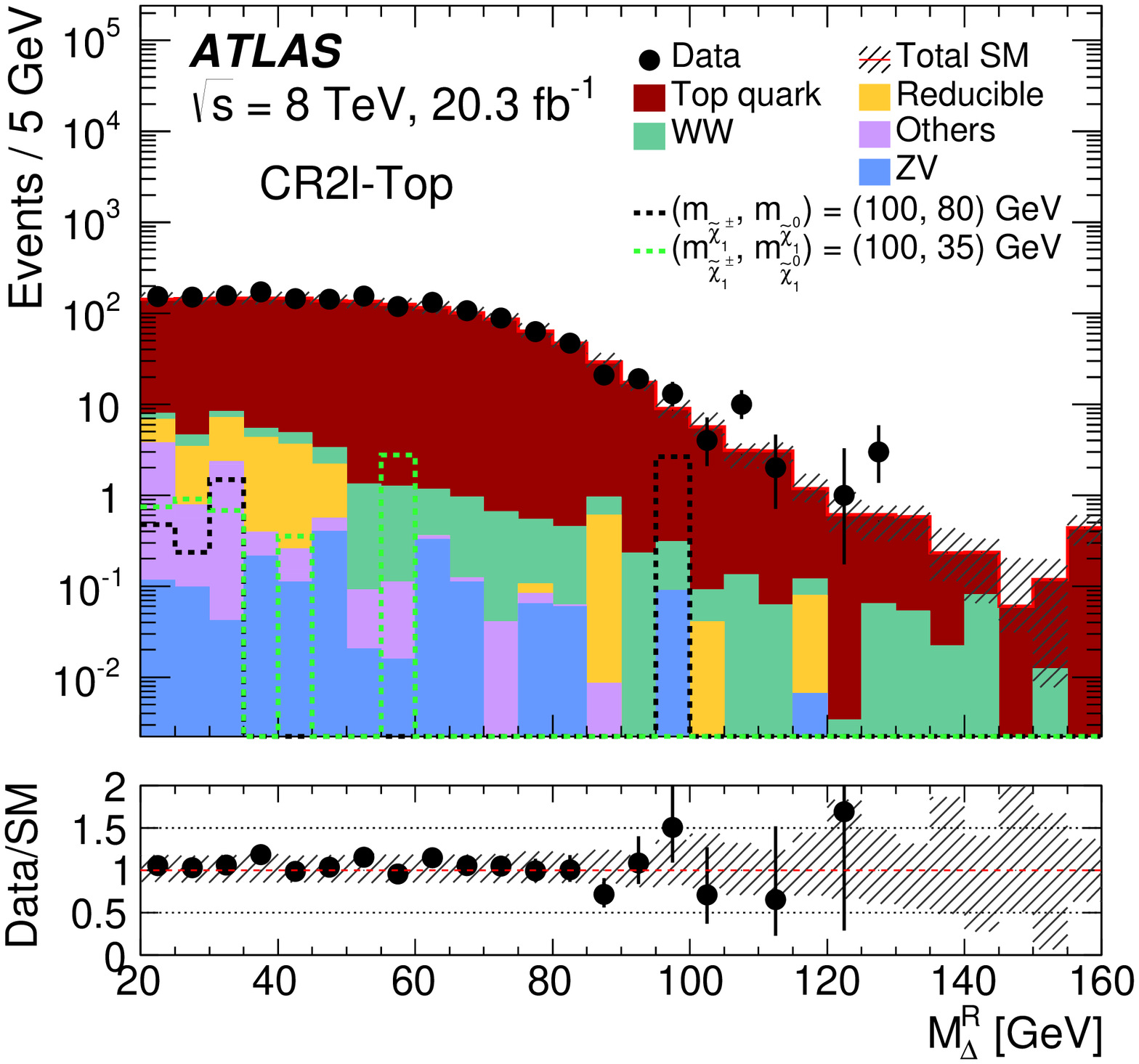}}
	\subfigure[]{\includegraphics[width=0.49\textwidth]{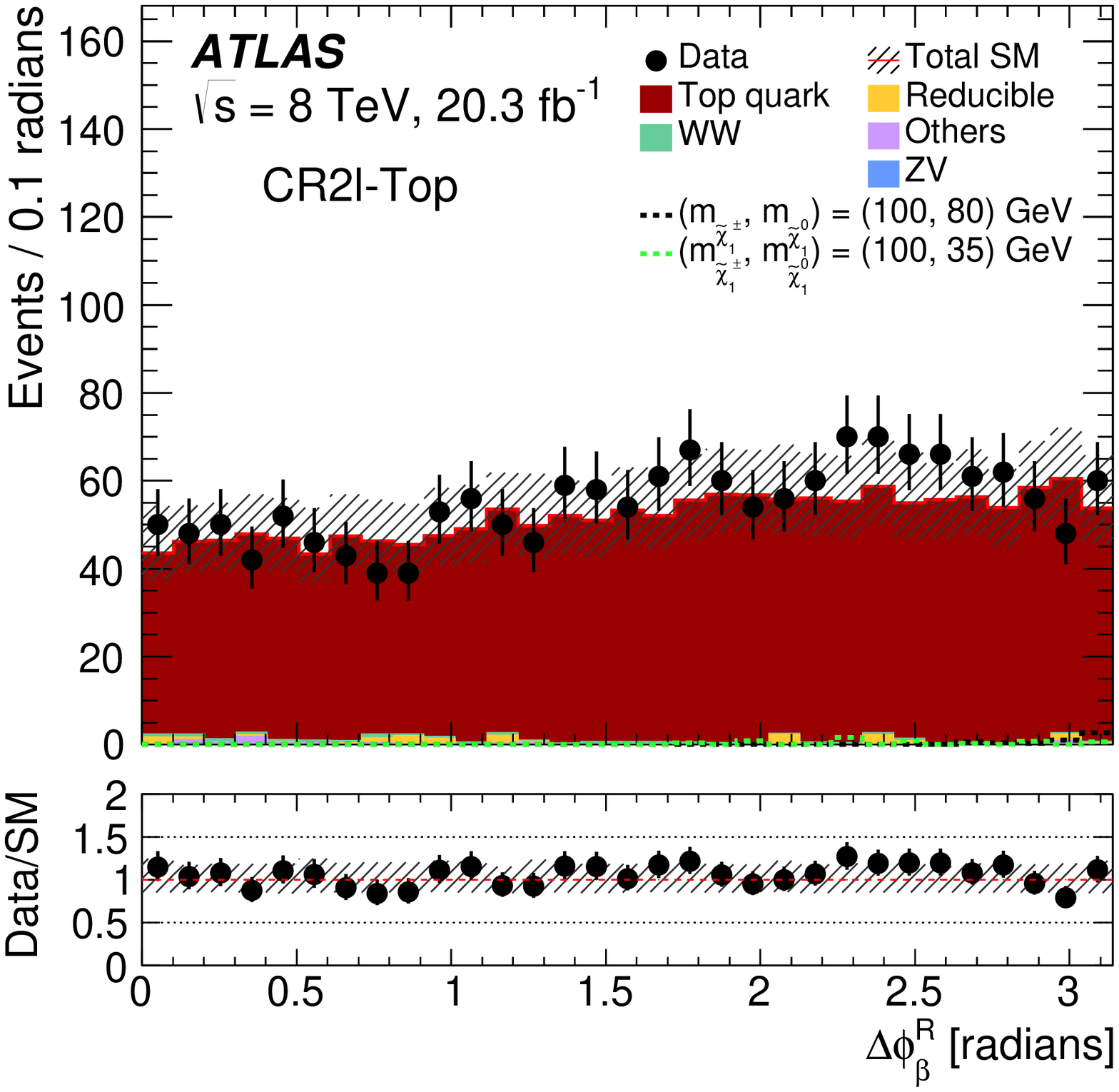}}
	\subfigure[]{\includegraphics[width=0.49\textwidth]{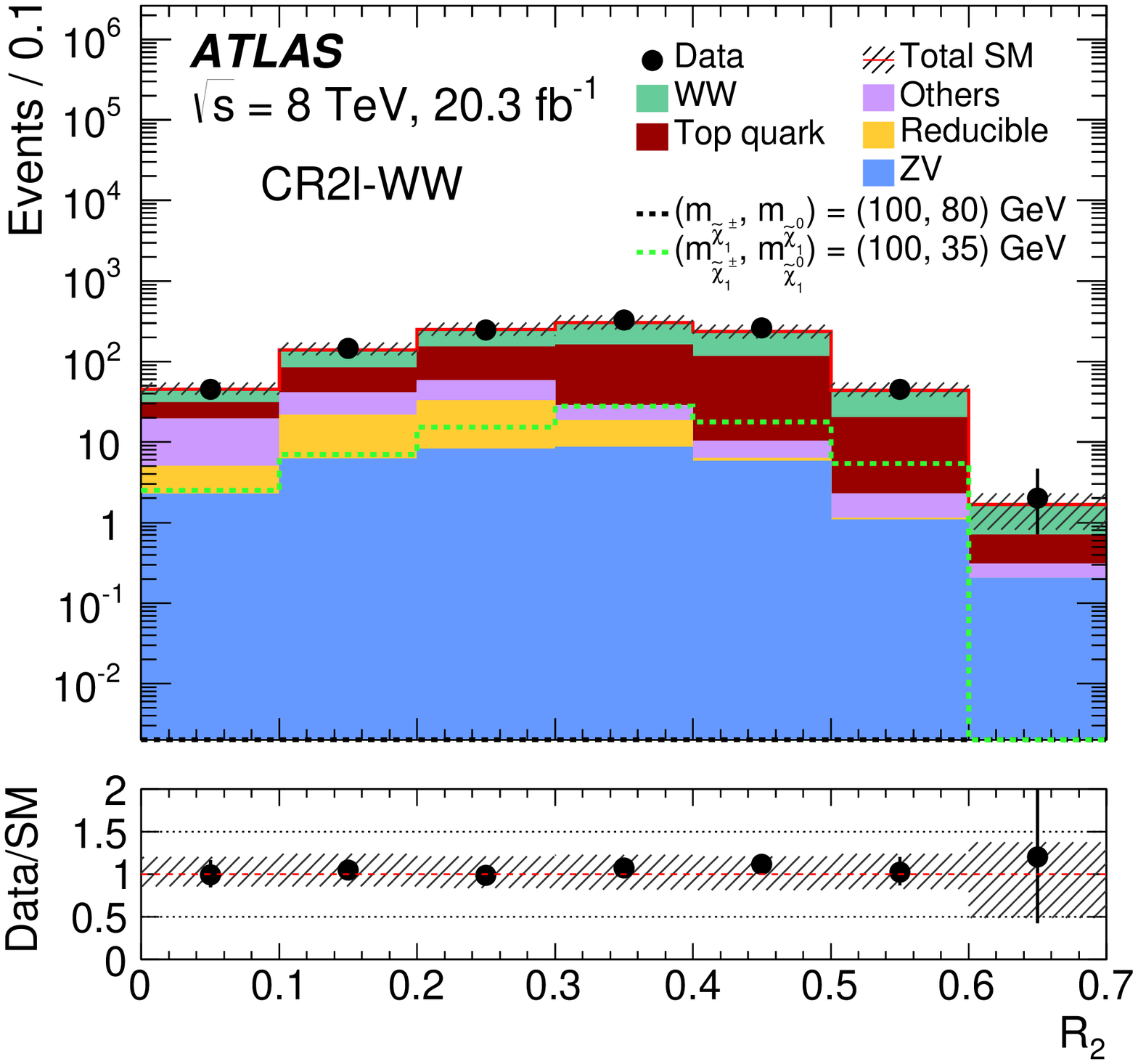}}
	\subfigure[]{\includegraphics[width=0.49\textwidth]{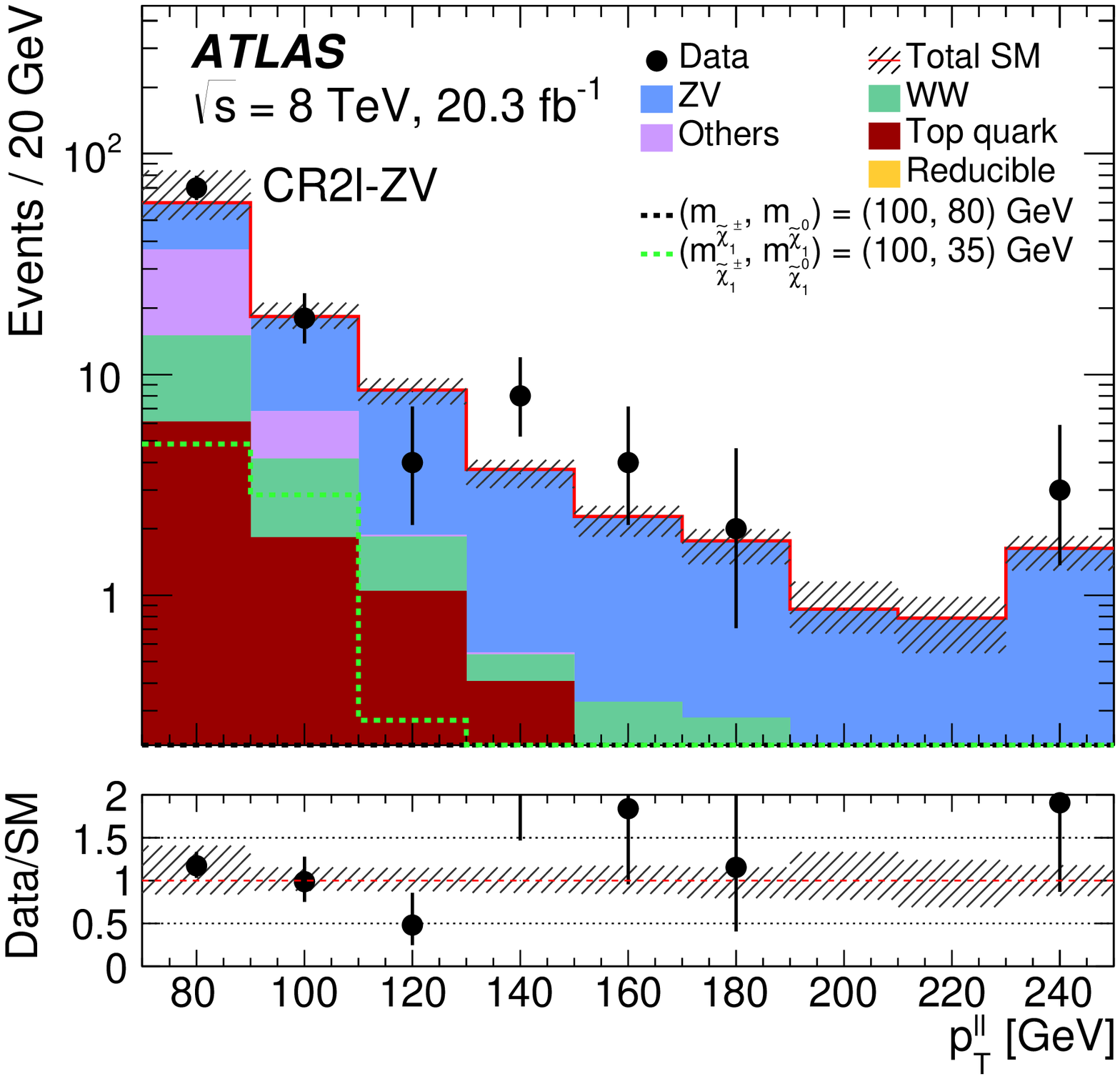}}
	\caption{\label{fig:2LOSCRfigs}
		Distributions in the opposite-sign, two-lepton control regions:
		(a) super-razor quantity \mDeltaR\ and (b) super-razor quantity \dPhiBr\ in the top CR,
		(c) ratio \rTwo\ in the $WW$ CR, and
		(d) transverse momentum of the two-lepton system \pTll\ in the $ZV$ CR.
		No data-driven normalization factors are applied to the distributions.
		The ``Others'' background category includes $Z$+jets and SM Higgs boson production.
		The hashed regions represent the total uncertainties on the background estimates.
		The rightmost bin of each plot includes overflow.
		The lower panel of each plot shows the ratio of data to the SM background prediction. 
		SM background prediction.
		Predicted signal distributions in simplified models are also shown.
		}
\end{figure}

The top CR is defined using the DF sample in order to suppress events from SM $Z$ boson production.
Events are required to have exactly one central light-flavor jet with \pt\,$>\,80\GeV$, no forward jet, and \mDeltaR\,$>20\GeV$.
At least one $b$-tagged jet is required to enrich the purity in top-quark production and ensure orthogonality to the SRs. 
Figures~\ref{fig:2LOSCRfigs}(a) and~\ref{fig:2LOSCRfigs}(b) show the \mDeltaR\ and \dPhiBr\ distributions in this CR, respectively.
The estimated signal contamination in this CR is less than 1\% for the signal models considered.

The $WW$ CR is also defined using the DF sample.
Events are required to have exactly one central light jet, no forward jet or $b$-tagged jet, \pTll\,$>70\GeV$, and \mDeltaR\,$>20\GeV$.
In order to ensure orthogonality to the SRs, \dPhiBr\,$<2$ is required.
Figure~\ref{fig:2LOSCRfigs}(c) shows the \rTwo\ distribution in this CR.
The estimated signal contamination in this CR is less than 20\% for the signal models considered.

The $ZV$ CR is defined using the SF samples, and by requiring exactly one central light jet, 
no forward jet or $b$-tagged jet, \pTll\,$>70\GeV$, \dPhiBr\,$>\,2$ and \mDeltaR\,$>20\GeV$.
In order to increase the purity in $ZV$ production, events with invariant mass of the reconstructed SFOS pair 
within 10$\GeV$ of the $Z$ boson mass are used. This requirement also ensures orthogonality to the SRs.
Figure~\ref{fig:2LOSCRfigs}(d) shows the \pTll\ distribution in this CR.
The estimated signal contamination in this CR is less than 10\% for the signal models considered.

A simultaneous likelihood fit to the top, $WW$ and $ZV$ CRs is performed to normalize the corresponding background estimates to obtain yields in the SR (as described in Section~\ref{sec:GenAna}). Table~\ref{tab:2LOSSRCRdefs} summarizes the definitions of the CRs, and Table~\ref{tab:2LOSCRyields} summarizes the numbers of observed and predicted events in these CRs, data/MC normalizations, and CR compositions obtained from the simultaneous fit.

\begin{table}[h!]
\centering
\caption{\label{tab:2LOSCRyields}
Numbers of observed and predicted events in the opposite-sign, two-lepton control regions, 
data/MC normalization factors, and composition of the CRs obtained from the background-only fit.
The ``Others'' background category includes $Z$+jets and SM Higgs boson production.
The $Z$+jets production is the dominant contribution to this category in the CR2$\ell$-$ZV$.
}\smallskip
\small{
\begin{tabular}{l | r | r | r}
\toprule
CR                  		& CR2$\ell$-Top   & CR2$\ell$-$WW$    & CR2$\ell$-$ZV$       \\
\midrule
Observed events     	& 1702   	& 1073    	& 109        \\
MC prediction       	&  1600$\pm$80   	& 1020$\pm$140    	& 98$\pm$14         	 \\
\midrule
Normalization       	& 1.06   	& 1.04   	& 1.19      	 \\
Total uncertainty   	& 0.07   	& 0.35   	& 0.42      	 \\
\midrule
Composition         	&        	&        	&               \\
\quad $WW$          	& 1\% 	& 43\% 	& 12\% 	 \\
\quad Top           	& 98\% 	& 41\% 	&   9\% 	 \\
\quad $ZV$          	&  <1\% 	&  4\% 	& 56\%  	 \\
\quad Reducible 	&  1\% 	&  5\%	&   <1\%	 \\
\quad Others         	&  <1\% 	&  7\%	& 22\%  	 \\
\bottomrule
\end{tabular}}
\end{table}

Systematic uncertainties affect the estimates of the backgrounds and signal event yields in the control and signal regions. 
A breakdown of the different sources of systematic uncertainty on the background estimate as described in Section~\ref{sec:commsyste} is shown in Table~\ref{tab:2LOSsyst}.
Generator modeling uncertainties are estimated by comparing the results from the
\POWHEGBOX\ and \Mcatnlo\ event generators for top-quark events, and 
\POWHEGBOX\ and \aMcAtNlo\ for $WW$ events, using \Herwig\ for parton showering in all cases.
Parton showering uncertainties are estimated in top-quark and $WW$ events by comparing \POWHEGBOX+ \Herwig\ with \POWHEGBOX+ \PYTHIA.
Both generator modeling and parton showering uncertainties are estimated for $ZV$ events by comparing \POWHEGBOX+ \PYTHIA\ to \Sherpa.
Top-quark samples are generated using \AcerMC+ \PYTHIA\ to evaluate the 
uncertainties related to the amount of initial- and final-state radiation~\cite{topjetveto}. 
The impact of the choice of renormalization and factorization scales is evaluated by varying these individually
between 0.5 and 2 times the nominal values in \POWHEGBOX\ for top-quark events and in \aMcAtNlo\ for diboson events.
The dominant contributions among the `Theory \& modeling' uncertainties come from the generator modeling and parton showering uncertainties.

\begin{table}[h]
\centering
\caption{\label{tab:2LOSsyst}
The dominant systematic uncertainties (in \%) on the total background estimated in the opposite-sign two-lepton signal regions.
Because of correlations between the systematic uncertainties and the fitted backgrounds, the total
uncertainty is different from the sum in quadrature of the individual uncertainties.}\smallskip
\small{
\begin{tabular}{l | c c c c }
\toprule
SR 						& \multicolumn{2}{c}{SR2$\ell$-1a} &  \multicolumn{2}{|c}{SR2$\ell$-1b}  \\
\midrule
$\ell$ flavor/sign 			& $\ell^{\pm}\ell^{\mp}$ &  $\ell^{\pm}\ell^{\prime\mp}$ & \multicolumn{1}{|c}{$\ell^{\pm}\ell^{\mp}$} &  $\ell^{\pm}\ell^{\prime\mp}$ \\
\midrule
Statistical uncertainty on MC samples				& 2\% & 6\% & \multicolumn{1}{|c}{4\%} & 10\% \\
Jet energy scale/resolution	&10\% & 9\% & \multicolumn{1}{|c}{13\%} & 11\% \\
Theory \& modeling 			& 22\% & 22\% & \multicolumn{1}{|c}{24\%} & 25\% \\
\midrule
Total 					& 23\% & 23\% & \multicolumn{1}{|c}{26\%} & 28\% \\
\bottomrule
\end{tabular}}
\end{table}

\FloatBarrier

\subsubsection{Results \label{sec:superrazor-results}}

\begin{figure}[h]
\centering
	\subfigure[]{\includegraphics[width=0.49\textwidth]{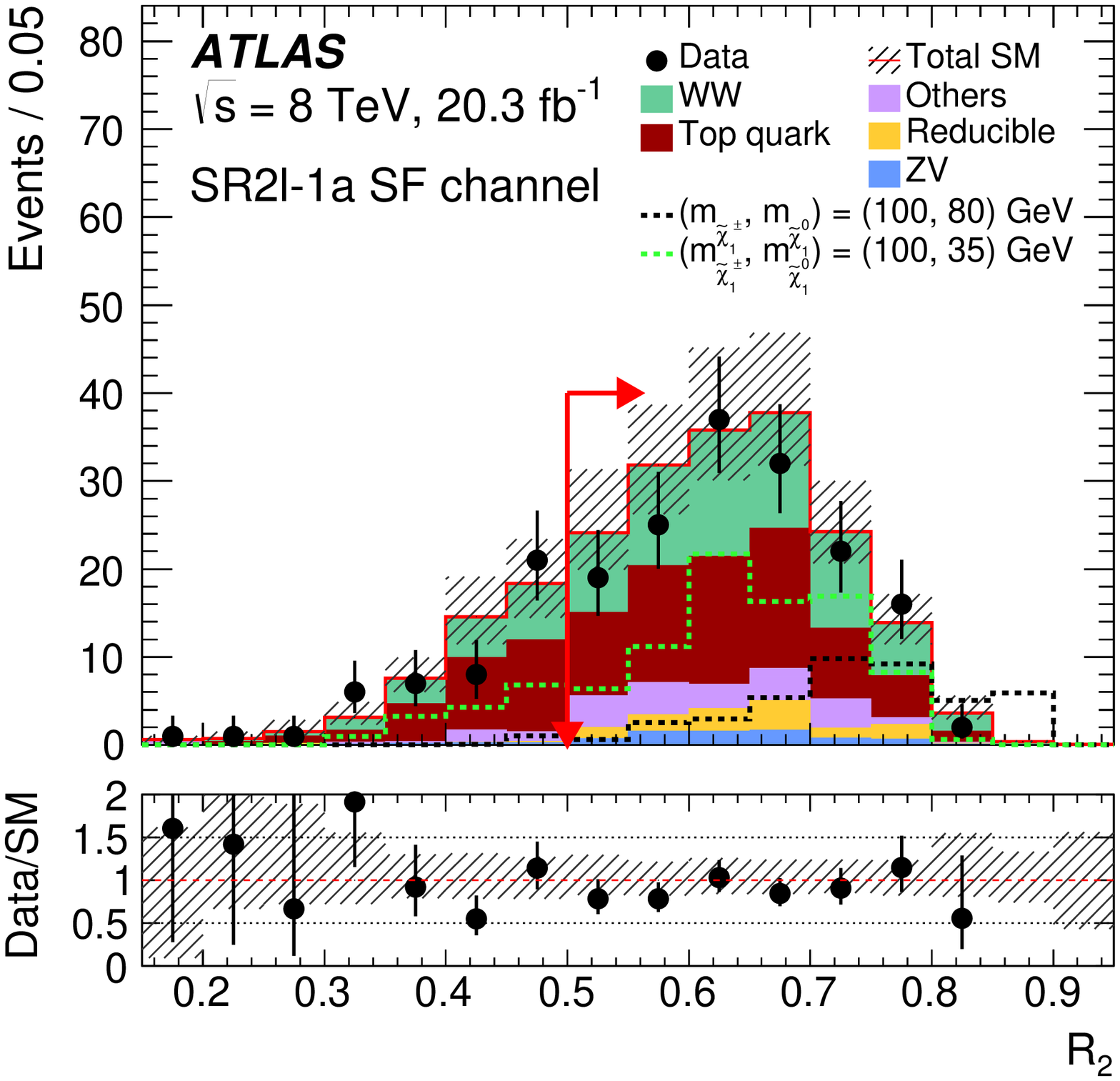}}
	\subfigure[]{\includegraphics[width=0.49\textwidth]{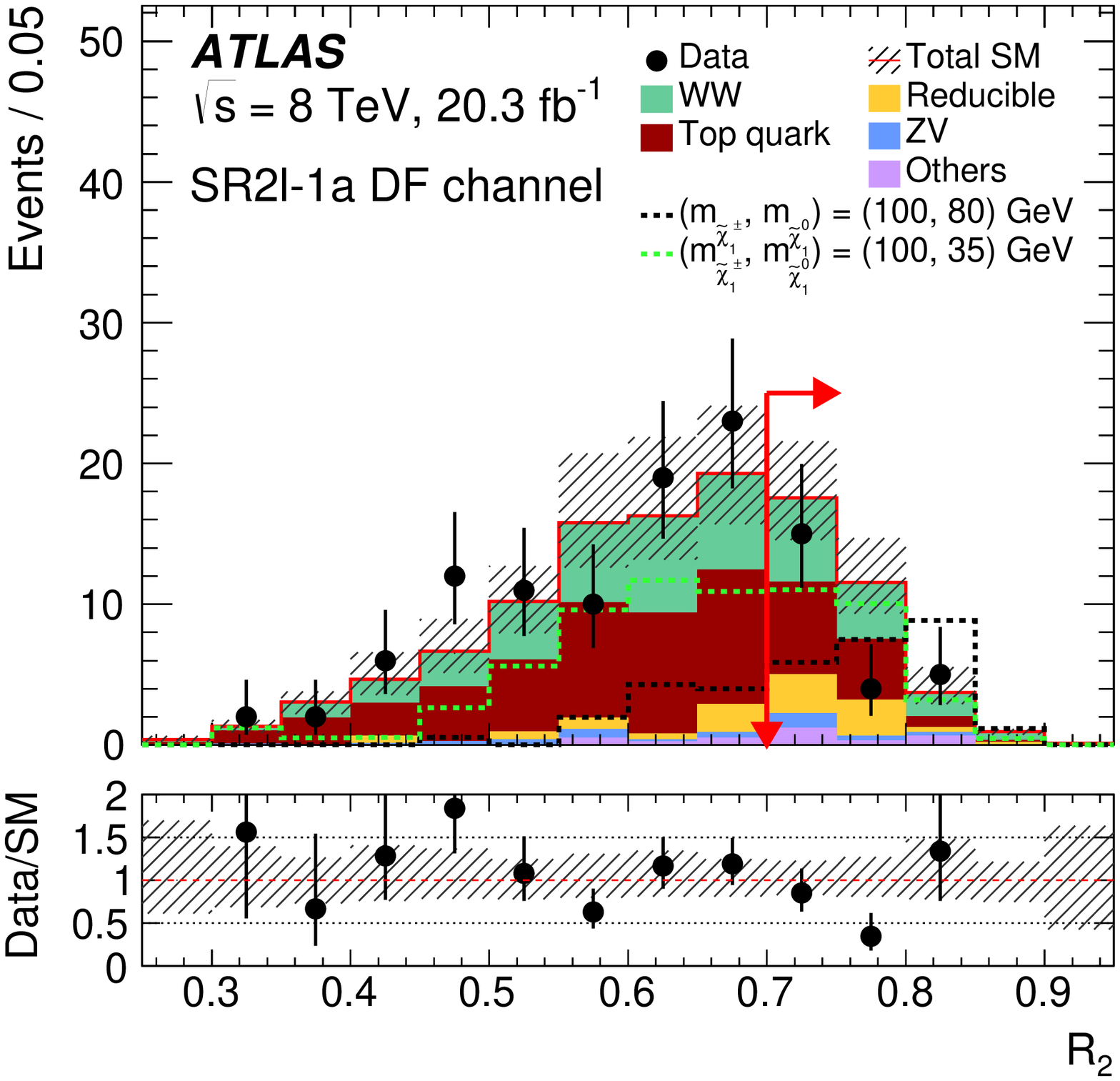}}
	\subfigure[]{\includegraphics[width=0.49\textwidth]{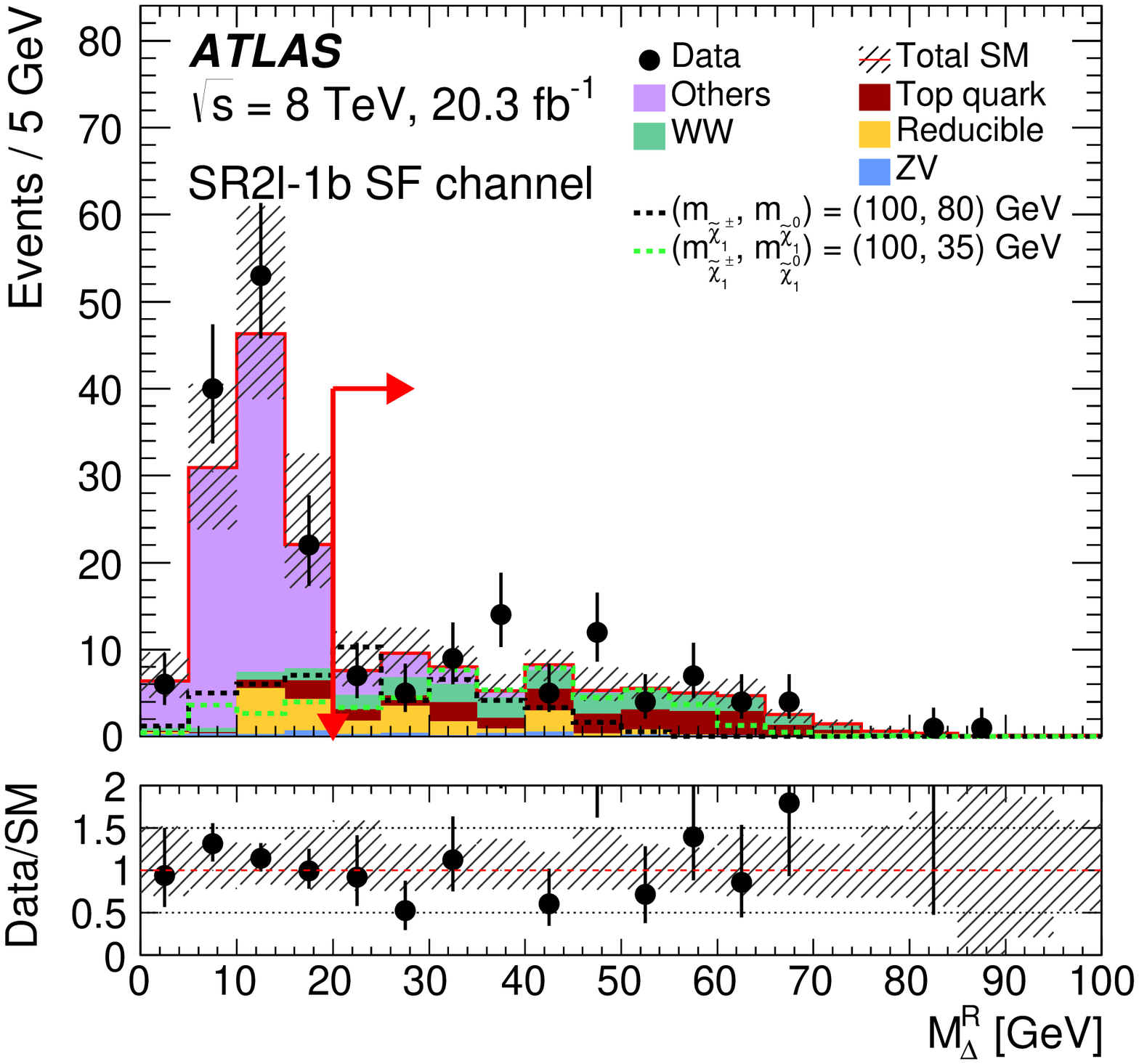}}
	\subfigure[]{\includegraphics[width=0.49\textwidth]{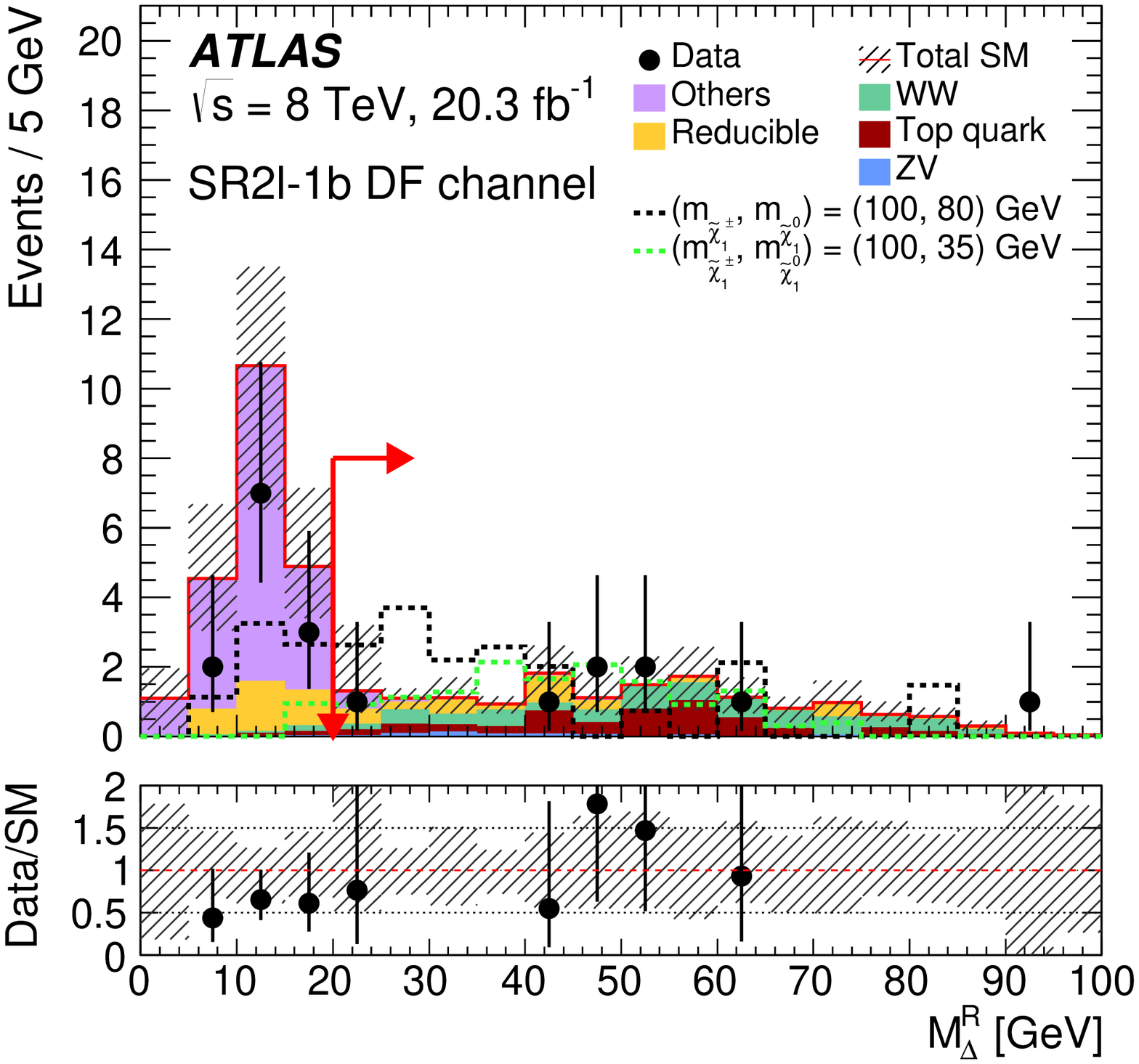}}
	\caption{
		Distributions of \rTwo\ in the (a) same flavor and (b) different flavor channels in SR2$\ell$-1a, and
		of \mDeltaR\ in the (c) same flavor and (d) different flavor channels in SR2$\ell$-1b, prior to the requirements on these variables.
		The ``Others'' background category includes $Z$+jets and SM Higgs boson production.
		Arrows indicate the limits on the values of the variables used to define the signal regions. 
                The lower panel of each plot shows the ratio of data to the SM background prediction. 
		The uncertainty band includes both the statistical and systematic uncertainties on the SM prediction. 
		The last bin in each distribution includes the overflow.
		Predicted signal distributions in simplified models are also shown.
\label{fig:2LOSSRsummary}}
\end{figure}

The observed number of events in each signal region is shown in Table~\ref{tab:2LOSSRyields} along with the background expectations and uncertainties, $p_0$-values, $S^{95}_{\rm exp}$, $S^{95}_{\rm obs}$, $\langle\epsilon{\rm \sigma}\rangle_{\rm obs}^{95}$, and the CL$_{b}$ values. 
Figures~\ref{fig:2LOSSRsummary}(a), \ref{fig:2LOSSRsummary}(b), \ref{fig:2LOSSRsummary}(c) and \ref{fig:2LOSSRsummary}(d) show the distributions of the quantities \rTwo\ and \mDeltaR\ in the SR2$\ell$-1a and SR2$\ell$-1b regions respectively, 
prior to the requirements on these variables.
For illustration, the distributions are also shown for two \chinoonep\chinoonem\ simplified models with $\slepL$-mediated decays and different mass splittings. 

\begin{table}[h]
\caption{\label{tab:2LOSSRyields}
Observed and expected number of events in the opposite-sign two-lepton signal regions.
The ``Others'' background category includes $Z$+jets and SM Higgs boson production.
The numbers of signal events are shown for the \chinoonep\chinoonem\ simplified models with $\slepL$-mediated decays and different \chinoonepm\ and \ninoone\ masses in $\GeV$.
The uncertainties shown include both statistical and systematic components.
Also shown are the model-independent limits calculated from the opposite-sign two-lepton signal region observations: the one-sided $p_0$ values; the expected and observed upper limits at 95\% CL on the number of beyond-the-SM events ($S^{95}_{\rm exp}$ and $S^{95}_{\rm obs}$) for each signal region, calculated using pseudoexperiments and the CL$_s$ prescription;  the observed 95\% CL upper limit on the signal cross-section times efficiency ($\langle\epsilon{\rm \sigma}\rangle_{\rm obs}^{95}$); and the CL$_{b}$ value for the background-only hypothesis.}
\centering
\small{
\begin{tabular}{l | c c | c c }
\toprule
SR 					& \multicolumn{2}{c|}{SR2$\ell$-1a} &  \multicolumn{2}{c}{SR2$\ell$-1b}  \\
\midrule
$\ell$ flavor/sign 		& $\ell^{\pm}\ell^{\mp}$ &  $\ell^{\pm}\ell^{\prime\mp}$ & $\ell^{\pm}\ell^{\mp}$ & $\ell^{\pm}\ell^{\prime\mp}$ \\
\midrule
Expected background	&  &  &  &              \\
\quad $WW$  			& $67 \pm 27$ & $12 \pm 5$ & $22 \pm 9$ & $5.7 \pm 2.4$             \\
\quad Top       			& $69 \pm 19$ & $12 \pm 4$ & $21 \pm 7$ &$5.0 \pm 2.0$             \\
\quad $ZV$    			& $7.3 \pm 3.4$ & $1.7 \pm 0.8$ & $2.4 \pm 1.5$ & $0.6 \pm 0.4$             \\
\quad Reducible 		& $12 \pm 6$ & $5.8 \pm 2.0$ & $10 \pm 4$ & $2.8 \pm 1.1$             \\
\quad Others    			& $18 \pm 5$ & $2.1 \pm 1.3$ & $9.4 \pm 3.4$ & $1.0 \pm 0.7$             \\
Total					& $173 \pm 23$ & $34 \pm 5$ & $65 \pm 9$ & $15.0 \pm 2.5$             \\
\midrule
Observed events 		& $153$ & $24$ & $73$ & $8$             \\
\midrule
Predicted signal 		&  &  &   & \\
$(m_{\chinoonepm},m_{\ninoone})=(100,35)$ & $81 \pm 16$ & $25 \pm 7$ & $44 \pm 8$ & $14 \pm 4$    \\
$(m_{\chinoonepm},m_{\ninoone})=(100,80)$ & $41 \pm 10$ & $23 \pm 6$ & $31 \pm 7$ & $18 \pm 5$      \\
\midrule
$p_0$ & 0.50 & 0.50 & 0.26 & 0.50 \\
$S_{\rm obs}^{95}$  & 35.7 & 9.3 & 30.8 & 5.6 \\
$S_{\rm exp}^{95}$ & ${46}^{+18}_{-12}$ & ${15}^{+6}_{-4}$ & ${25}^{+10}_{-7}$ & ${9.4}^{+4.2}_{-2.8}$ \\
$\langle\epsilon{\rm \sigma}\rangle_{\rm obs}^{95}$[fb]  &  1.76 & 0.46 & 1.52 & 0.27 \\
CL$_{b}$ &   0.22 & 0.09 & 0.73 & 0.07 \\
\bottomrule
\end{tabular}}
\end{table}

\FloatBarrier

\subsection{Searches with two same-sign light leptons  \label{sec:samesignLL}}

In compressed mass scenarios, one or more of the three leptons from $\chinoonepm\ninotwo$ production may have momentum too low to be reconstructed. 
Therefore, the search for $\chinoonepm\ninotwo$ production using two same-sign leptons can complement the three-lepton search documented in Ref.~\cite{Aad:2014nua} and extend the reach for small mass splittings. 
The search for same-sign lepton pairs is preferable to opposite-sign pairs, due to the comparatively small SM background.
A multivariate analysis technique is used here to discriminate between signal and backgrounds.

\subsubsection{Event selection}
Events are selected using the basic reconstruction, object and event selection criteria described in Section~\ref{sec:evtreco}. 
In addition, if tagged light leptons form an SFOS pair with $\msfos\,$$<\,$12$ \GeV$, both leptons in the pair are rejected. 
Signal electrons with $\pt\,$$<\,$60$\GeV$ have a tightened track (calorimeter) isolation of 7\% (13\%) of the electron $\pt$ applied, whereas for electrons with $\pt\,$$>\,$60$\GeV$, a track isolation requirement of 4.2$\GeV$ (7.8$\GeV$) is used. 
For signal muons, the track (calorimeter) isolation requirement is tightened to 6\% (14\%) of the muon $\pt$ for $\pt\,$$<\,$60$\GeV$, and 4.2$\GeV$ (8.4$\GeV$) otherwise. 
The stricter lepton isolation requirements are optimized to suppress the reducible SM backgrounds with semileptonically decaying $b$/$c$-hadrons, which are an important background in this search.

Events must have exactly two light leptons with the same charge, $e^{\pm}e^{\pm}$, $\mu^{\pm}\mu^{\pm}$ or $e^{\pm}\mu^{\pm}$ and satisfy the symmetric or asymmetric dilepton trigger criteria, as described in Section~\ref{sec:evtreco}. 
Eight BDTs are independently trained to define eight signal regions optimized for four mass splitting scenarios, $m(\ninotwo)$-$m(\ninoone)$ = 20, 35, 65, 100$\GeV$, referred to as $\Delta$M20, $\Delta$M35, $\Delta$M65 and $\Delta$M100 respectively, each with and without the presence of a central light jet with $\pt\,$$>\,$20$\GeV$, referred to as ISR and no-ISR. 
For the BDT training, signal scenarios of $\chinoonepm\ninotwo$ production with $\slepL$-mediated decays are used, where the slepton mass is set at 95\% between the $\chinoonepm$ and the $\ninoone$ masses. 
Seven variables are considered as input for the BDT training procedure: $\mttwo$, $\pt^{\ell\ell}$, $\metrel$, $\Ht$, $\mtlone$, $\mtltwo$ and $\Delta\phi(\ell,\ell)$. 
Three further variables are also considered for the ISR signal regions: $\Delta\phi(\met, \rm jet 1)$ and the ratios $\metrel$/$\pt^{\rm jet 1}$ and $\pt^{\rm lep 1}$/$\pt^{\rm jet 1}$. 
These variables exploit the kinematic properties of a compressed mass SUSY system, with and without a high-\pt\ ISR jet. 
The MC simulation samples are compared to data for these variables and their correlations to ensure that they are modeled well.

For the training and testing of the BDT, the signal and background samples are split into two halves, including those backgrounds estimated from data as described in Section~\ref{sec:SSMVAbg}. 
The eight signal region definitions are shown in Table~\ref{tab:SSSRDef}.
Since the selection on the BDT output, $t_{\mathrm{cut}}$, is independent for each SR, the overlap between SRs with looser and tighter selections is small.

\begin{table}[h]
  \centering
  \caption{Same-sign, two-lepton MVA signal region BDT requirements, targeting  $\chinoonepm\ninotwo$ production with small mass splittings between the $\chinoonepm/\ninotwo$ and LSP. 
The selection on the BDT output, $t_{\mathrm{cut}}$, is independent for each SR.
\label{tab:SSSRDef}  }
\small{
    \begin{tabular}{ c r c c c c | c}
      \toprule
\multicolumn{2}{c}{Common} & \multicolumn{5}{c}{$\ell^{\pm}\ell^{\pm}$ pair, $b$-jet veto} \\
\midrule
& & SR $\Delta$M20 & SR $\Delta$M35 & SR $\Delta$M65 & SR $\Delta$M100 & VR \\
      \midrule
      ISR & $t_{\mathrm{cut}}$ & $>0.071$ & $>0.087$ & $>0.103$ & $>0.119$ & $-0.049-0.051$ \\
      no-ISR & $t_{\mathrm{cut}}$ & $>0.071$ & $>0.087$ & $>0.135$ & $>0.135$ & $-0.049-0.051$  \\
      \bottomrule
    \end{tabular} }
\end{table}

\subsubsection{Background determination \label{sec:SSMVAbg}}

Several SM processes produce events with two same-sign signal leptons. 
The SM background processes are  classified as irreducible background if they lead to events with two real, prompt, same-sign leptons, 
reducible background if the event has at least one fake or non-prompt lepton, 
or ``charge flip'' if the event has one lepton with mismeasured charge. 

Irreducible processes include diboson ($W^{\pm}W^{\pm}$, $WZ$, $ZZ$), triboson ($VVV$), $\ttbar V$, $tZ$ and Higgs boson production and are determined using the corresponding MC samples. 
The reducible $W\gamma$ process is estimated with MC simulation samples; other reducible processes are estimated with the matrix method, similar to that described in Section~\ref{sec:matrixmethod}.

In this implementation of the matrix method, the fake and non-prompt lepton misidentification probabilities are measured in control regions that are kinematically close and similar in composition to the signal regions. 
The regions where the misidentification probabilities are measured are required to have large $\Ht$ ($\Ht\,$$>\,$50$\GeV$) and large transverse mass using the leading lepton ($\mt\,$$>\,$50$\GeV$). 
The contamination from signal events in these measurement regions is $<$1\%. 
The charge-flip, irreducible, and $W\gamma$ backgrounds are subtracted from the control regions before calculating lepton misidentification probabilities. 

Charge-flip processes include sources of opposite-sign prompt leptons for which the charge of one lepton is mismeasured ($Z$, $\ttbar$, $W^{+}W^{-}$). 
In the relevant momentum range the muon charge-flip background is found to be negligible.
Control samples of $e^+e^-$ and $e^{\pm}e^{\pm}$ with invariant mass near the $Z$ boson mass (75$<\,$$\mll\,$$<\,$100$\GeV$) are used to extract the electron charge-flip rate. 
A small background due to misidentified jets is subtracted by interpolating the mass sidebands and subtracting them from the observed data events. 
A likelihood fit is used that takes the numbers of $e^+e^-$ and $e^{\pm}e^{\pm}$ pairs observed in the charge-flip control regions as input. 
The charge-flip probability is a free parameter of the fit and is extracted as a function of the electron $\pt$ and $\eta$.  
The charge-flip background event yield is found by applying the charge-flip probability to control regions in data with the same kinematic requirements as the signal and validation regions, but with opposite-sign light lepton pairs. 
The contamination from fake and non-prompt leptons, and from signal events, is negligible in the $e^+e^-$ and $e^{\pm}e^{\pm}$ control regions.

Generator modeling uncertainties for the diboson processes are estimated by comparing the results from the \POWHEGBOX\ and \Mcatnlo\ event generators, while parton showering uncertainties are estimated by comparing \Mcatnlo+\Herwig\ with \Mcatnlo+\PYTHIA.
The impact of the choice of renormalization and factorization scales is evaluated by varying these individually
between 0.5 and 2 times the nominal values in \aMcAtNlo\ for diboson events.

\begin{figure}[h]
\centering
\subfigure[]{\includegraphics[width=0.49\textwidth]{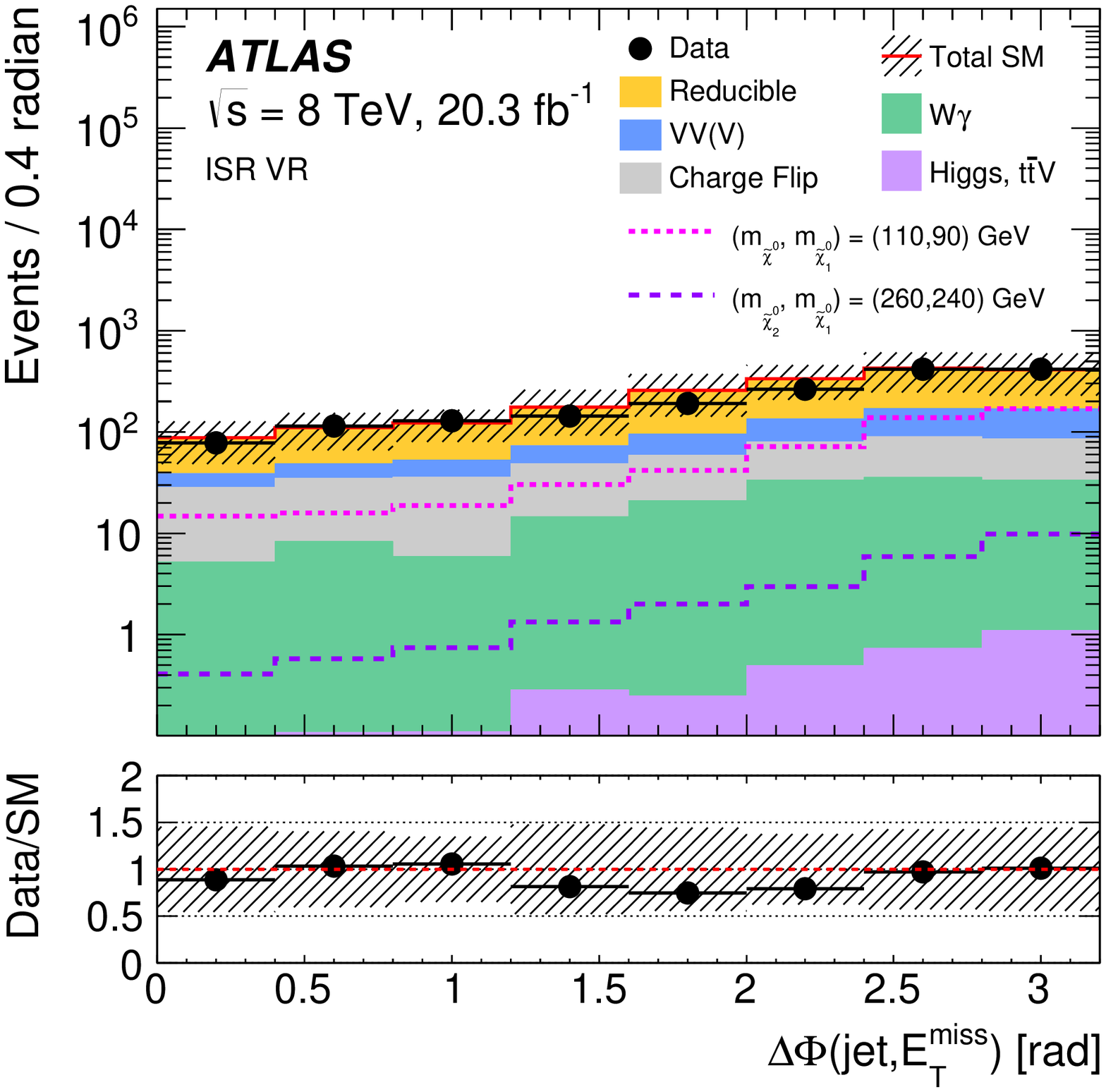}}
\subfigure[]{\includegraphics[width=0.49\textwidth]{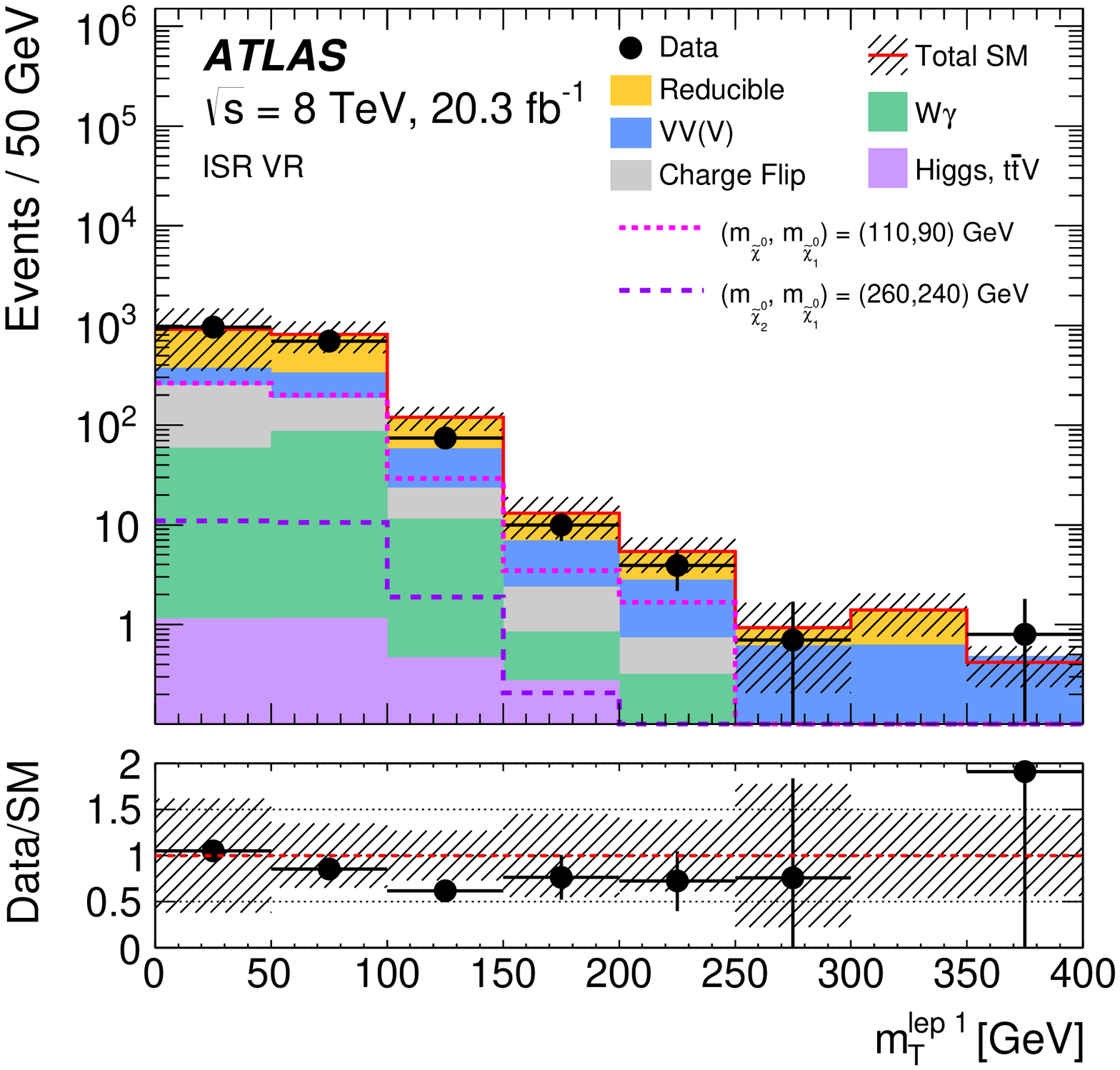}}
\subfigure[]{\includegraphics[width=0.49\textwidth]{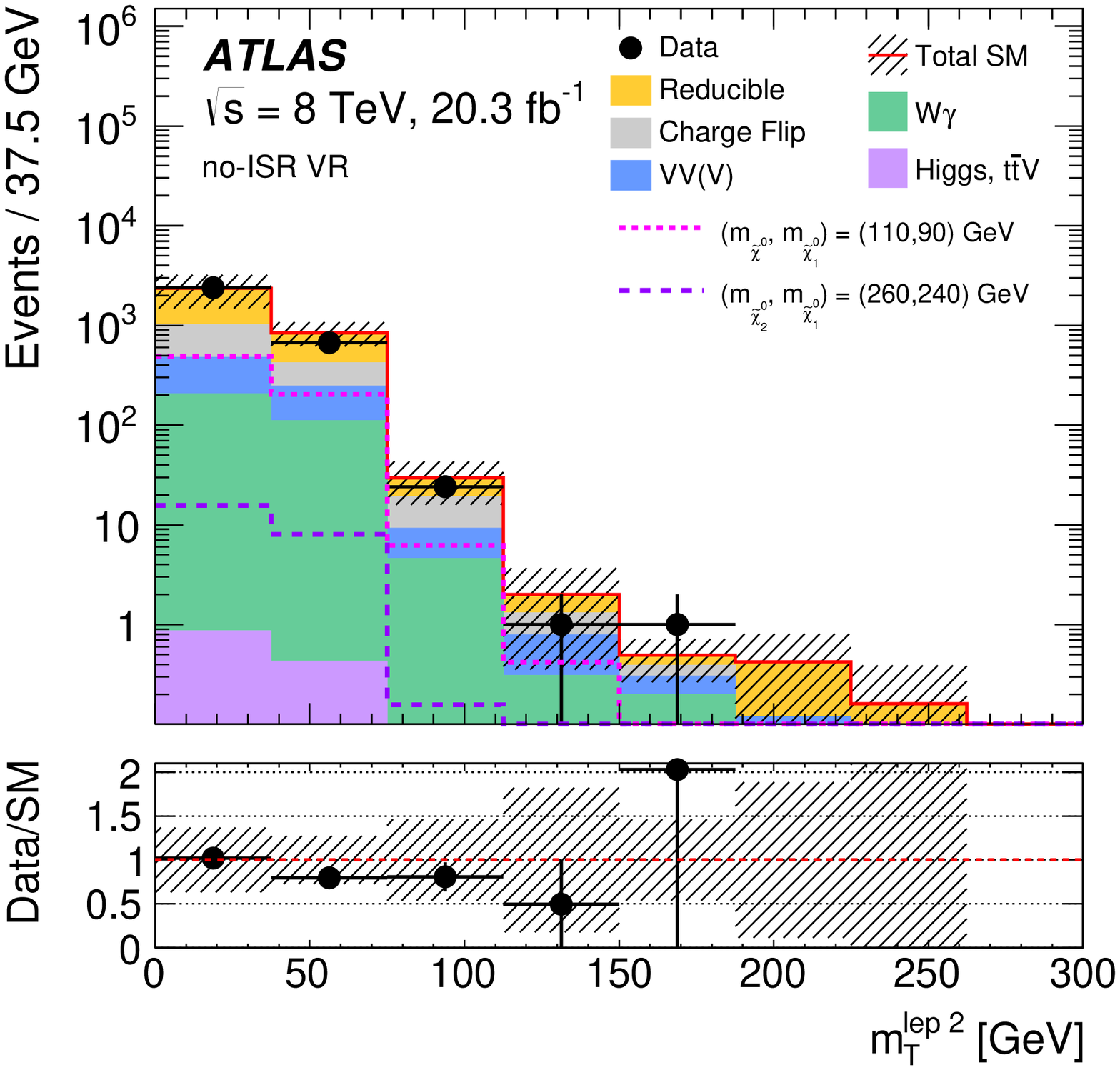}}
\subfigure[]{\includegraphics[width=0.49\textwidth]{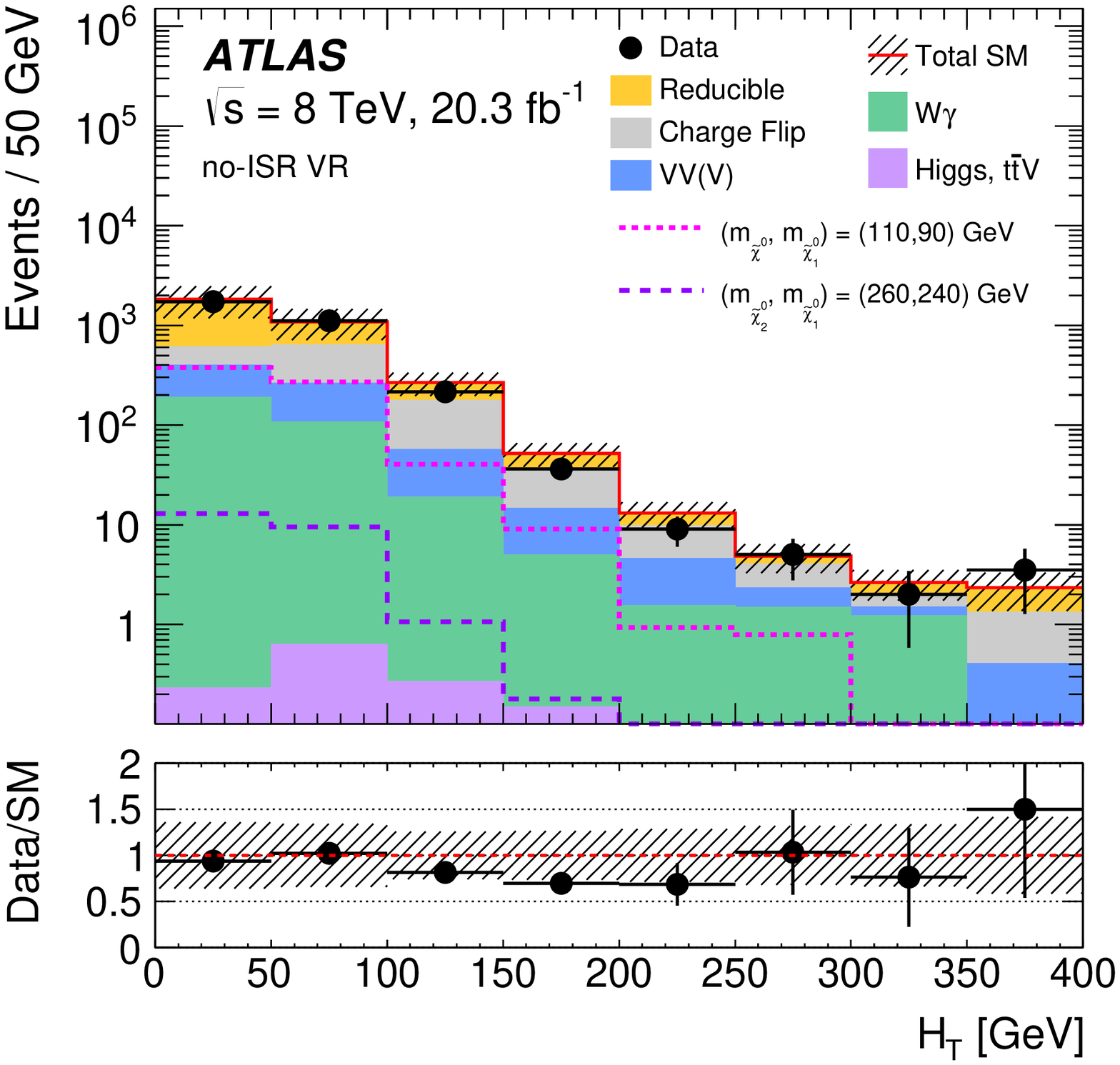}}
\caption{
For events in a selection of the same-sign, two-lepton MVA validation regions: (a) separation in $\phi$ between the leading jet and the $\met$, $\Delta\phi(\mathrm{jet}, \met)$ and (b) transverse mass using the leading lepton $\mtlone$ in the ISR VR, (c) transverse mass using the second leading lepton $\mtltwo$ and (d) scalar sum $\Ht$ of the $\pt$ of the leptons and jets in the non-ISR VR. 
The lower panel of each plot shows the ratio of data to the SM background prediction. 
The uncertainty band includes both the statistical and systematic uncertainties on the SM prediction. 
\label{fig:SSMVAVR-summary}}
\end{figure}

To test the background prediction methods, two validation regions with looser selection on the BDT output than the SRs are defined; the definitions are shown in Table~\ref{tab:SSSRDef}. 
The light-lepton flavor content ($ee$, $\mu\mu$, or $e\mu$) is checked separately in each validation region. 
Table~\ref{tab:SSMVAVR-results} and Figures~\ref{fig:SSMVAVR-summary}(a), \ref{fig:SSMVAVR-summary}(b), \ref{fig:SSMVAVR-summary}(c), and \ref{fig:SSMVAVR-summary}(d) show the agreement between data and expectation in the validation regions.

\begin{table}[htbp]
\centering
\caption{
The expected and observed yields in the same-sign, two-lepton MVA validation regions, 
separated into $ee$ events, $e\mu$ events and $\mu\mu$ events.  
The uncertainties shown include both statistical and systematic components.
\label{tab:SSMVAVR-results} }
\small{
\begin{tabular}{ lccc|ccc}
\hline
 & \multicolumn{3}{c}{VR ISR} &  \multicolumn{3}{|c}{VR no-ISR}  \\
\hline
 & $ee$ & $e\mu$ & $\mu\mu$ & $ee$ & $e\mu$ & $\mu\mu$  \\
\hline
Reducible background & $260 \pm 140$ & $670 \pm 330$ & $160 \pm 110$ & $410 \pm 190$ & $1100 \pm 400$ & $310 \pm 170$ \\
Charge-flip & $289 \pm 15$ & $15.0 \pm 1.2$ &- & $711 \pm 34$ & $28.1 \pm 2.0$ &  - \\
Diboson & $58 \pm 23$ & $155 \pm 37$ & $110 \pm 26$ & $678 \pm 25$ & $199 \pm 34$ & $154 \pm 34$ \\
Higgs & $0.42 \pm 0.30$ & $0.7 \pm 0.5$ & $0.7 \pm 0.5$ & $0.23 \pm 0.18$ & $0.6 \pm 0.4$ & $0.50 \pm 0.33$ \\
$\ttbar V$ & $0.23 \pm 0.18$ & $0.7 \pm 0.4$ & $0.44 \pm 0.29$ & $0.01 \pm 0.022$ & $0.01 \pm 0.022$ & $0.01 \pm 0.022$ \\ 
$W\gamma$ & $61 \pm 25$ & $94 \pm 23$ & $1.0 \pm 0.9$ & $120 \pm 50$ & $200 \pm 40$ & $2.3 \pm 2.0$ \\
\hline 
Total & $670 \pm 140$ & $940 \pm 330$ & $270 \pm 120$ & $1300 \pm 200$ & $1500 \pm 400$ & $470 \pm 180$ \\
\hline 
Data & $585$ & $799$ & $363$ & $1134$ & $1349$ & $612$ \\
\hline
\end{tabular}}
\end{table}

\FloatBarrier

\subsubsection{Results}

The observed number of events in each signal region is shown in Table~\ref{tab:SSMVASRmodelind} along with the background expectation and uncertainties, $p_0$-values, $S^{95}_{\rm exp}$, $S^{95}_{\rm obs}$, $\langle\epsilon{\rm \sigma}\rangle_{\rm obs}^{95}$, and the CL$_{b}$ values. 
No significant excess with respect to the SM expectation is observed. 
The sizes and sources of uncertainty on the background estimation in the signal regions are shown in Table~\ref{tab:2LSSMVAsyst}, where the dominant sources of uncertainty are the statistical uncertainty on the reducible background estimation, the statistical uncertainty on the MC simulation samples, and the uncertainty related to the choice of generator for the $WZ$ MC simulation sample. 

\begin{table}[h]
\caption{
The model-independent limits calculated from the same-sign two-lepton MVA signal region observations:  
the observed 95\% CL upper limit on the signal cross-section times efficiency ($\langle\epsilon{\rm \sigma}\rangle_{\rm obs}^{95}$); 
the expected and observed upper limits at 95\% CL on the number of beyond-the-SM events ($S^{95}_{\rm exp}$ and $S^{95}_{\rm obs}$) for each signal region, calculated using pseudoexperiments and the CL$_s$ prescription; 
the CL$_{b}$ value for the background-only hypothesis; 
and the one-sided $p_0$ values.
\label{tab:SSMVASRmodelind}}
\centering
\small{
  \renewcommand\arraystretch{1.3}
\begin{tabular}{lrccccccc}
\toprule
SR              &          & $N_{\mathrm{exp}}$ & $N_{\mathrm{obs}}$ & $\langle\epsilon{\rm \sigma}\rangle_{\rm obs}^{95}$[fb]  &  $S_{\rm obs}^{95}$  & $S_{\rm exp}^{95}$ & CL$_{b}$ & $p_0$  \\
\midrule

      $\Delta$M20 & ISR $ee$ & 3.2$\pm$0.9 & 5 & $0.36$ &  $7.3$  & ${5.4}^{+2.2}_{-1.2}$ & $0.81$ &  $0.19$\\
	& $e\mu$     & 9.7$\pm$2.8 & 9 & $0.44$ &  $8.9$  & ${9.0}^{+3.5}_{-2.5}$ & $0.47$ & $0.50$\\
      & $\mu\mu$     & 4.3$\pm$2.6 & 5 &  $0.47$ &  $9.5$  & ${8.8}^{+2.8}_{-1.8}$ & $0.63$ &  $0.44$ \\ \cline{2-9}
      & no-ISR $ee$   & 28$\pm$5 & 23 &$0.57$ &  $11.6$  & ${14}^{+6}_{-4}$  & $0.27$ & $0.50$\\
      & $e\mu$   & 25$\pm$8 & 29 & $1.08$ &  $21.9$  & ${19}^{+7}_{-5}$ & $0.68$ &  $0.33$  \\ 
      & $\mu\mu$   & 7.6$\pm$1.7 & 12 & $0.59$ &  $12.1$  & ${8.0}^{+2.7}_{-2.0}$ & $0.90$ &  $0.10$  \\ 
      \midrule
      $\Delta$M35 &  ISR $ee$   & 3.9$\pm$1.2 & 1 & $0.17$ &  $3.5$  & ${4.9}^{+2.3}_{-1.1}$ & $0.09$ &  $0.50$\\
      & $e\mu$   & 6.5$\pm$1.8 & 10 & $0.53$ &  $10.8$  & ${7.4}^{+3.2}_{-1.9}$ & $0.85$ &  $0.14$ \\
      & $\mu\mu$   & 5.4$\pm$2.1 & 5 & $0.37$ &  $7.6$  & ${7.6}^{+2.7}_{-1.7}$ & $0.51$  &  $0.50$\\\cline{2-9}
      & no-ISR $ee$   & 23$\pm$5 & 19 &  $0.56$ &  $11.4$  & ${13.4}^{+4.8}_{-3.4}$ & $0.30$ &  $0.50$ \\
      & $e\mu$   & 46$\pm$11 & 39 & $0.94$ &  $19.0$  & ${22}^{+8}_{-6}$ & $0.32$ &  $0.50$\\
      & $\mu\mu$   & 27$\pm$10 & 21 & $0.79$ &  $15.9$  & ${17.6}^{+2.4}_{-4.0}$ & $0.34$ &  $0.50$\\
      \midrule
      $\Delta$M65 & ISR $ee$   & 1.7$\pm$0.8 & 4 & $0.36$ &  $7.3$  & ${4.7}^{+1.8}_{-0.8}$ & $0.90$ &  $0.09$ \\
      & $e\mu$   & 2.4$\pm$0.8 & 4 & $0.33$ &  $6.7$  & ${5.0}^{+1.9}_{-1.3}$ & $0.54$ & $0.34$\\ 
      & $\mu\mu$   & 1.4$\pm$0.6 & 2 &  $0.24$ &  $4.9$  & ${4.1}^{+1.6}_{-0.6}$ & $0.70$ &  $0.30$\\\cline{2-9}
      & no-ISR $ee$   & 1.2$\pm$0.6 & 0 & $0.11$ &  $2.1$  & ${3.4}^{+1.3}_{-0.4}$ & $0.20$ &  $0.50$ \\
      & $e\mu$   & 1.3$\pm$0.5 & 2 & $0.24$ &  $4.9$  & ${4.1}^{+1.4}_{-0.7}$ & $0.73$ &  $0.26$  \\
      & $\mu\mu$   & 1.5$\pm$0.5 & 2 &  $0.23$ &  $4.7$  & ${4.1}^{+1.6}_{-0.8}$ & $0.68$ &  $0.32$  \\
      \midrule
      $\Delta$M100 & ISR $ee$    & 0.9$\pm$0.6 & 0 & $0.13$ &  $2.6$  & ${3.06}^{+1.25}_{-0.09}$ & $0.29$ &  $0.50$ \\
      & $e\mu$     & 0.57$\pm$0.29 & 0 & $0.14$ &  $2.9$  & ${3.00}^{+1.20}_{-0.10}$& $0.29$ &  $0.50$ \\
      & $\mu\mu$     & 0.38$\pm$0.35 & 0 & $0.15$ &  $3.0$  & ${3.15}^{+0.96}_{-0.11}$  & $0.38$ &  $0.50$ \\\cline{2-9}
      & no-ISR $ee$   & 0.31$\pm$0.22 & 0 & $0.16$ &  $3.2$  & ${2.99}^{+0.78}_{-0.05}$ & $0.38$ &  $0.50$\\
      & $e\mu$   & 0.55$\pm$0.30 & 1 & $0.19$ &  $3.9$  & ${3.33}^{+0.93}_{-0.22}$ & $0.75$ &  $0.27$  \\ 
      & $\mu\mu$   & 0.25$\pm$0.21 & 0 &$0.16$ &  $3.2$  & ${2.94}^{+0.73}_{-0.09}$ & $0.37$ &  $0.50$ \\
\bottomrule
\end{tabular}}
\end{table}

\begin{table}
\centering
\caption{Overview of the dominant systematic uncertainties on the background estimates in
  the same-sign, two-lepton MVA signal regions.  The percentages show the sizes of the uncertainty relative to the total expected background; the range shows the variation among the flavor channels. \label{tab:2LSSMVAsyst}}
\footnotesize{
\begin{tabular}{l rr rr rr rr }
\toprule
 & \multicolumn{2}{c}{SR $\Delta$M20} & \multicolumn{2}{c}{SR $\Delta$M35} & \multicolumn{2}{c}{SR $\Delta$M65} & \multicolumn{2}{c}{SR $\Delta$M100} \\
& ISR & no-ISR & ISR & no-ISR & ISR & no-ISR & ISR & no-ISR \\
\midrule
Reducible background & & & & & & & & \\
\multicolumn{1}{l}{~~~~- Fake lepton composition} & 7--14\%  & 15--20\% & 4--14\%  & 5--17\%  & 5--17\%  & 21\%  & 9--24\% & 20--22\%  \\
\multicolumn{1}{l}{~~~~- Real lepton subtraction} & 13--32\%  & 12--25\%  & 10--20\%  & 18--26\%  & 8--18\%  & 26\% & 15--32\% & 22--33\%  \\
\multicolumn{1}{l}{~~~~- Statistical uncertainty on data} & 5--8\%  & 9--12\%  & 3--7\%  & 4--8\%  & 3--9\%  & 9\% & 5--11\% & 9--11\%  \\
Statistical uncertainty on MC samples & 15--37\% & 7--12\% & 15--28\%  & 8--16\%  &15--43\%  & 16--32\% & 30--45\% &  35--74\% \\
Choice of generator for $WZ$ & 9--17\% & 4--20\% & 15--17\%  & 5--11\%  & 13--20\%& 6--21\%  &  3--27\% & 4--20\%   \\
Choice of generator for $W\gamma$ & 2--3\% & 3--7\% & 2\% & 4--8\% & 3--9\% & - & - & - \\
Jet energy resolution           &  1--18\% & 1--7\%  & 1--7\%  & 6--12\%  & 1--10\% & 1--6\% & 5--70\% & 4--35\% \\
\midrule
Total & 28--60\% & 18--32\% & 28--39\% & 22--37\% & 33--47\% & 33--50\% & 51--92\% & 55--84\% \\
\bottomrule
\end{tabular}}
\end{table}

\FloatBarrier

\subsection{Searches with three light leptons  \label{sec:threelep}}

Previous searches for $\chinoonepm\ninotwo$ production using the three-lepton final state are extended here to increase the sensitivity to compressed SUSY scenarios. 
The three-lepton analysis presented here probes $\ninotwo$--$\ninoone$ mass splittings below 25$\GeV$ using low-$\pt$ leptons and ISR jets.

\subsubsection{Event selection}

Events are selected as described in Section~\ref{sec:evtreco}. 
In addition, signal muons with $\pt\,$$<\,$15$\GeV$ have tightened track and calorimeter isolation requirements of 7\% of the muon $\pt$. 
The stricter muon isolation requirements suppress SM backgrounds with semileptonically decaying $b$/$c$-hadrons, which are larger for muons rather than electrons due to the lower muon-$\pT$ threshold. 
Events must satisfy a single-lepton, dilepton, or trilepton trigger.

Four signal regions are defined with exactly three light leptons, all with $\pT\,$$<\,$$30\,$$\GeV$, and at least one SFOS pair present among the leptons. 
All signal regions veto events with $b$-tagged jets to reduce the $\ttbar$ SM background and events with 8.4$\,$$<\,$$\msfos\,$$<\,$10.4$\GeV$ to suppress backgrounds with leptonic $\Upsilon$ decays. 
The three-lepton signal region selections are summarized in Table~\ref{tab:3LSRdefs}. 
 
The first two signal regions, SR3$\ell$-0a and SR3$\ell$-0b, closely follow the selection in Ref.~\cite{Aad:2014nua}, using $\met$, $\mt$ and $\msfos$ selections. 
SR3$\ell$-0a and SR3$\ell$-0b are defined with $\met\,$$>\,$50$\GeV$ and 30$\,$$<\,$$\mlll\,$$<\,$60$\GeV$ to reject diboson processes. 
Events with a jet with $\pt\,$$>\,$50$\GeV$ are vetoed to be disjoint from the ISR signal region. 
The first signal region, SR3$\ell$-0a, targets the smallest $\ninotwo$--$\ninoone$ mass splittings by selecting events with $\minmsfos$ between 4 and 15$\GeV$. 
In addition, SR3$\ell$-0a requires small $\mt$ to reduce the $WZ$ SM background. 
The second signal region, SR3$\ell$-0b, targets the slightly larger $\ninotwo$--$\ninoone$ mass splittings by selecting events with $\minmsfos$ between 15 and 25$\GeV$. 

The third and fourth signal regions, SR3$\ell$-1a and SR3$\ell$-1b, both require the presence of a $\pt\,$$>\,$50$\GeV$ jet to target signal production with ISR. 
The leptons from a compressed SUSY decay chain would have too low $\pT$ to be reconstructed; however, due to the recoil against the high-$\pt$ ISR jet, all three leptons can be boosted enough to meet the selection requirements. 
The third signal region, SR3$\ell$-1a, targets the smallest $\ninotwo$--$\ninoone$ mass splittings and selects events with 5$\,$$<\,$$\minmsfos\,$$<\,$15$\GeV$. 
Here the leading jet is required to be back-to-back in the transverse plane with the $\met$, $\Delta\phi(\met,\,\rm jet \, 1)\,$$>\,$2.7 rad, and the ratio of leading lepton $\pt$ to the  jet $\pT$ is required to be small, $\pt^{\rm lep\, 1} / \pt^{\rm jet\, 1}\,$$<\,$0.2, to suppress the diboson and $\ttbar$ backgrounds. 
The fourth signal region, SR3$\ell$-1b, targets the slightly larger $\ninotwo$--$\ninoone$ mass splittings by selecting events with 15$\,$$<\,$$\minmsfos\,$$<\,$25$\GeV$. 
To suppress the $WZ$ and $\ttbar$ backgrounds in SR3$\ell$-1b, the angle between the $\met$ and the three-lepton system is required to be large, $\Delta\phi(\met,\,3\ell)\,$$>\,$0.7$\pi$ rad.

\begin{table}[h]
\centering
 \caption{The selection requirements for the three-lepton signal regions, targeting  $\chinoonepm\ninotwo$ production with small mass splittings between the $\chinoonepm/\ninotwo$ and LSP. . \label{tab:3LSRdefs}}
\footnotesize{
 \begin{tabular}{l  c| c|   c| c }
  \toprule
      & \multicolumn{4}{c}{Common}   \\
\midrule
$\ell$ flavor/sign      & \multicolumn{4}{c}{$\ell^{\pm}\ell^{\mp}\ell$, $\ell^{\pm}\ell^{\mp}\ell'$} \\
$\pt^{ \rm lep \, 1}$ & \multicolumn{4}{c}{$<\,$30$\GeV$}  \\
$b$-jet              & \multicolumn{4}{c}{veto}  \\
$\met$               & \multicolumn{4}{c}{$>\,$50$\GeV$}  \\
$\msfos$             & \multicolumn{4}{c}{veto 8.4--10.4$\GeV$}  \\
\midrule
  SR                 & SR3$\ell$-0a   & SR3$\ell$-0b      &     SR3$\ell$-1a & SR3$\ell$-1b \\
\midrule
Central jets                 & \multicolumn{2}{c}{no jets $\pt\,$$>\,$50$\GeV$}  & \multicolumn{2}{|c}{$\geq\,$1 jet $\pt\,$$>\,$50$\GeV$} \\
$\minmsfos$          & 4--15$\GeV$  & 15--25$\GeV$    & 5--15$\GeV$    & 15--25$\GeV$ \\

Other                & 30$\,$$<\,$$\mlll\,$$<\,$60$\GeV$ & 30$\,$$<\,$$\mlll\,$$<\,$60$\GeV$ & $\Delta\phi(\met,\,\rm jet \, 1)\,$$>\,$2.7 rad  & $\Delta\phi(\met,\,3\ell)\,$$>\,$0.7$\pi$ rad \\
                     & $\mt\,$$<\,$20$\GeV$ &    &   $\pt^{\rm lep\, 1} / \pt^{\rm jet \, 1}\,$$<\,$0.2  &  \\

\bottomrule
 \end{tabular}
}
\end{table}

\subsubsection{Background determination}

Several SM processes produce events with three signal leptons. 
The SM background processes are  classified as irreducible background if they lead to events with three or more real leptons, 
or as reducible background if the event has at least one fake or non-prompt lepton.  
The predictions for irreducible and reducible backgrounds are tested in validation regions. 
For this search, irreducible processes include diboson ($WZ$ and $ZZ$), $VVV$, $\ttbar V$, $tZ$ and Higgs boson production and are determined from MC simulation samples.

Reducible processes include single- and pair-production of top quarks, $WW$ production and a single $W$ or $Z$ boson produced in association with jets or photons. 
The dominant reducible background component is $\ttbar$, followed by $Z$+jets. 
The reducible background is estimated using the matrix method, similar to that described in Section~\ref{sec:matrixmethod}. 
In this implementation of the matrix method, the highest-$\pT$ signal electron or muon is taken to be real and only the second and third leptons are used in the matrix method. 
Simulation studies show that neglecting the case that the leading lepton is non-prompt or fake is valid in more than 95\% of the events.

The uncertainty on the reducible background includes the MC statistical uncertainty on the weights for the process-dependent misidentification probabilities, 
the uncertainty on the correction factors for the misidentification probability, the statistical uncertainty on the data events to which the matrix equation is applied and the statistical uncertainty from the misidentification probability measured in simulation.

The systematic uncertainty related to the theoretical modeling of the $WZ$ and $ZZ$ backgrounds is assessed by comparing MC estimates with data in dedicated regions. 
The $WZ$ region requires three light leptons with $\pt\,$$>\,$30$\GeV$, an SFOS pair among the three leptons, 30$<\,$$\met\,$$<\,$50$\GeV$ and one jet with $\pt\,$$>\,$50$\GeV$. 
Events with an SFOS pair or three-lepton invariant mass within 10$\GeV$ of the $Z$ boson mass are vetoed.
The $ZZ$ region is defined with four light leptons with $\pt\,$$>\,$10$\GeV$, two SFOS pairs with invariant mass within 10$\GeV$ of the $Z$ boson mass and $\met\,$$<\,$50$\GeV$. 
This approach for estimating the systematic uncertainties is used here instead of the MC-based approach discussed in Section~\ref{sec:commsyste}.
The $WZ$ and $ZZ$ MC simulation samples are both found to agree with observations in the dedicated regions within 15\%, which is applied as a systematic uncertainty in the three-lepton validation and signal regions.

\begin{figure}[h]
\centering
\subfigure[]{\includegraphics[width=0.49\textwidth]{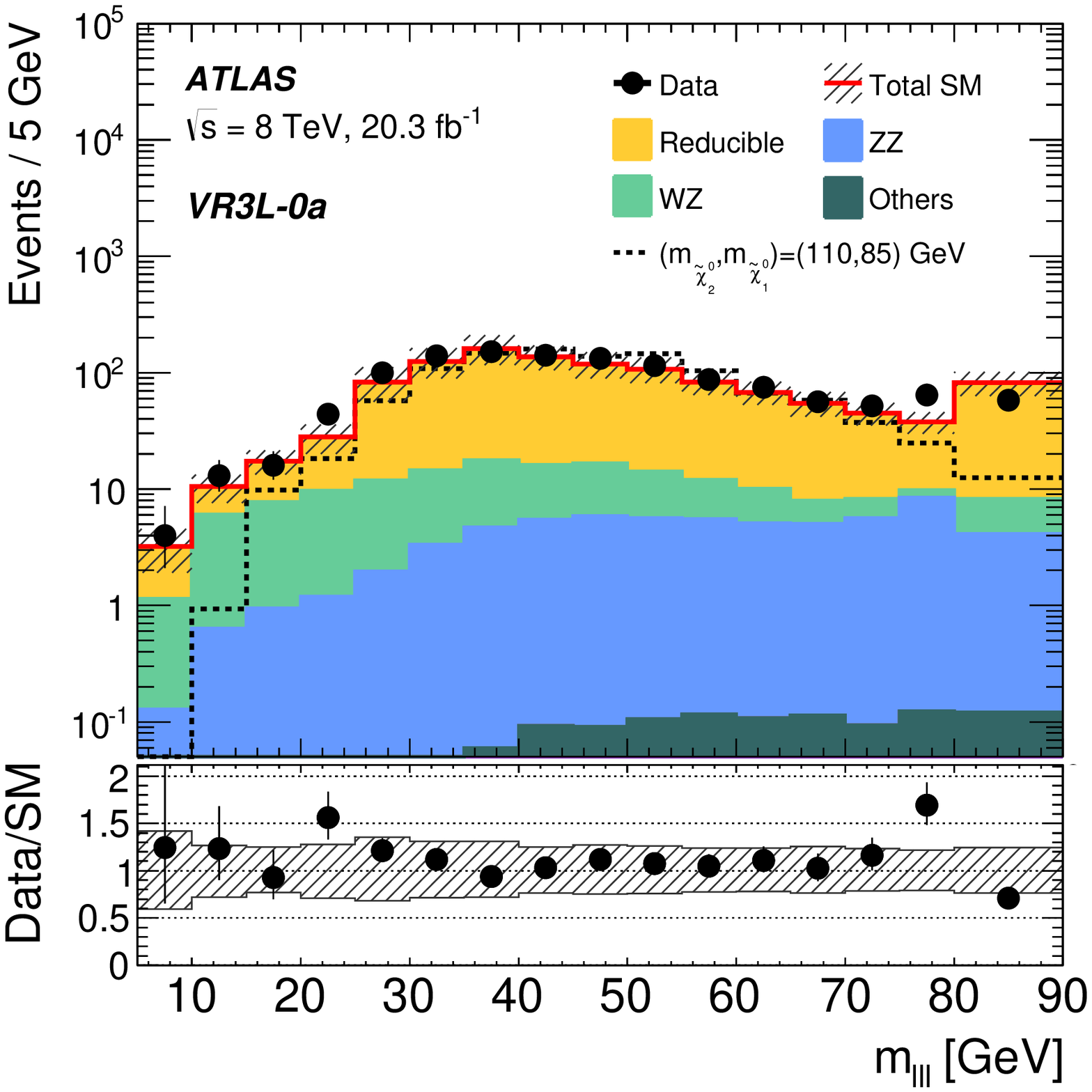}}
\subfigure[]{\includegraphics[width=0.49\textwidth]{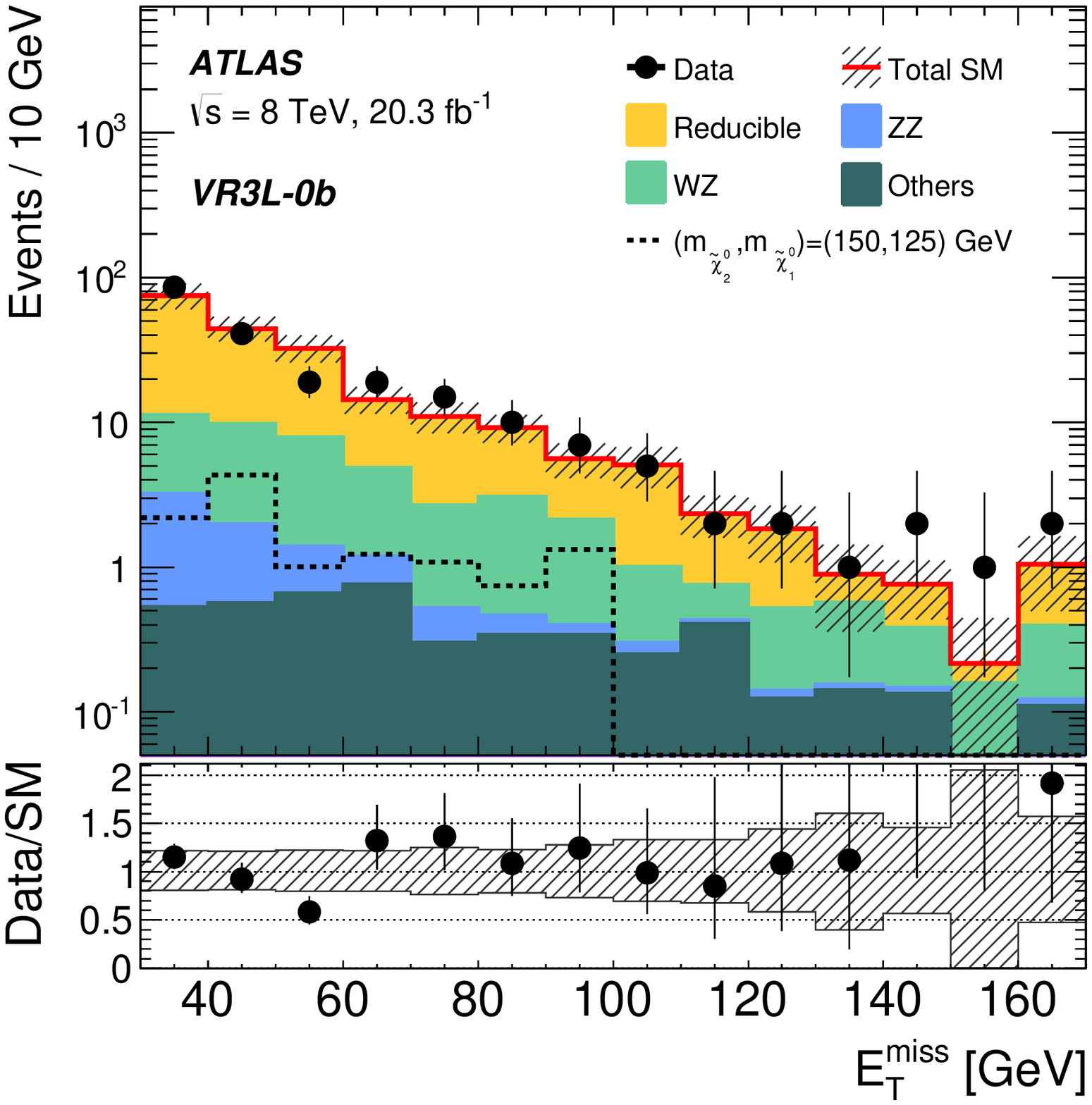}}
\subfigure[]{\includegraphics[width=0.49\textwidth]{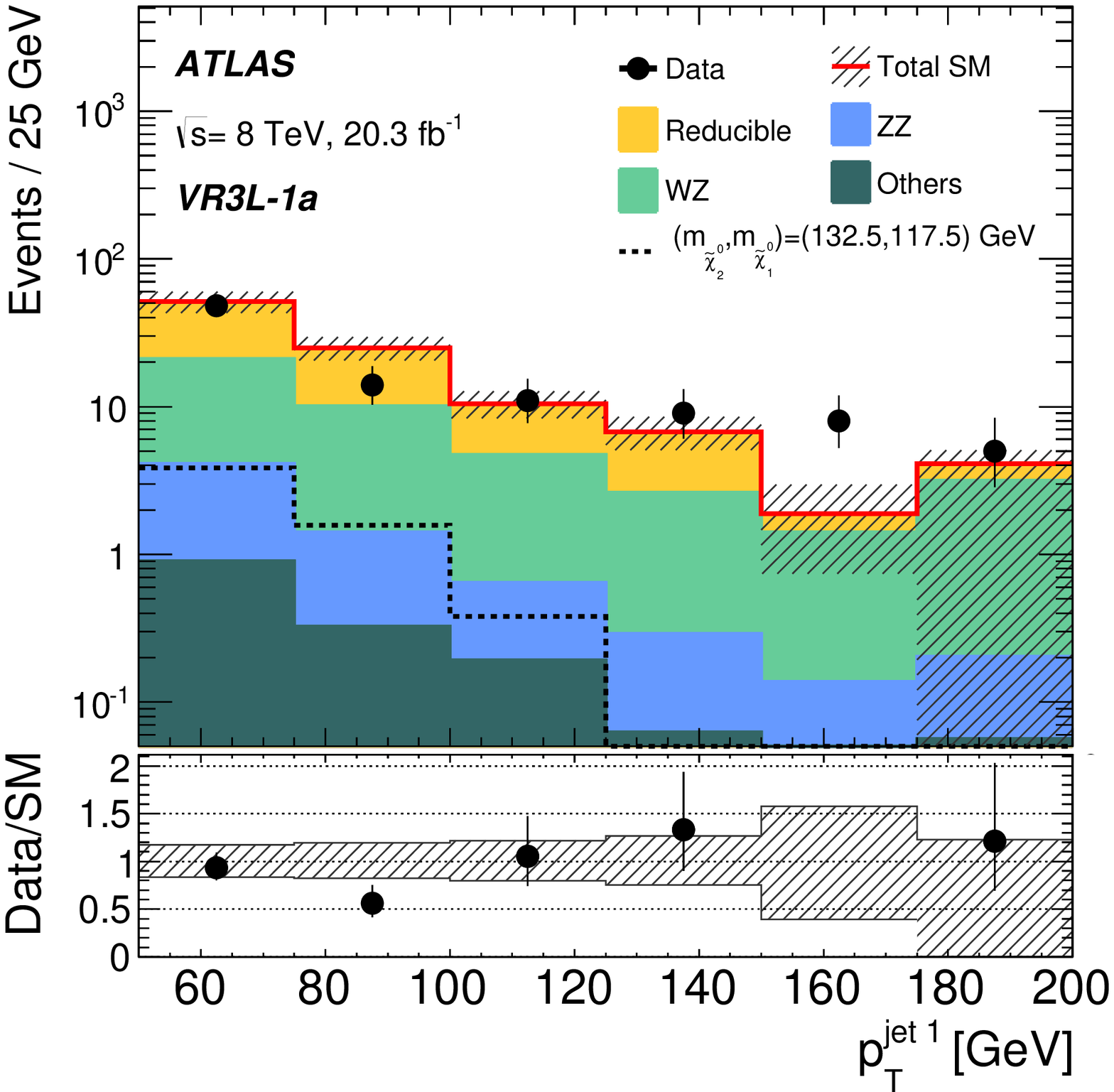}}
\subfigure[]{\includegraphics[width=0.49\textwidth]{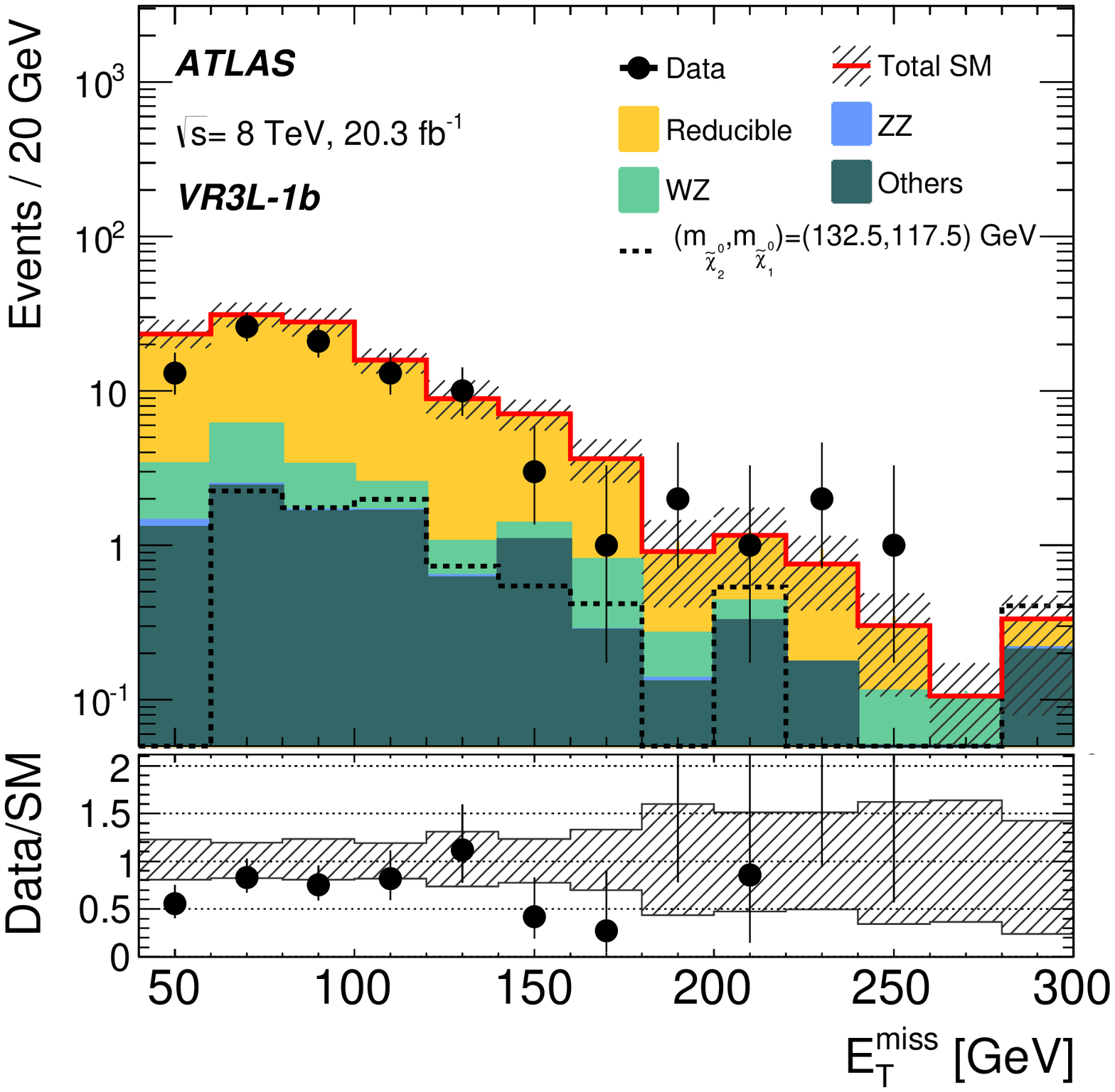}}
\caption{
Distributions in the three-lepton validation regions: (a) three-lepton invariant mass $\mlll$ in VR3$\ell$-0a, (b) $\met$ in VR3$\ell$-0a, (c) transverse momentum of the leading jet $\pt^{\rm jet \, 1}$ in VR3$\ell$-1a, and (d) $\met$ in VR3$\ell$-1a. 
The ``Others'' background category includes $t\bar{t}V$+$t{Z}$, $VVV$ and SM Higgs boson production. 
The lower panel of each plot shows the ratio of data to the SM background prediction. 
The uncertainty band includes both the statistical and systematic uncertainties on the SM prediction. 
The last bin in each distribution includes the overflow. 
\label{fig:3LVR-summary}}
\end{figure}

The background predictions are tested in validation regions that are defined to be adjacent to, yet disjoint from, the signal regions. 
Low-$\met$ validation regions (``a'' regions) and high-$\met$ + $b$-jet validation regions (``b'' regions) are defined to target different background processes. 
The definition of the regions and the targeted processes are shown in Table~\ref{tab:3LVRdefs}. 
In the three-lepton validation regions, the observed data counts and SM expectations are in good agreement within statistical and systematic uncertainties, as shown in Table~\ref{tab:3LbkgVR} and Figures~\ref{fig:3LVR-summary}(a), \ref{fig:3LVR-summary}(b), \ref{fig:3LVR-summary}(c), and \ref{fig:3LVR-summary}(d). 

\begin{table}[h]
\centering
 \caption{The selection requirements for the three-lepton validation regions. 
The ``$Z$ boson'' requirement is defined as $\msfos$ in the range 81.2--101.2$\GeV$. 
\label{tab:3LVRdefs}}
\small{
 \begin{tabular}{l  c| c|   c| c }
  \toprule
      & \multicolumn{4}{c}{Common}   \\
\midrule
$\ell$ flavor/sign      & \multicolumn{4}{c}{$\ell^{\pm}\ell^{\mp}\ell$, $\ell^{\pm}\ell^{\mp}\ell'$} \\
$\minmsfos$          & \multicolumn{4}{c}{$>\,$4$\GeV$} \\
$\msfos$             & \multicolumn{4}{c}{veto 8.4--10.4$\GeV$}  \\
\midrule
  SR                 & VR3$\ell$-0a   & VR3$\ell$-0b      &     VR3$\ell$-1a & VR3$\ell$-1b \\
\midrule
Central jets                 & \multicolumn{2}{c|}{no jets $\pt\,$$>\,$50$\GeV$}  & \multicolumn{2}{c}{$\geq\,$1 jet $\pt\,$$>\,$50$\GeV$} \\
$N_{b-\mathrm{jets}}$           & 0 & 1 & 0 & 1 \\
$\met$               & $<\,$30$\GeV$ & $>\,$30$\GeV$ & $<\,$50$\GeV$ & $>\,$50$\GeV$ \\
$Z$ boson            & veto & -- & veto & veto \\
$\pt^{ \rm lep \, 1}$   & $<\,$30$\GeV$ & -- & -- & --  \\
\midrule
Target Process  & & & &  \\
~~~~Irreducible &   $WZ$ & $WZ$ & $WZ$& $WZ$ \\
~~~~Reducible   & $Z$+jets, $\Upsilon$ &  $\ttbar$ & $Z$+jets & $\ttbar$ \\

\bottomrule
 \end{tabular}
}
\end{table}

\begin{table}[h]
  \centering
  \caption{Estimated and observed yields in the three-lepton validation regions. 
  The uncertainties shown include both statistical and systematic components. 
  The ``Others'' background category includes $\ttbar V$, $VVV$ and SM Higgs boson production. \label{tab:3LbkgVR}}
  \small{
  \renewcommand\arraystretch{1.3}
    \begin{tabular}{ c   cccc }
\toprule
  & VR3$\ell$-0a & VR3$\ell$-0b & VR3$\ell$-1a & VR3$\ell$-1b \\
\midrule
$WZ$ & $108 \pm 20$  & $35 \pm 7$  & $36 \pm 7$  & $9.7^{+2.0}_{-2.2}$  \\
$ZZ$ & $63 \pm 11$  & $5.9 \pm 1.3$  & $5.2 \pm 1.1$  & $0.33^{+0.08}_{-0.07}$  \\
Reducible & $990^{+300}_{-270}$  & $159^{+40}_{-35}$  &  $56 \pm 16$  & $102^{+23}_{-19}$  \\
Others & $1.0\pm 0.8$  & $4.8 \pm 1.7$  & $1.5 \pm 0.6$  & $9.9^{+3.4}_{-3.5}$  \\
\midrule
Total SM  & $1160^{+300}_{-280}$ & $200 \pm 40$  & $99 \pm 17$  & $122^{+24}_{-20}$  \\
Data &  $1247$ & $212$ & $95$ & $93$    \\
\bottomrule
   \end{tabular}
 }
\end{table}

\subsubsection{Results}

\begin{figure}[h]
\centering
\subfigure[]{\includegraphics[width=0.49\textwidth]{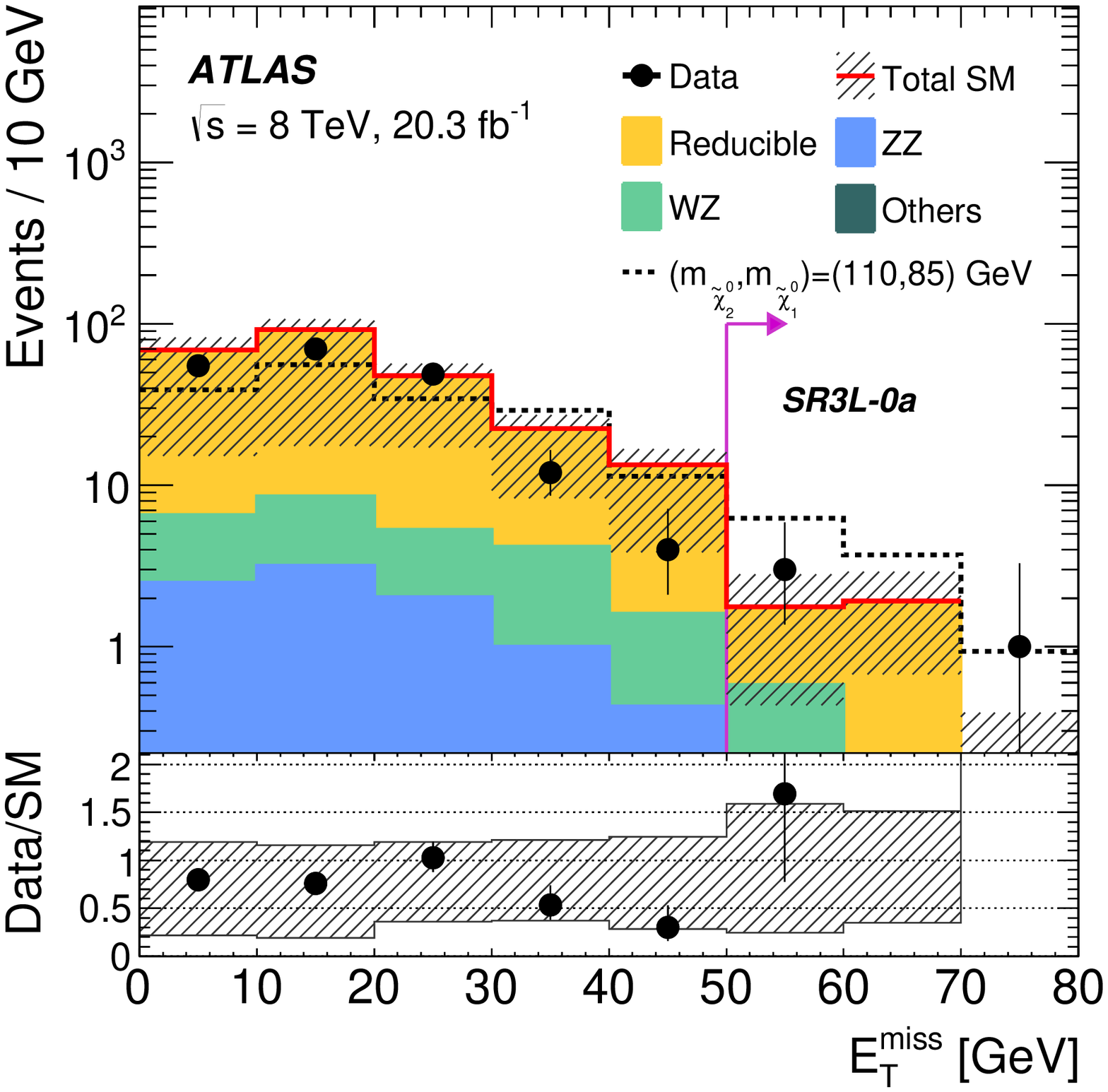}}
\subfigure[]{\includegraphics[width=0.49\textwidth]{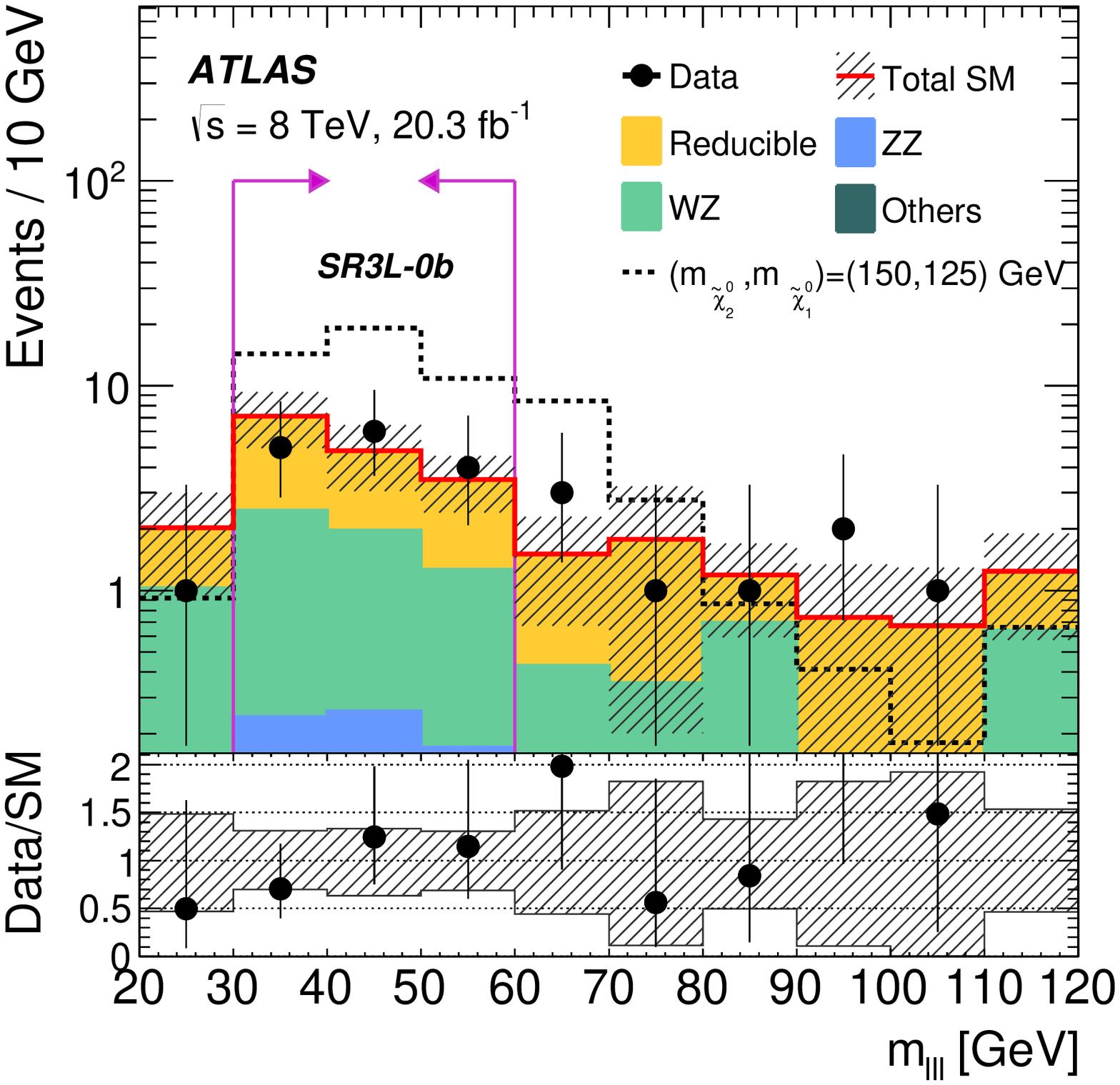}}
\subfigure[]{\includegraphics[width=0.49\textwidth]{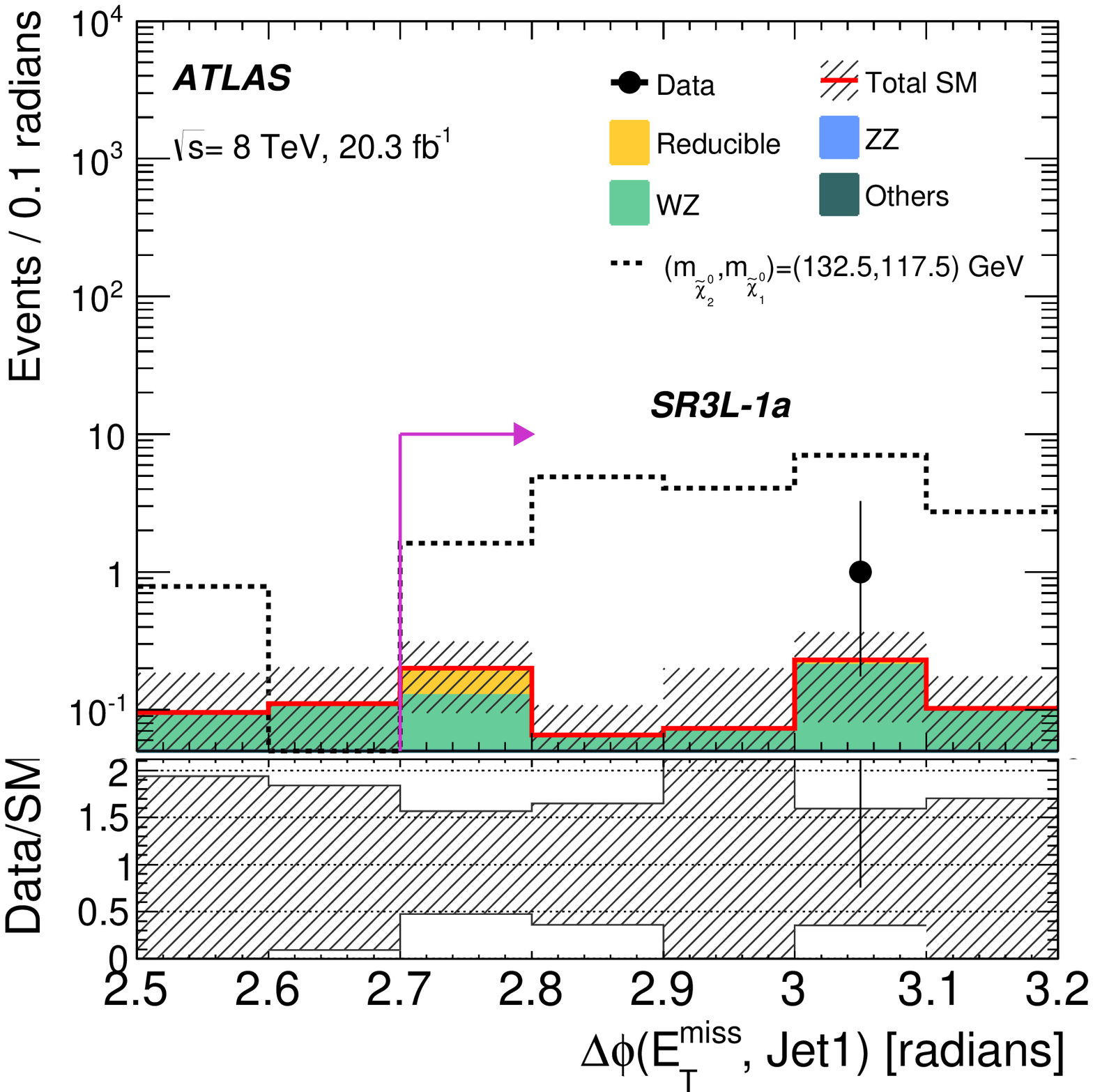}}
\subfigure[]{\includegraphics[width=0.49\textwidth]{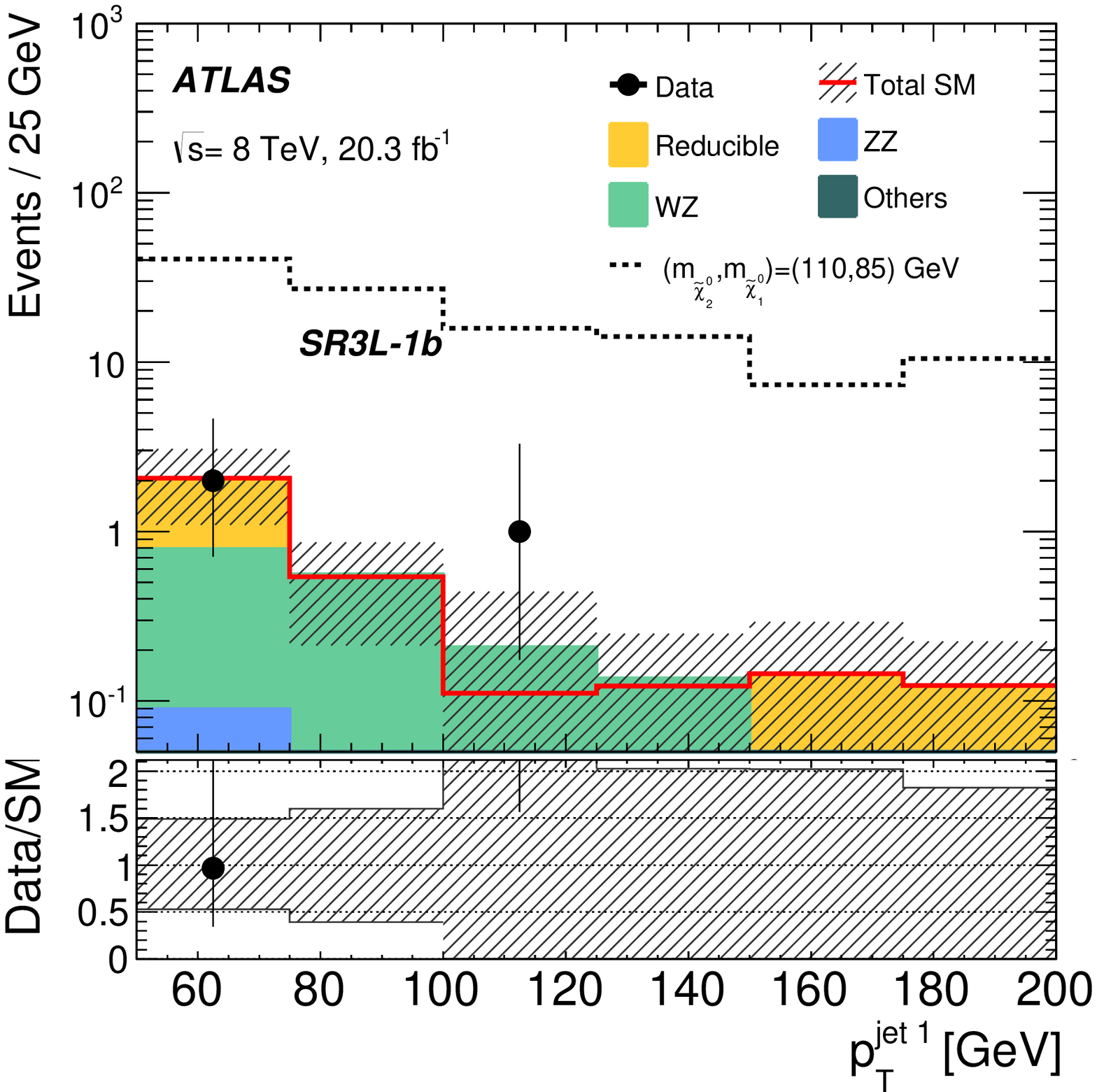}}
\caption{
Distributions in the three-lepton signal regions: (a) $\met$ in SR3$\ell$-0a, (b) $\mlll\,$ in SR3$\ell$-0b, (c) $\Delta\phi(\met,\,\rm jet \, 1)$ in SR3$\ell$-1a, and (d) $\pt^{\rm jet \, 1}$ in SR3$\ell$-1b. 
All are shown prior to the requirements on these variables.
The ``Others'' background category includes $t\bar{t}V$+$t{Z}$, $VVV$ and SM Higgs boson production.
The lower panel of each plot shows the ratio of data to the SM background prediction. 
Arrows indicate the limits on the values of the variables used to define the signal regions.
The uncertainty band includes both the statistical and systematic uncertainties on the SM prediction. 
The last bin in each distribution includes the overflow.
\label{fig:3LSR-summary}}
\end{figure}

The observed number of events in each signal region is shown in Table~\ref{tab:3LSRresults} along with the background expectations and uncertainties, $p_0$-values, $S^{95}_{\rm exp}$, $S^{95}_{\rm obs}$, $\langle\epsilon{\rm \sigma}\rangle_{\rm obs}^{95}$, and the CL$_{b}$ values. 
The sizes and sources of uncertainty on the background estimation in the three-lepton signal regions are shown in Table~\ref{tab:3Lsyst}, where the dominant sources of uncertainty are the statistical uncertainty on the data for the reducible background estimate, and the uncertainty on the electron and muon misidentification probabilities. 
Figures~\ref{fig:3LSR-summary}(a), \ref{fig:3LSR-summary}(b), \ref{fig:3LSR-summary}(c) and \ref{fig:3LSR-summary}(d) show the distributions of the quantities $\met$, $\mlll\,$, $\Delta\phi(\met,\,\rm jet \, 1)$ and $\pt^{\rm jet \, 1}$ in SR3$\ell$-0a, SR3$\ell$-0b, SR3$\ell$-1a and SR3$\ell$-1b regions respectively, prior to the requirements on these variables. 
For illustration, the distributions are also shown for a $\chinoonepm \ninotwo$ scenario with $\slepL$-mediated decays, where the slepton mass is set halfway between the $\chinoonepm$ and the $\ninoone$ masses. 

\begin{table}[h]
  \caption{Expected and observed yields in the three-lepton signal regions. The uncertainties shown include both statistical and systematic components. The ``Others'' background category includes $\ttbar V$, $VVV$ and SM Higgs boson production.
Also shown are the model-independent limits calculated from the three-lepton signal region observations: the one-sided $p_0$-values; the expected and observed upper limits at 95\% CL on the number of beyond-the-SM events ($S^{95}_{\rm exp}$ and $S^{95}_{\rm obs}$) for each signal region, calculated using pseudoexperiments and the CL$_s$ prescription;  the observed 95\% CL upper limit on the signal cross-section times efficiency ($\langle\epsilon{\rm \sigma}\rangle_{\rm obs}^{95}$); and the CL$_{b}$ value for the background-only hypothesis.
\label{tab:3LSRresults}}
  \centering
  \small{
  \renewcommand\arraystretch{1.3}
    \begin{tabular}{ l   c c c c }\toprule
            & SR3$\ell$-0a & SR3$\ell$-0b  & SR3$\ell$-1a & SR3$\ell$-1b \\
\midrule
$WZ$ & $0.59^{+0.47}_{-0.32}$  & $5.0^{+1.5}_{-1.2}$  & $0.54^{+0.20}_{-0.19}$  & $1.6 \pm 0.4$  \\
$ZZ$ & $0.23^{+0.09}_{-0.07}$  & $0.66 \pm 0.16$  & $0.024 \pm 0.013$  & $0.10^{+0.05}_{-0.04}$   \\
Reducible & $2.8^{+1.5}_{-2.2}$  & $9.7^{+3.1}_{-3.6}$  & $0.09 \pm 0.08$  & $1.4^{+1.0}_{-1.1}$   \\
Others & $0.0033^{+0.0036}_{-0.0033}$ & $0.07 \pm 0.05$ & $0.013 \pm 0.010$ & $0.038 \pm 0.021$ \\
\midrule
Total SM & $3.7^{+1.6}_{-2.2}$  & $15.4^{+3.5}_{-3.9}$  & $0.67^{+0.22}_{-0.21}$  & $3.1^{+1.1}_{-1.2}$   \\
Data  & $4$   & $15$    & $ 1 $    & $ 3 $    \\
\midrule
$p_0$ & 0.47 & 0.50 & 0.36 & 0.50 \\
$S_{\rm obs}^{95}$  &  8.3 & 12.6 & 4.0 & 6.1 \\
$S_{\rm exp}^{95}$ & ${8.2}^{+1.7}_{-2.2}$ & ${12.6}^{+5.2}_{-3.0}$ & ${3.8}^{+0.6}_{-0.3}$ & ${6.0}^{+2.1}_{-1.3}$ \\
$\langle\epsilon{\rm \sigma}\rangle_{\rm obs}^{95}$ [fb]  &  0.41 & 0.62 & 0.20 & 0.30 \\
CL$_{b}$ &   0.59 & 0.50 & 0.69 & 0.54 \\
\bottomrule
   \end{tabular}
 }
\end{table}

\begin{table}
\centering
\caption{
Breakdown of the dominant systematic uncertainties on background estimates in the three-lepton signal regions.
The percentages show the size of the uncertainty relative to the total expected background.
\label{tab:3Lsyst}}
\small{
\begin{tabular}{lrrrr}
\toprule
Source of uncertainty        & \multicolumn{1}{c}{SR3$\ell$-0a} & \multicolumn{1}{c}{SR3$\ell$-0b}  & \multicolumn{1}{c}{SR3$\ell$-1a}  & \multicolumn{1}{c}{SR3$\ell$-1b}          \\
\midrule
Reducible background & & & & \\
~~~~- statistical uncertainty                  &    $34\%$     &    $14\%$     &    $11\%$      &   $30\%$   \\
~~~~- muon misidentification probability       &    $30\%$     &    $11\%$     &    $<1\%$      &   $11\%$   \\
~~~~- electron misidentification probability   &    $21\%$     &    $10\%$     &    $2\%$       &    $9\%$   \\
~~~~- heavy-flavor relative contribution       &    $22\%$     &    $5\%$      &    $<1\%$      &    $2\%$   \\
~~~~- light-flavor relative contribution       &    $23\%$     &    $4\%$      &    n/a          &   $<1\%$   \\
~~~~- conversion relative contribution         &    $2\%$      &    $6\%$      &    $<1\%$      &   $10\%$   \\
$\met$ soft-term scale                         &    $12\%$     &    $7\%$      &    $<1\%$      &    $1\%$   \\
Statistical uncertainty on MC samples          &    $4\%$      &    $3\%$      &    $25\%$      &   $10\%$   \\
Theoretical modeling of $WZ$                   &    $2\%$      &    $5\%$      &    $12\%$      &    $8\%$   \\
Cross-section                                  &    $2\%$      &    $2\%$      &    $6\%$       &    $4\%$   \\
\midrule
Total & $59\%$ & $25\%$ & $33\%$ & $39\%$ \\
\bottomrule
\end{tabular}}
\end{table}

\FloatBarrier

\section{Same-sign chargino-pair production via vector-boson fusion \label{sec:samesignvbf}}
This section presents a search for the same-sign chargino-pair production via VBF with subsequent $\slepL$-mediated
chargino decays into final states with two same-sign light leptons, at least two jets and \met.
Although the cross-section for VBF production is significantly lower than that for direct production,
the two additional jets in the event provide a means to separate the signal from the background for compressed spectra scenarios, and complement
the direct production searches that use low-momentum leptons and ISR jets.  

\subsection{Event selection}
Events are selected using the basic reconstruction, object and event selection criteria described in Section~\ref{sec:evtreco}. 
In addition, signal muons with $\pt\,$$<\,$15$\GeV$ have tightened isolation requirements as in the three-lepton analysis described in Section~\ref{sec:threelep}. 
A tighter isolation is needed for muons rather than electrons due to the lower $\pT$ threshold for muons. 
The stringent lepton isolation suppresses the dominant reducible background processes.
Events are required to satisfy an \met\ trigger. 

One signal region, SR2$\ell$-2, is defined with exactly two same-sign light leptons, 
at least two jets (central light or forward) and large missing transverse momentum \met\,$>\,120\GeV$.
In order to select events that originate from VBF production, the highest-\pt\ jet (jet$\,$1) and the second highest-\pt\ jet (jet$\,$2) are required to 
have large invariant mass, \mjj$\,>\,350\GeV$, be well separated in pseudorapidity, \dEtajj$\,>1.6$, and  be in opposite sides of the detector, $\eta^{\mathrm{jet\,1}}\cdot\eta^{\mathrm{jet\,2}}\,<\,0$. 
The last requirement greatly reduces the SM background originating from non-VBF diboson and Higgs boson production. 
The residual SM background originating from diboson and top-quark production is minimized by requiring the events to have
no $b$-tagged jets, moderate invariant mass of the two leptons (\mll$\,<\,100\GeV$), small stransverse mass (\mttwo$\,<\,40\GeV$) and a high-$\pT$ jet (\ptjone$\,>\,95\GeV$). 
In addition, requirements are made on the ratios of the jet $\pT$, $\met$, $\pT^{jj}$ and $\pTll$. The SR definition is summarized in Table~\ref{tab:2LSSVBFSRdefs}. 

\begin{table}[h]
\centering
\caption{The selection requirements for the same-sign, two-lepton VBF signal region, targeting $\chinoonepm\chinoonepm$ production via VBF with small mass splittings between the $\chinoonepm$ and LSP. \label{tab:2LSSVBFSRdefs}}
\small{
\begin{tabular}{c | c }
\toprule
 & \multicolumn{1}{c}{SR2$\ell$-2} \\
\midrule
$\ell$ flavor/sign 	& $\ell^{\pm}\ell^{\pm}$, $\ell^{\pm}\ell^{\prime\pm}$ \\
Jets  			& $\ge\,2$ \\
Central $b$-jets  	& veto  \\
\met\ [$\GeV$] 		& $>\,120$ \\
\mttwo\ [$\GeV$] 		& $<\,40$ \\
\mll\ [$\GeV$] 		& $<\,100$ \\
\ptjone\ [$\GeV$] 		& $>\,95$ \\
\mjj\ [$\GeV$] 		& $>\,350$ \\
$\eta^{\mathrm{jet1}}\cdot\eta^{\mathrm{jet2}}$ & $<\,0$ \\
\dEtajj 			& $>\,1.6$ \\
\pTll/\met 			& $<\,0.4$ \\
\ptjone/\met 		& $<\,1.9$ \\
\pTll/$\pt^{jj}$ 		& $<\,0.35$ \\
\bottomrule
\end{tabular} }
\end{table}

\FloatBarrier

\subsection{Background determination}
Several SM processes lead to events with two same-sign signal leptons. 
The irreducible background is dominated by diboson production, which is estimated using MC simulation samples. 
The dominant reducible background component is from $W$+jets production, followed by $t\bar{t}$ production, and these are estimated using a data-driven technique called the ``fake factor method'', similar to that described in Ref.~\cite{ATLAS:2014aga}. 
The production of $W\gamma$ is also an important background component, and is modeled using MC simulation samples. 
The charge-flip background is estimated by applying data-driven corrections to the MC simulation samples, following the procedure outlined in Section~\ref{sec:SSMVAbg}.

The fake factor method estimates the contributions from processes that produce one or two fake or non-prompt leptons using 
data events that contain one signal lepton and one lepton failing to satisfy the signal lepton requirements.  
These events are scaled by a ``fake factor'' to predict the reducible background in the signal region. 
The fake factor is defined as the ratio of events with two signal leptons to events with one signal lepton and one lepton failing the signal lepton requirements. 
It is measured in data using a control sample of jets faking leptons in $Z\rightarrow \ell\ell$ events. 
The SM background process dependence of the fake factor is studied using simulation, and no strong dependence is observed. 
Residual differences are covered by assigning a 30\% uncertainty, independent of the lepton \pt, to the fake factor.
The uncertainty on the reducible background estimate ranges from 37\% to 42\%, depending on the channel ($ee$, $\mu\mu$ or $e\mu$), 
and is dominated by the prompt lepton contamination in the control sample and the uncertainty on the extrapolation of fake factors into the signal region.

The contributions from diboson processes are estimated using MC simulation samples. 
\Sherpa\ is used to produce all diboson samples, taking into account both the strong and electroweak production of associated jets.
The $W^{\pm}W^{\pm}$+2jets and $WZ$+2jets processes are normalized to NLO cross-sections using corrections evaluated in dedicated VBF fiducial regions at the parton level.
The corrections are calculated separately for strong and electroweak jet production. 
For the $W^{\pm}W^{\pm}$+2jets production, the fiducial cross-section is calculated using \POWHEGBOX+\PYTHIA~\cite{Nason:2004rx,Frixione:2007vw,Alioli:2010xd} and the fiducial region is defined to be identical to the signal region at the parton level, except for the lepton isolation requirement.
For the $WZ$+2jets production, the fiducial cross-sections are calculated using \VBFnlo-2.7.0~\cite{Arnold:2011wj}. 
Since it is not possible to define a fiducial region that is identical to the signal region using \VBFnlo-2.7.0, a looser set of requirements is imposed.
The generator modeling uncertainty is estimated by comparing \POWHEGBOX+\PYTHIA\ with \VBFnlo-2.7.0 for $W^{\pm}W^{\pm}$+2jets production, and parton showering uncertainties are estimated by comparing \POWHEGBOX+\Herwig\ with \POWHEGBOX+\PYTHIA. 
The impact of the choice of renormalization and factorization scales is evaluated by varying each between 0.5 and 2 times the nominal values. 
The uncertainties due to the PDFs are evaluated using 90\% CL CT10 PDF eigenvectors.
Finally, the interference between the strong and electroweak jet production is studied at LO accuracy using \Sherpa\
and is found to have a negligible effect on the combined fiducial cross-section in the signal region. 

\begin{figure}[h]
\centering
\subfigure[]{\includegraphics[width=0.49\textwidth]{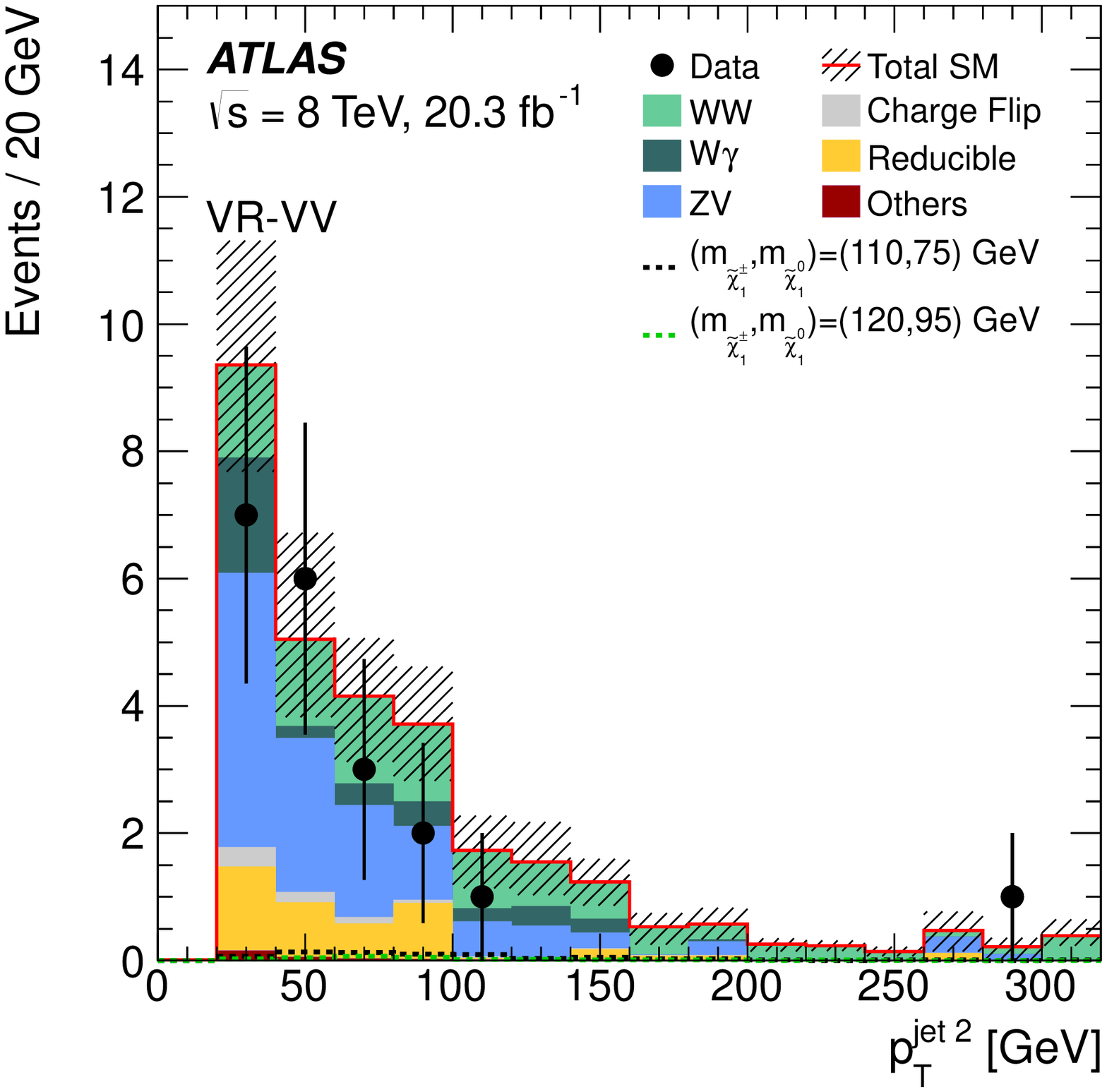}}
\subfigure[]{\includegraphics[width=0.49\textwidth]{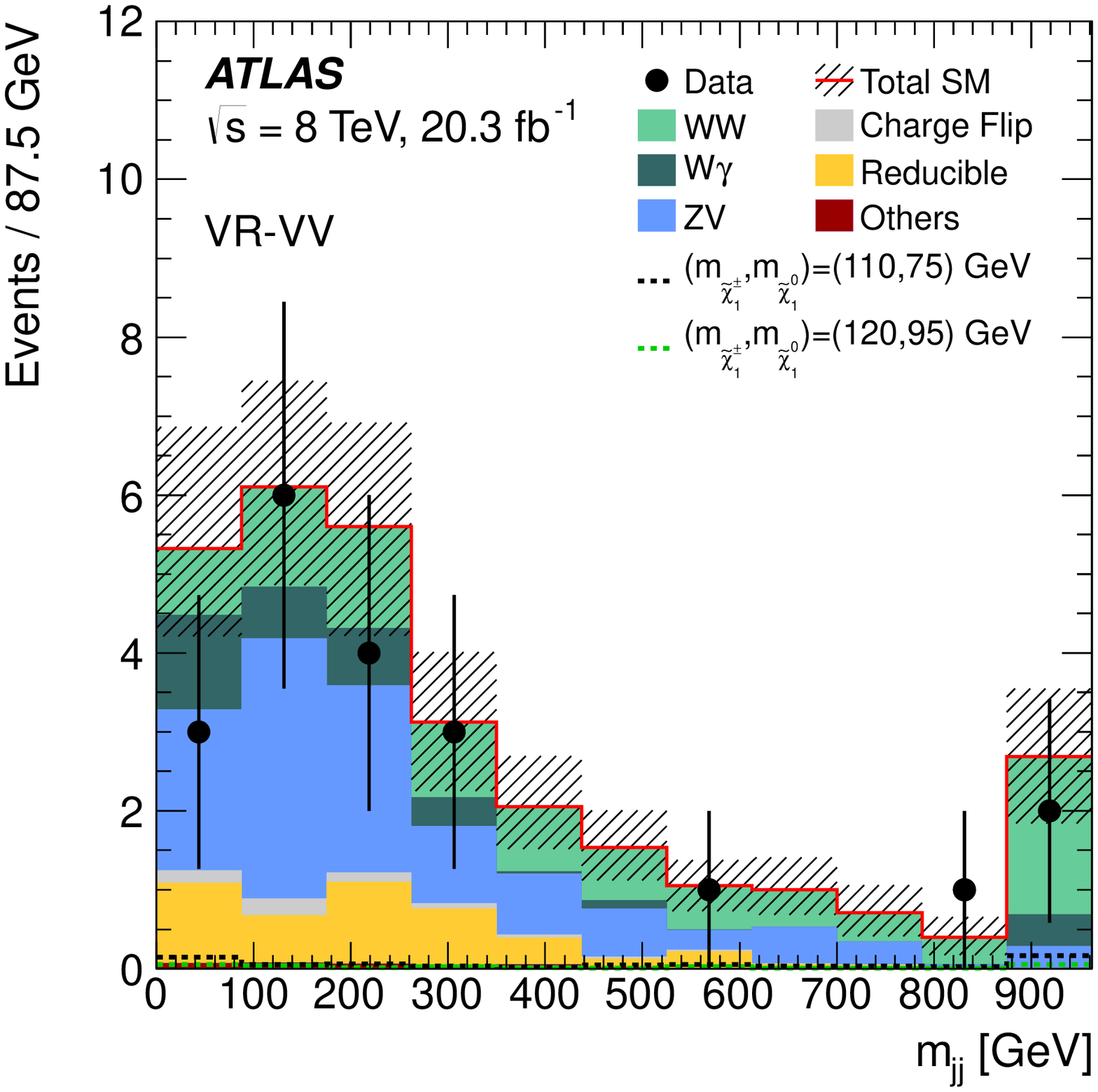}}
\subfigure[]{\includegraphics[width=0.49\textwidth]{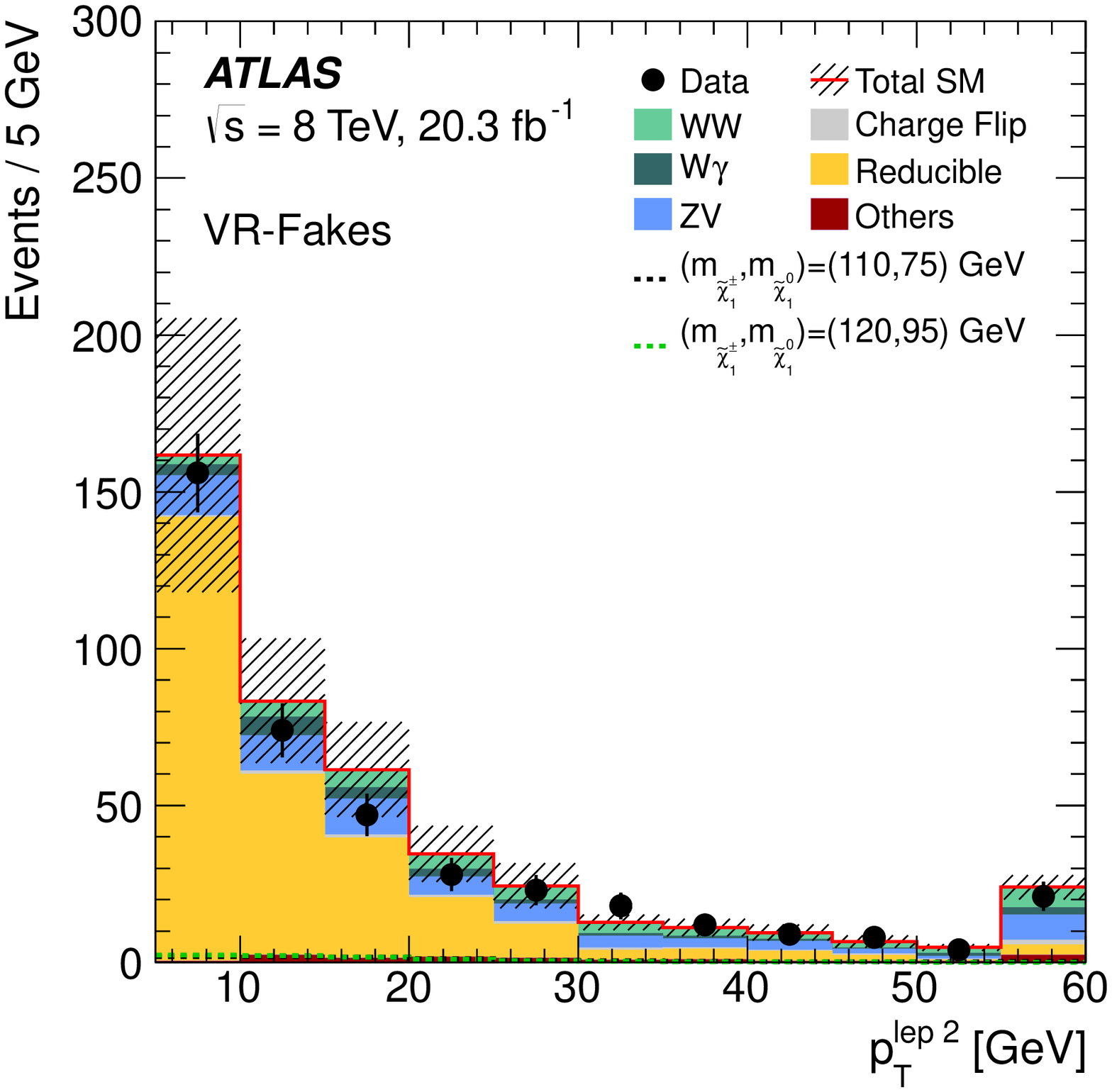}}
\subfigure[]{\includegraphics[width=0.49\textwidth]{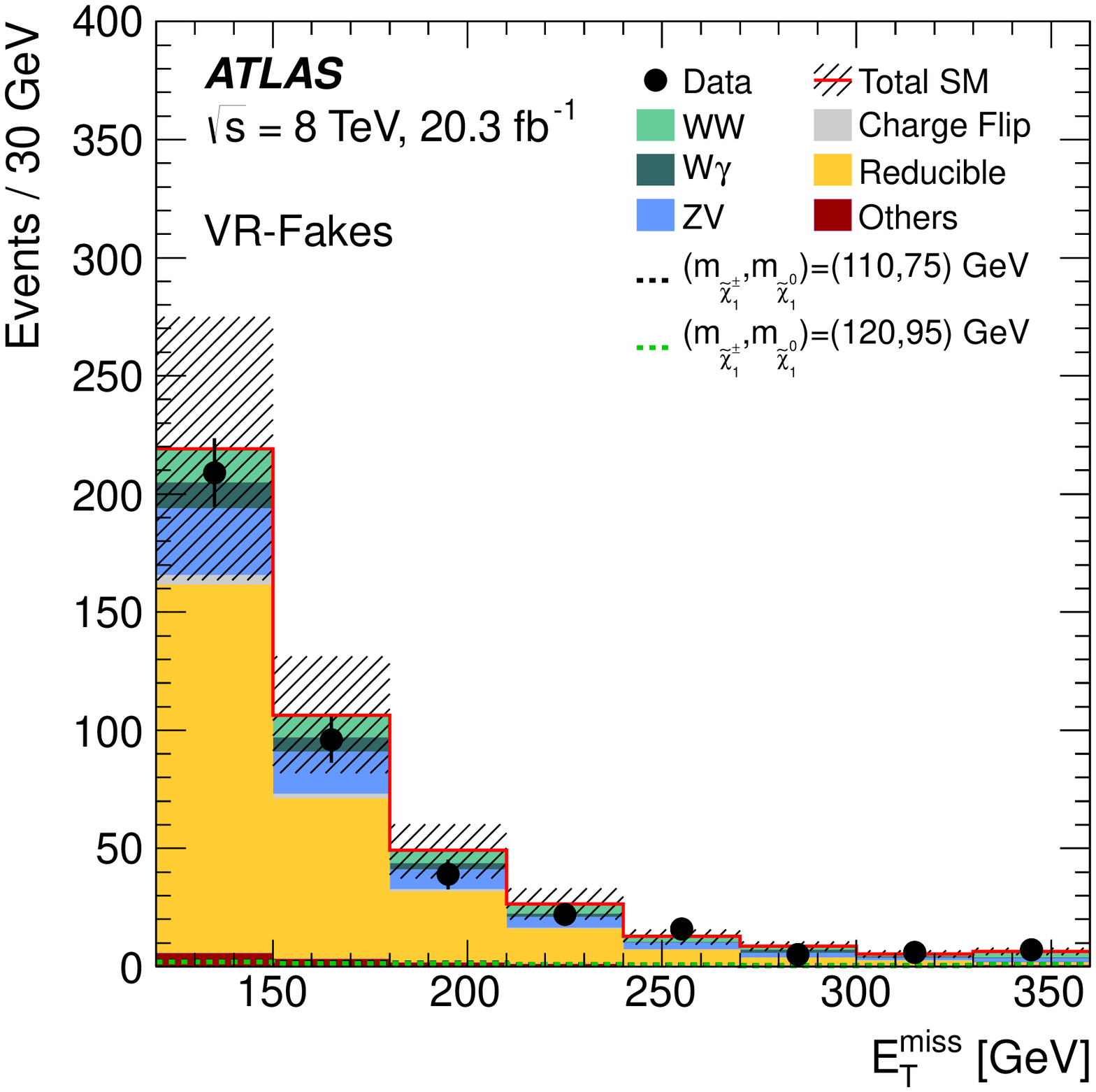}}
\caption{
For events in the same-sign VBF validation region VR-Fakes, the (a) transverse momentum of the second leading jet \ptjtwo\ and (b) invariant mass of the two leading jets \mjj\ in VR-$VV$, and (c) transverse momentum of the second leading lepton $\pt^{\rm lep 2}$ and (d) \met. 
The ``Others'' background category includes $t\bar{t}V$+$t{V}$, $VVV$ and SM Higgs boson production.
The uncertainty band includes both the statistical and systematic uncertainties on the SM prediction. 
The last bin in each distribution includes the overflow.
\label{fig:SSVBFSRplots}}
\end{figure}

The background predictions are tested in VRs that are defined to be as kinematically close to the SR as possible.
The first VR, VR-Fakes, is defined with two signal light leptons, large \met\ and at least two jets to test backgrounds with
fake and non-prompt leptons modeled by the fake factor method.
The second VR, VR-$VV$, adopts the same requirements as the VR-Fakes, in addition to higher lepton-\pt\ thresholds and a $b$-jet veto
that allow it to test the MC modeling of the diboson background.
By definition, the VRs are not disjoint from the SR, but have negligible overlaps. 
The overlap between the VR-Fakes (VR-$VV$) and the SR is 2.4\% (0.2\%)
and the largest signal contamination is 1.9\% (0.9\%) of the total expected background in the VR-Fakes (VR-$VV$). 
The definitions of the validation regions are shown in Table~\ref{tab:2LSSVBFVRdefs}, along with the targeted processes.
The yields in the VRs are shown in Table~\ref{tab:SSVBFSR-results}, where the background expectation is in good agreement with the observed data, within the total uncertainties. 
Figures~\ref{fig:SSVBFSRplots}(a), \ref{fig:SSVBFSRplots}(b), \ref{fig:SSVBFSRplots}(c), and \ref{fig:SSVBFSRplots}(d) show the distributions of $\pT^{\rm lep 2}$ and $m_{jj}$ in VR-$VV$, along with $\pT^{\rm lep 2}$ and $\met$ in VR-Fakes, with good agreement observed.

\begin{table}[h]
\centering
\caption{The selection requirements for the same-sign, two-lepton VBF validation regions. \label{tab:2LSSVBFVRdefs}}
\begin{tabular}{c | c | c }
\toprule
\multicolumn{3}{c}{Common} \\
\midrule
$\ell$ flavor/sign & \multicolumn{2}{c}{$\ell^{\pm}\ell^{\pm}$, $\ell^{\pm}\ell^{\prime\pm}$} \\ 
\met\ [$\GeV$] & \multicolumn{2}{c}{$>\,120$} \\ 
Jets & \multicolumn{2}{c}{$\ge\,2$} \\
\midrule
 & VR-$VV$ & VR-Fakes \\
\midrule
$\pt^{\rm lep 1}$ [$\GeV$] 		& $>40\GeV$ & -- \\
$\pt^{\rm lep 2}$ [$\GeV$] 		& $>40\GeV$ & -- \\
Central $b$-jets  	& veto & -- \\
\midrule
Target process & Dibosons & Non-prompt and fake leptons \\
\bottomrule
\end{tabular}
\end{table}

\subsection{Results}

\begin{figure}[h]
\centering
\subfigure[]{\includegraphics[width=0.49\textwidth]{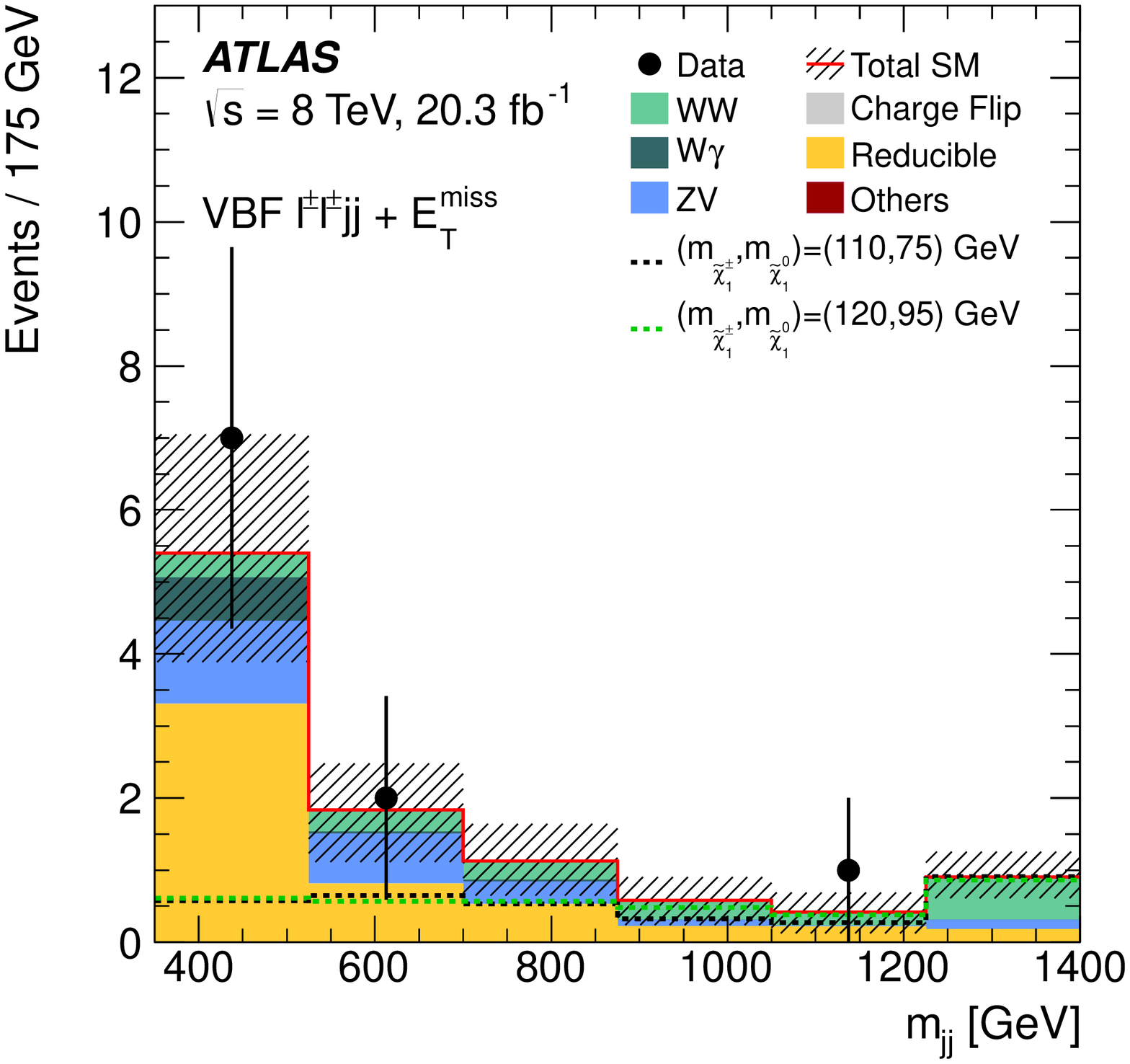}}
\subfigure[]{\includegraphics[width=0.49\textwidth]{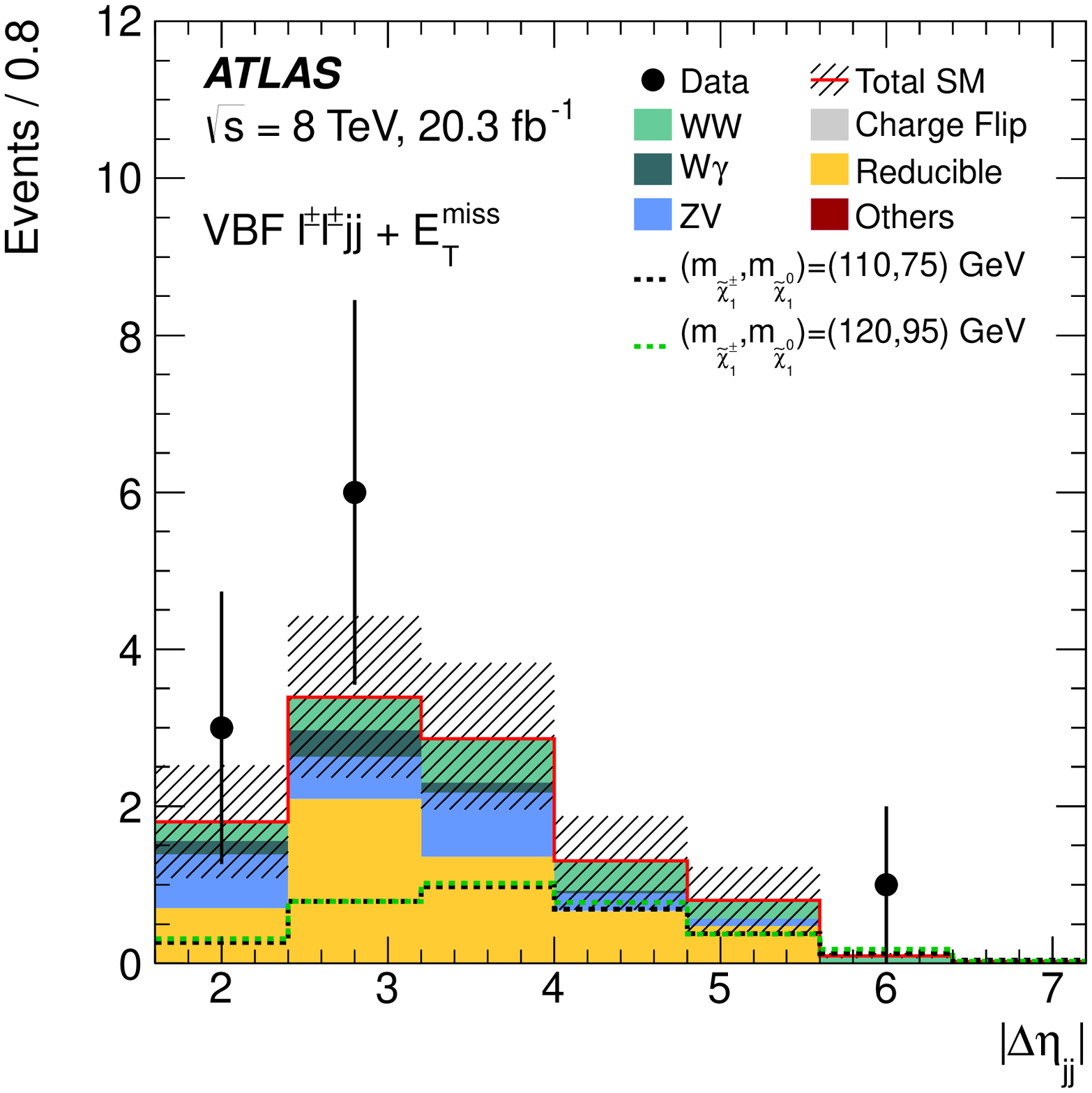}}
\subfigure[]{\includegraphics[width=0.49\textwidth]{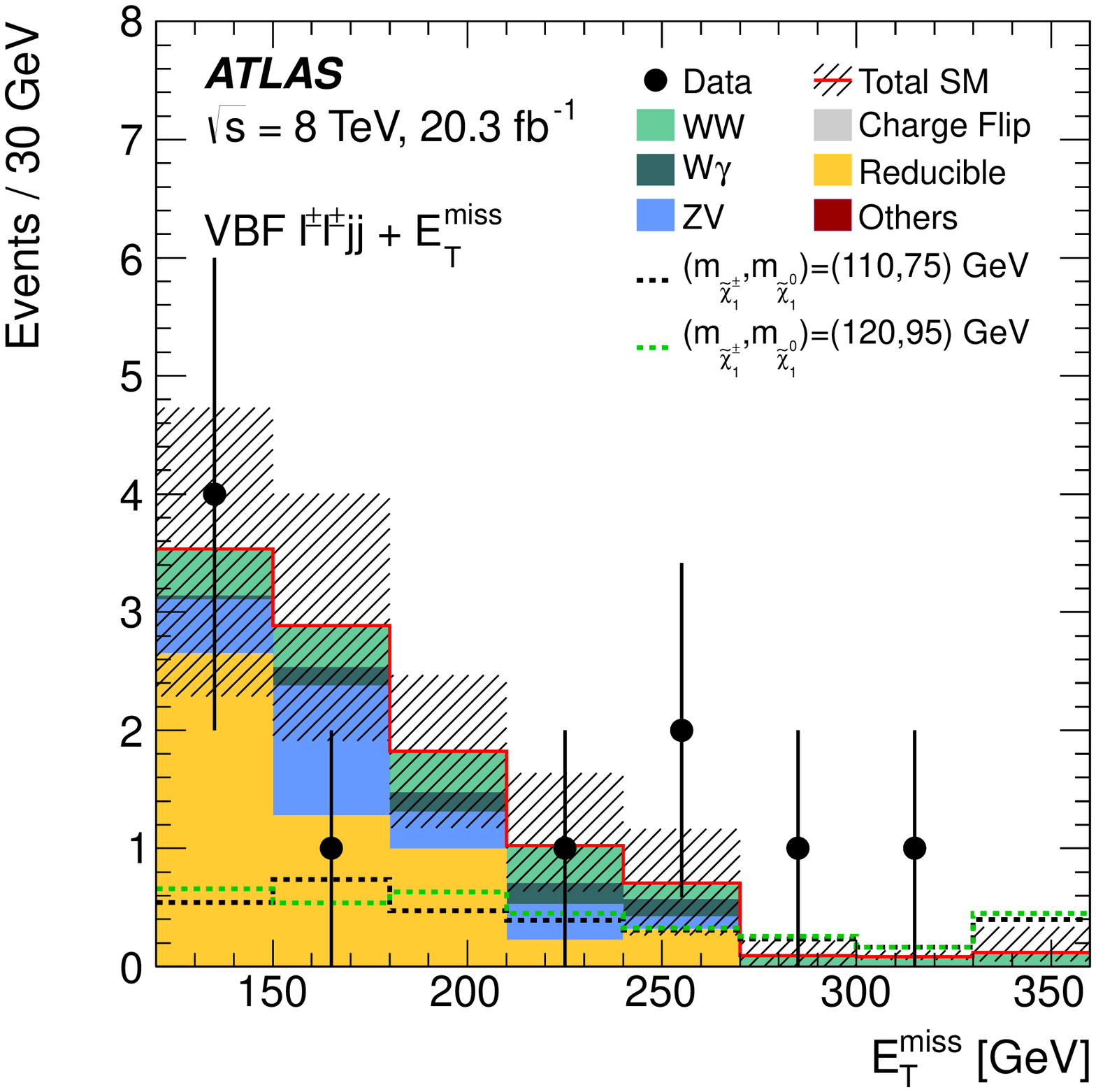}}
\subfigure[]{\includegraphics[width=0.49\textwidth]{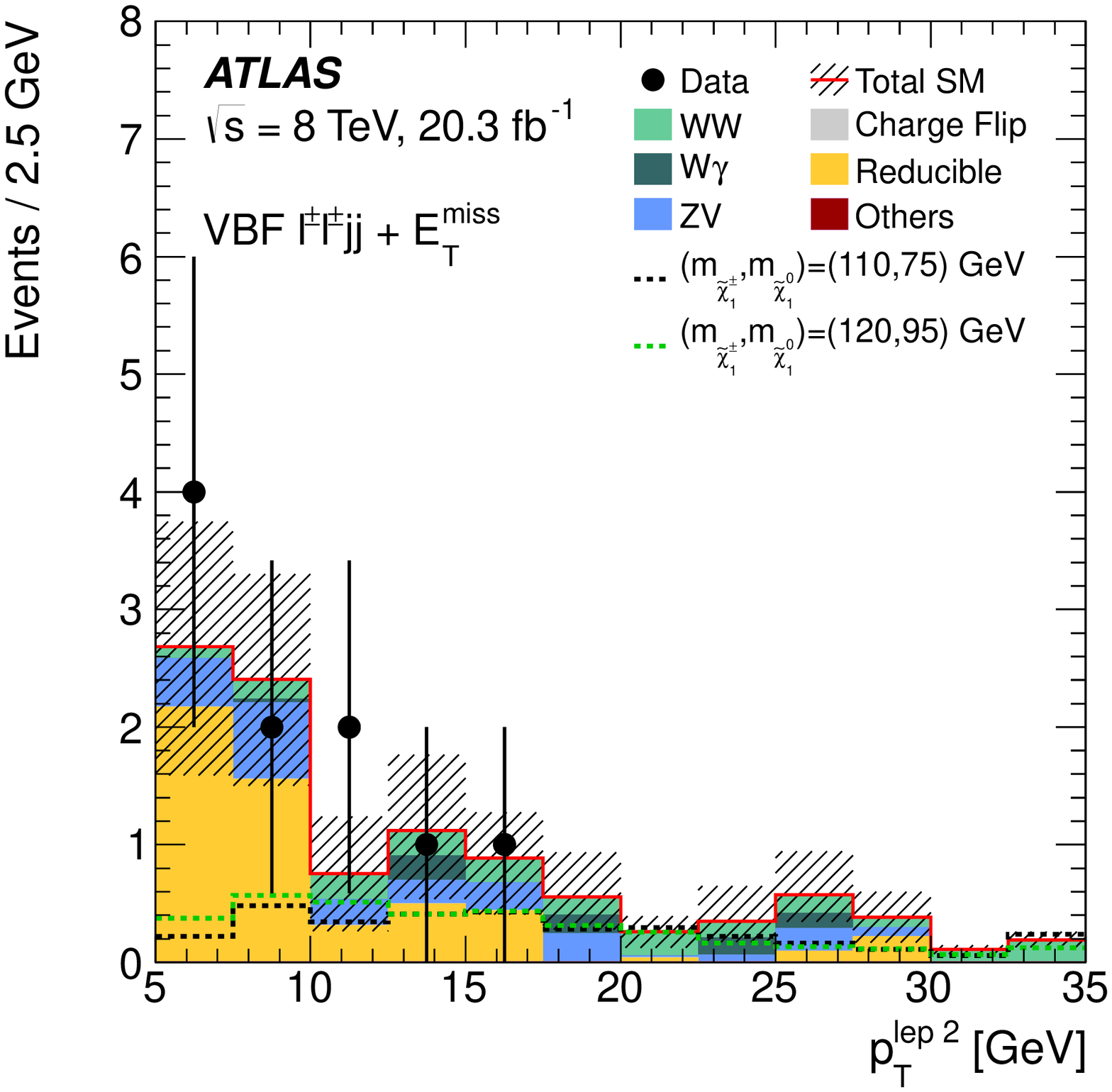}}
\caption{
For events in the same-sign VBF signal region, the (a) \mjj, (b) separation in $\eta$ between the two leading jets \dEtajj, (c) \met, and (d) \ptltwo\ in SR2$\ell$-2. 
The ``Others'' background category includes $t\bar{t}V$+$t{Z}$, $VVV$ and SM Higgs boson production.
The uncertainty band includes both the statistical and systematic uncertainties on the SM prediction. 
The last bin in each distribution includes the overflow.
\label{fig:SSVBFSRsummary}}
\end{figure}

The observed number of events in the signal region is shown in Table~\ref{tab:SSVBFSR-results} along with the background expectation and uncertainties, $p_0$-value, $S^{95}_{\rm exp}$, $S^{95}_{\rm obs}$, $\langle\epsilon{\rm \sigma}\rangle_{\rm obs}^{95}$, and the CL$_{b}$ value. 
No significant excess with respect to the SM expectation is observed. 
A breakdown of the different sources of systematic uncertainty in the signal region, including those described in Section~\ref{sec:commsyste}, is shown in Table~\ref{tab:SSVBFsyst}. 
Figures~\ref{fig:SSVBFSRsummary}(a), \ref{fig:SSVBFSRsummary}(b), \ref{fig:SSVBFSRsummary}(c), and \ref{fig:SSVBFSRsummary}(d) show the distributions of the quantities \mjj, \dEtajj, $\met$ and $\pt^{\rm lep 2}$ in the signal region.

\begin{table}[h]
\caption{\label{tab:SSVBFSR-results}
Observed and expected number of events in the same-sign, two-lepton VBF validation and signal regions.
The numbers of signal events are shown for the \chinoonepm\chinoonepm\ 
VBF simplified model with $\slepL$-mediated decays, with the \chinoonepm\ and \ninoone\ masses in $\GeV$.
The uncertainties shown include both statistical and systematic components.
The model-independent limits are also shown: 
the one-sided $p_0$ value; 
the expected and observed upper limit at 95\% CL on the number of beyond-the-SM events ($S^{95}_{\rm exp}$ and $S^{95}_{\rm obs}$) for the signal region, 
calculated using pseudoexperiments and the CL$_s$ prescription;  
the observed 95\% CL upper limit on the signal cross-section times efficiency ($\langle\epsilon{\rm \sigma}\rangle_{\rm obs}^{95}$); 
and the CL$_{b}$ value for the background-only hypothesis.
}
\centering
\small{
\begin{tabular}{l | c | c | c }
\toprule
 					& VR-$VV$ & VR-Fakes & SR2$\ell$-2   \\
\midrule
$\ell$ flavor/sign 		& $\ell^{\pm}\ell^{\pm}$, $\ell^{\pm}\ell^{\prime\pm}$ & $\ell^{\pm}\ell^{\pm}$, $\ell^{\pm}\ell^{\prime\pm}$  & $\ell^{\pm}\ell^{\pm}$, $\ell^{\pm}\ell^{\prime\pm}$  \\
\midrule
Expected background	&  & & \\
\rule{0pt}{2.5ex}\quad $W^{\pm}W^{\pm}$		& $8.9^{+1.0}_{-1.1}$   	&  $41 \pm 13$ 		& $1.95^{+0.21}_{-0.23}$  \\
\rule{0pt}{2.5ex}\quad $W\gamma$		        & $3.5 \pm 0.8$   		&  $22.8^{+4.2}_{-2.5}$  	& $0.67^{+0.52}_{-0.31}$  \\
\rule{0pt}{2.5ex}\quad $WZ$    				& $11.0 \pm 3.0$   		&  $65 \pm 16$ 		& $2.3^{+0.8}_{-0.9}$        \\
\rule{0pt}{2.5ex}\quad $ZZ$    				& $0.65^{+0.20}_{-0.19}$  &   $1.7 \pm 0.4$		& $0.05^{+0.11}_{-0.17}$  \\
\rule{0pt}{2.5ex}\quad Reducible 			& $4.0 \pm 2.2$   		&  $280 \pm 100$		& $5.2 \pm 2.0$          \\
\rule{0pt}{2.5ex}\quad Charge-flip			& $0.7 \pm 0.7$    &  $8 \pm 4 $ & $0.03^{+0.04}_{-0.02}$         \\
\rule{0pt}{2.5ex}\quad Others			& $0.32^{+0.07}_{-0.06}$    &  $13.6 \pm 1.5 $ & $0.013 \pm 0.007$         \\
\rule{0pt}{2.5ex}Total						& $29 \pm 5$  &  $430 \pm 100$ & $10.3 \pm 2.3$            \\
\midrule
Observed events 						&  $20$    &  $400$ &  $10$         \\
\midrule
Predicted signal 		&  &  &          \\
$(m_{\chinoonepm},m_{\ninoone})=(120,95)$ & $0.25 \pm 0.03$   & $8.32 \pm 0.19$ & $3.47 \pm 0.12$        \\
\midrule
\rule{0pt}{2.5ex}$p_0$  			& --- & --- & $0.50$\\
\rule{0pt}{2.5ex}$S_{\rm obs}^{95}$  	& --- & --- & $8.4$ \\
\rule{0pt}{2.5ex}$S_{\rm exp}^{95}$ 	& --- & --- & $8.7^{+3.9}_{-2.5}$ \\
\rule{0pt}{2.5ex}$\langle\epsilon{\rm \sigma}\rangle_{\rm obs}^{95}$[fb]  & --- & --- & $0.41$ \\
\rule{0pt}{2.5ex}CL$_{b}$ 			& --- & --- & $0.47$ \\
\bottomrule
\end{tabular}}
\end{table}

\begin{table}[h]
\centering
\caption{\label{tab:SSVBFsyst}
The dominant systematic uncertainties on the background estimates for the same-sign, two-lepton VBF signal region.
The percentages show the size of the uncertainty relative to the total expected background.
Because of correlations between the systematic uncertainties, the total
uncertainty is different from the sum in quadrature of the individual uncertainties.
}
\small
\begin{tabular}{l|c}
\toprule
 Source of uncertainty & SR2$\ell$-2 \\
\midrule
Fake factor closure test  & $13\%$ \\
Statistical uncertainty on the reducible background & $11\%$      \\
$WZ$+2jets scale and PDF & $5\%$     \\
Statistical uncertainty on $WZ$+2jets  & $4\%$\\
Statistical uncertainty on the electron fake factor & $3\%$ \\
Jet energy resolution	& $3\%$     \\
Statistical uncertainty on $W^{\pm}W^{\pm}$+2jets & $3\%$\\
$W^{\pm}W^{\pm}$+2jets scale and PDF & $1\%$\\
\midrule
Total & $21\%$ \\
\bottomrule
\end{tabular}
\end{table}

\FloatBarrier


\section{Interpretation of results  \label{sec:interpretation}}

Previous ATLAS searches for EW SUSY production~\cite{Aad:2014vma,Aad:2014nua,Aad:2014iza,Aad:2014yka,Aad:2015jqa} are combined with the new analyses presented in Sections~\ref{sec:taumvachannel}--\ref{sec:samesignvbf}. 
The combined results are interpreted in the SUSY models discussed in Section~\ref{sec:susysignals}. 
The analyses combined for each SUSY model are shown in Table~\ref{tab:analysesVSmodels}. 
Limits in the simplified models targeted by the analysis presented in the previous sections are presented in Sections~\ref{sec:Interp_stau}--\ref{sec:Interp_c1n2}. 
A summary is provided in Section~\ref{sec:summaryplots}, including the limits previously obtained from the ATLAS searches for $\chinoonep\chinoonem$ production with $WW$-mediated decays~\cite{Aad:2014vma}, $\chinoonepm\ninotwo$ production with $WZ$-mediated decays~\cite{Aad:2014nua} and $\chinoonepm\ninotwo$ production with $Wh$-mediated decays~\cite{Aad:2015jqa}.  
Finally, limits on phenomenological models are presented in Sections~\ref{sec:interp_pMSSM}--\ref{sec:Interp_gmsb}. 
For these models, the new searches presented in this article are not included, since they target very specific areas of parameter space and their sensitivity is small.

Exclusion limits are calculated by statistically combining results from a number of disjoint signal regions. 
In general, the analyses in Table~\ref{tab:analysesVSmodels} are mutually exclusive by design (the exceptions are indicated in the table), using the lepton multiplicity and charge, and are statistically combined. 
Where overlapping signal regions exist within an analysis, the signal region with the best-expected exclusion is used. 
During the combinations, all experimental uncertainties are treated as correlated between regions and processes, with the exception of the experimental uncertainties on data-driven backgrounds, which are correlated between regions only. 
Theoretical uncertainties on the irreducible background and signal are treated as correlated between regions, while statistical uncertainties are treated as uncorrelated between regions and processes.  
For the exclusion limits, the observed and expected 95\% CL limits are calculated using asymptotic formulas for each SUSY model point, taking into account the theoretical and experimental uncertainties on the SM background and the experimental uncertainties on the signal. 
Where the three-lepton~\cite{Aad:2014nua} analysis is used in the combination, 95\% CL limits are calculated using pseudoexperiments as the asymptotic approximation becomes inappropriate where the expected and observed yields are close to zero. 
The impact of the theoretical uncertainties on the signal cross-section is shown for the observed mass limit; where quoted in the text, mass limits refer to the $-1\sigma$ variation on the observed limit.

\begin{sidewaystable}[h]
\centering
\caption{ Searches used to probe each of the models described in Section~\ref{sec:susysignals}. 
\label{tab:analysesVSmodels}}
\small{
\renewcommand\arraystretch{1.4}
\hspace*{-0.05\textwidth}\begin{tabular}{l | c | c | c | c | c | c | c | c | c | c }
\toprule
Model & $Wh$~\cite{Aad:2015jqa} & 2$\ell^{\,\dagger}$~\cite{Aad:2014vma} & 2$\tau^{\ast}$~\cite{Aad:2014yka} & 3$\ell^{\,\diamondsuit}$~\cite{Aad:2014nua} & 4$\ell$~\cite{Aad:2014iza} & 2$\tau$ MVA$^{\ast}$ & SR2$\ell$-1$^\dagger$ & SS MVA$^{\S}$ & SR3$\ell$-0/1$^{\diamondsuit}$ & SR2$\ell$-2$^{\S}$ \\
\midrule
\stau\stau &&&\checkmark&&&\checkmark&&&& \\
\midrule
\chinoonep\chinoonem\ via $\tilde{\ell}_{L}$ with $x=0.5$ &&\checkmark&&&&&\checkmark&&& \\
\chinoonep\chinoonem\ via $\tilde{\ell}_{L}$ with variable $x$ &&\checkmark&&&&&&&& \\
\chinoonep\chinoonem\ via $WW$  &&\checkmark&&&&&&&& \\
\chinoonepm\chinoonepm\ via VBF &&&&&&&&&&\checkmark \\
\midrule
\chinoonepm\ninotwo\ via $\tilde{\tau}_{L}$ &&&\checkmark&\checkmark&&&&&& \\
\chinoonepm\ninotwo\ via $\tilde{\ell}_{L}$ with $x=0.5$ &&&&\checkmark&&&&\checkmark&\checkmark& \\
\chinoonepm\ninotwo\ via $\tilde{\ell}_{L}$ with variable $x$ &&&&\checkmark&&&&&& \\
\chinoonepm\ninotwo\ via $WZ$ &&\checkmark&&\checkmark&&&&&& \\
\chinoonepm\ninotwo\ via $Wh$ &\checkmark&\checkmark&&\checkmark&&&&&& \\
\midrule
\ninotwo\ninothree\ via $\tilde{\ell}_{L}$ with $x=0.5$ &&&&\checkmark&\checkmark&&&&& \\
\ninotwo\ninothree\ via $\tilde{\ell}_{L}$ with variable $x$ &&&&\checkmark&\checkmark&&&&& \\
\midrule
pMSSM &\checkmark&\checkmark&&\checkmark&&&&&& \\
NUHM2 &&\checkmark&&\checkmark&\checkmark&&&&& \\
GMSB &&&&&\checkmark&&&&& \\
\bottomrule
\multicolumn{11}{l}{$\dagger$ The opposite-sign, two-lepton signal regions in Ref.~\cite{Aad:2014vma} and Section~\ref{sec:superrazor} overlap.} \\
\multicolumn{11}{l}{$\ast$ The two-tau signal regions in Ref.~\cite{Aad:2014yka} and Section~\ref{sec:taumvachannel} overlap.} \\
\multicolumn{11}{l}{$\diamondsuit$ The three-lepton signal regions in \Ref.~\cite{Aad:2014nua} and Section~\ref{sec:threelep} overlap.} \\
\multicolumn{11}{l}{$\S$ The same-sign, two-lepton signal regions in Section~\ref{sec:samesignLL} and Section~\ref{sec:samesignvbf} overlap.} \\
\end{tabular}}
\end{sidewaystable}

\FloatBarrier

\subsection{Direct stau production \label{sec:Interp_stau}}
\begin{figure}[h]
  \centering
  \includegraphics[width=1.\textwidth]{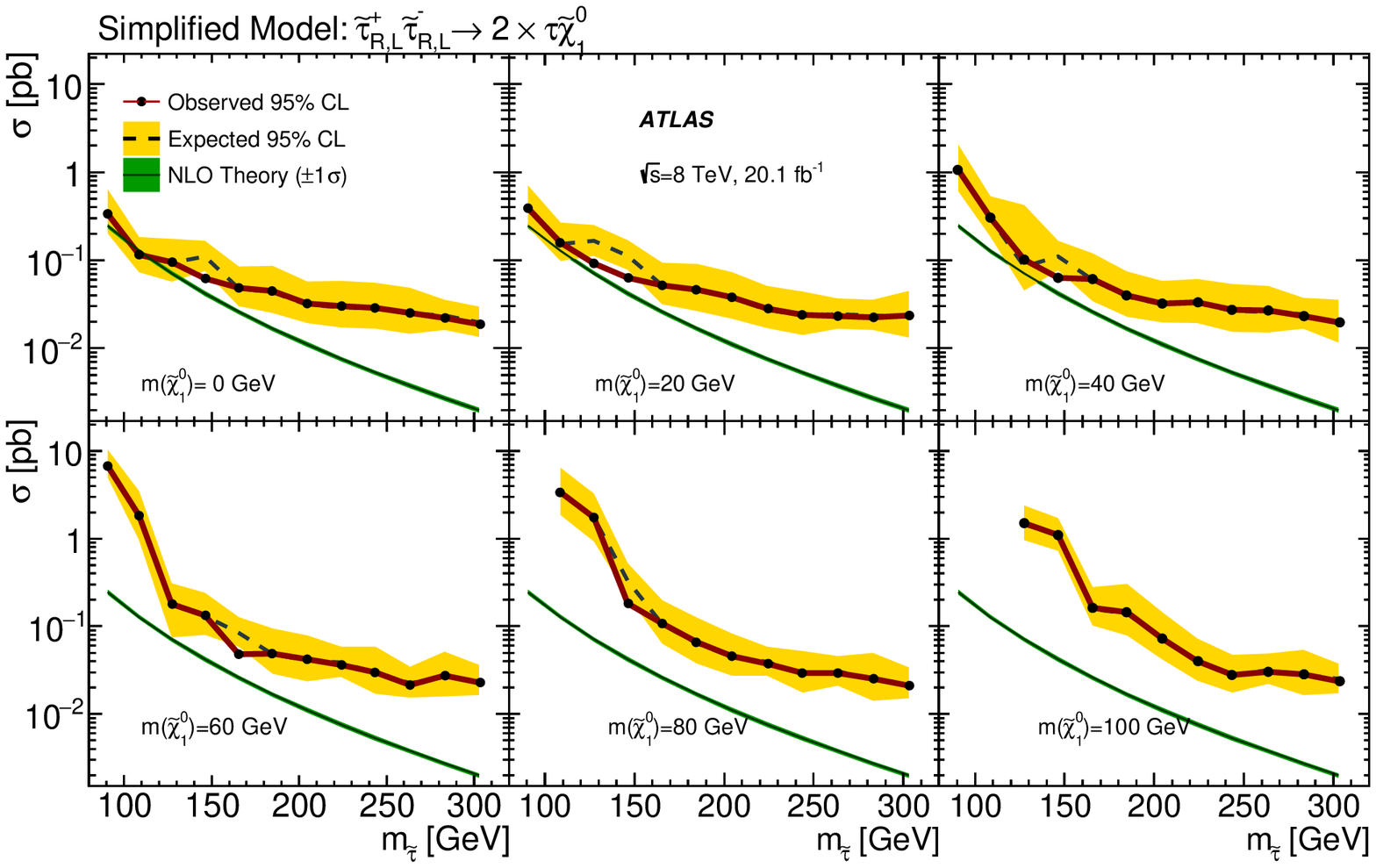}
  \caption{The 95$\%$ CL exclusion limits on the cross-section for production of left- and right-handed stau pairs for various \ninoone\ masses. The NLO theoretical cross-section for left and right-handed stau pair production is also shown.   \label{fig:InterpDirectStau} }

\end{figure}

The combination of the two-tau MVA results in Section~\ref{sec:taumvachannel} with the simple cut-based analysis from Ref.~\cite{Aad:2014yka} is used to set limits on the direct production of stau pairs. 
For each signal point, the signal region with the best expected limit is used. 
The upper limits on the cross-section for direct stau production are shown in Figure~\ref{fig:InterpDirectStau} for combined $\stauL\stauL$ and $\stauR\stauR$ production, where the observed limit is nearly always above the theoretical prediction. 
One scenario of combined $\stauL\stauL$ and $\stauR\stauR$ production is excluded, where the $\stauR$ mass is 109$\GeV$ and the $\ninoone$ is massless. 
For this scenario, cross-sections above 0.115 pb are excluded, where the theoretical cross-section at NLO is 0.128 pb. 
No scenarios can be excluded where only $\stauR\stauR$ production or $\stauL\stauL$ production is considered.
Cross-sections above 0.06 (0.21) pb are excluded for $\stauR\stauR$ ($\stauL\stauL$) production with a $\stauR$ ($\stauL$) mass of 109$\GeV$ and a massless $\ninoone$, where the theoretical cross-section at NLO is 0.04 (0.09) pb.
For this scenario [$m(\stauR)=109\GeV$, $m(\ninoone)=0\GeV$], the expected yields from $\stauR\stauR$ production are larger than from $\stauL\stauL$ in the signal region, making the experimental limits stronger for $\stauR\stauR$ production. 
However, for other mass points the experimental limit is generally weaker for $\stauR\stauR$ production due to the lower production cross-section.
These limits on direct production of stau pairs improve upon the previous limits in Ref.~\cite{Aad:2014yka}, particularly for stau masses below $\sim$150$\GeV$.

\subsection{Direct chargino production~\label{sec:Interp_c1c1}}

\begin{figure}[h]
 \centering
  \subfigure[]{ \includegraphics[width=0.49\textwidth]{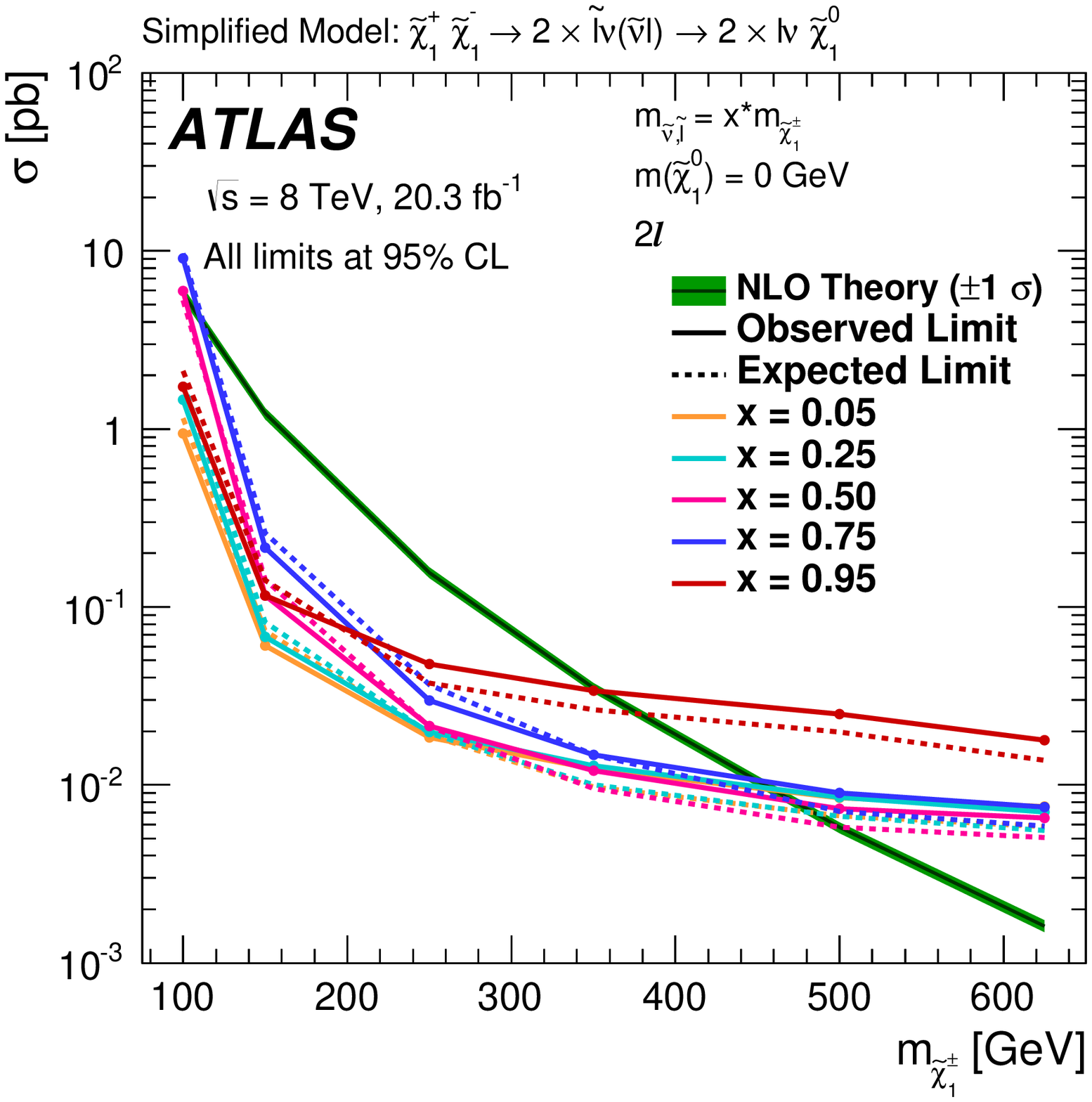}}
  \subfigure[]{ \includegraphics[width=0.49\textwidth]{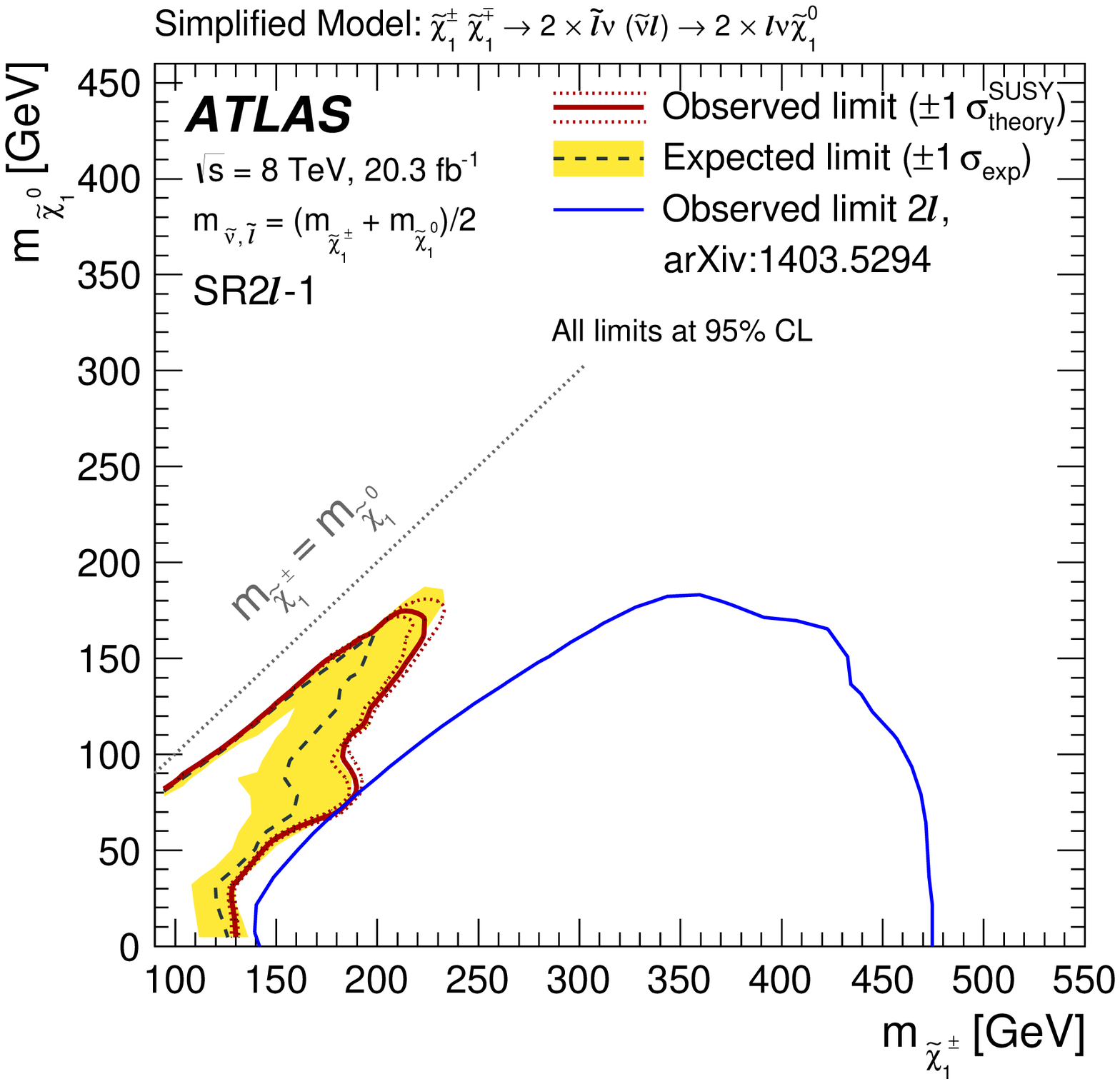}}
  \caption{ The 95$\%$ CL exclusion limits on $\chinoonep\chinoonem$ production with $\slepL$-mediated decays, (a) where the $\ninoone$ is massless and the intermediate slepton mass is set to 5$\%$, 25$\%$, 50$\%$, 75$\%$ and 95$\%$ of the $\chinoonepm$ mass, and (b) as a function of the $\chinoonepm$ and $\ninoone$ masses, where the slepton mass is halfway between the $\ninotwo$ and $\ninoone$ masses. The limits in (a) are set using the 2$\ell$ analysis from Ref.~\protect\cite{Aad:2014vma}, while the limits in (b) use the opposite-sign, two-lepton analysis from this article.  The limit from Ref.~\protect\cite{Aad:2014vma} is also shown in (b).
\label{fig:InterpC1C1Slep} }
\end{figure}

The opposite-sign, two-lepton analysis in Ref.~\cite{Aad:2014vma} is used to reinterpret the limits on $\chinoonep\chinoonem$ production decaying through sleptons, where the slepton mass is varied between the $\chinoonepm$ and $\ninoone$ masses. 
Scenarios where the slepton mass is 5$\%$, 25$\%$, 50$\%$, 75$\%$ and 95$\%$ of the $\chinoonepm$ mass are studied for a massless $\ninoone$, and the limits are shown in Figure~\ref{fig:InterpC1C1Slep}(a). 
For the majority of the $\chinoonepm$ masses considered, the slepton mass does not have a significant effect on the sensitivity and $\chinoonepm$ masses are excluded up to $\sim$500$\GeV$.
The sensitivity is reduced for a very small mass splitting between the chargino and the slepton ($x$ = 0.95), as in this case leptons from the
$\chinoonepm\rightarrow\snu\ell$ decays have low momentum, making these events difficult to reconstruct in the two lepton final state.

Limits are also set in the $\chinoonep\chinoonem$ scenario with $\slepL$-mediated decays, with slepton masses set halfway between the $\chinoonepm$ and the $\ninoone$ masses, where both the $\chinoonepm$ and the $\ninoone$ masses are varied. 
Figure~\ref{fig:InterpC1C1Slep}(b) shows the opposite-sign, two-lepton analysis presented in Section~\ref{sec:superrazor}, which provides new sensitivity to compressed scenarios for $\chinoonepm$ masses below $\sim$220$\GeV$. 
The 2$\ell$ analysis in Ref.~\cite{Aad:2014vma} continues to dominate the sensitivity to scenarios with large mass splittings, excluding $\chinoonepm$ masses up to $\sim$465$\GeV$.

\begin{figure}[h]
  \centering  
  \subfigure[]{\includegraphics[width=0.49\textwidth]{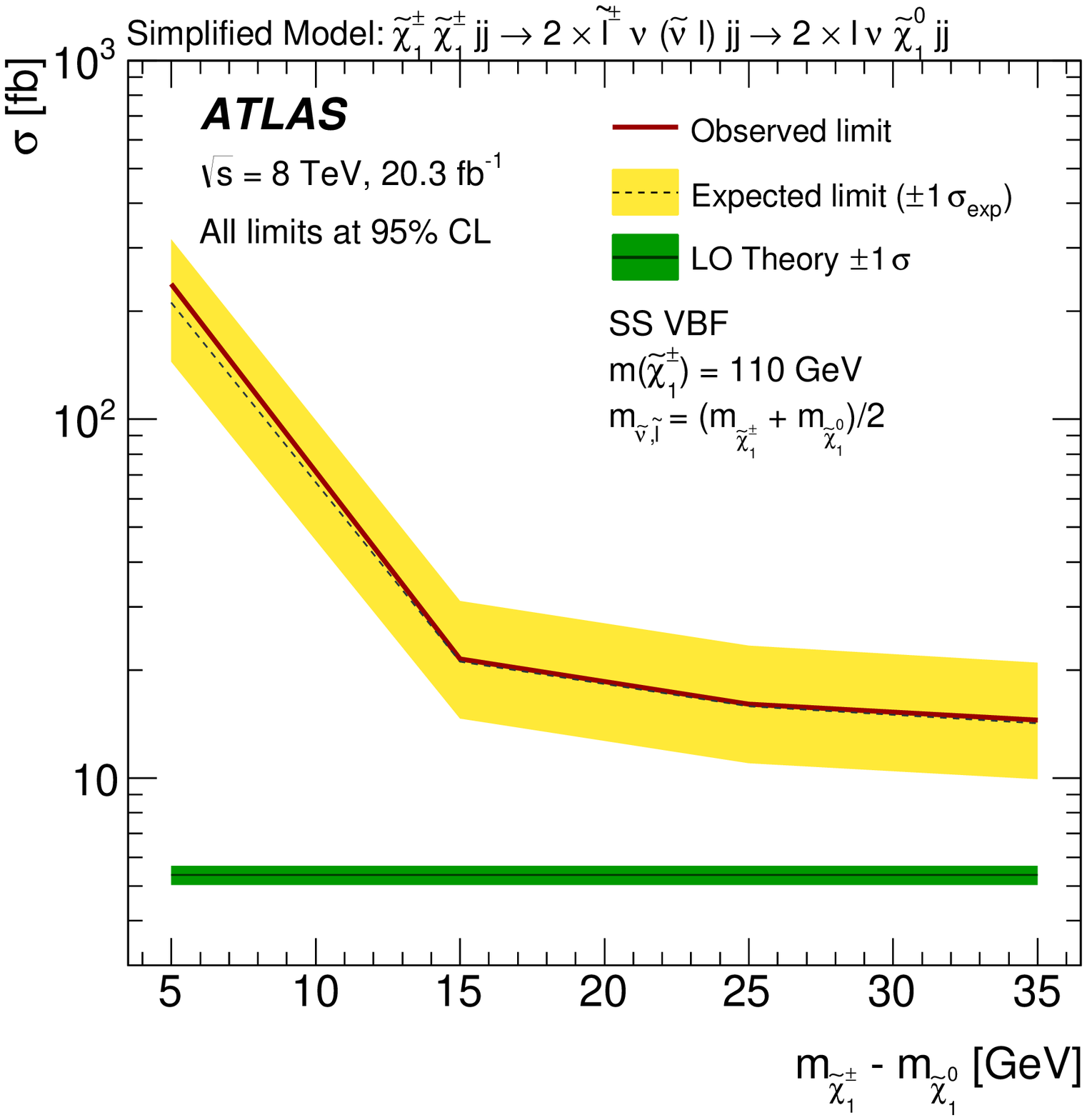}}
  \subfigure[]{\includegraphics[width=0.49\textwidth]{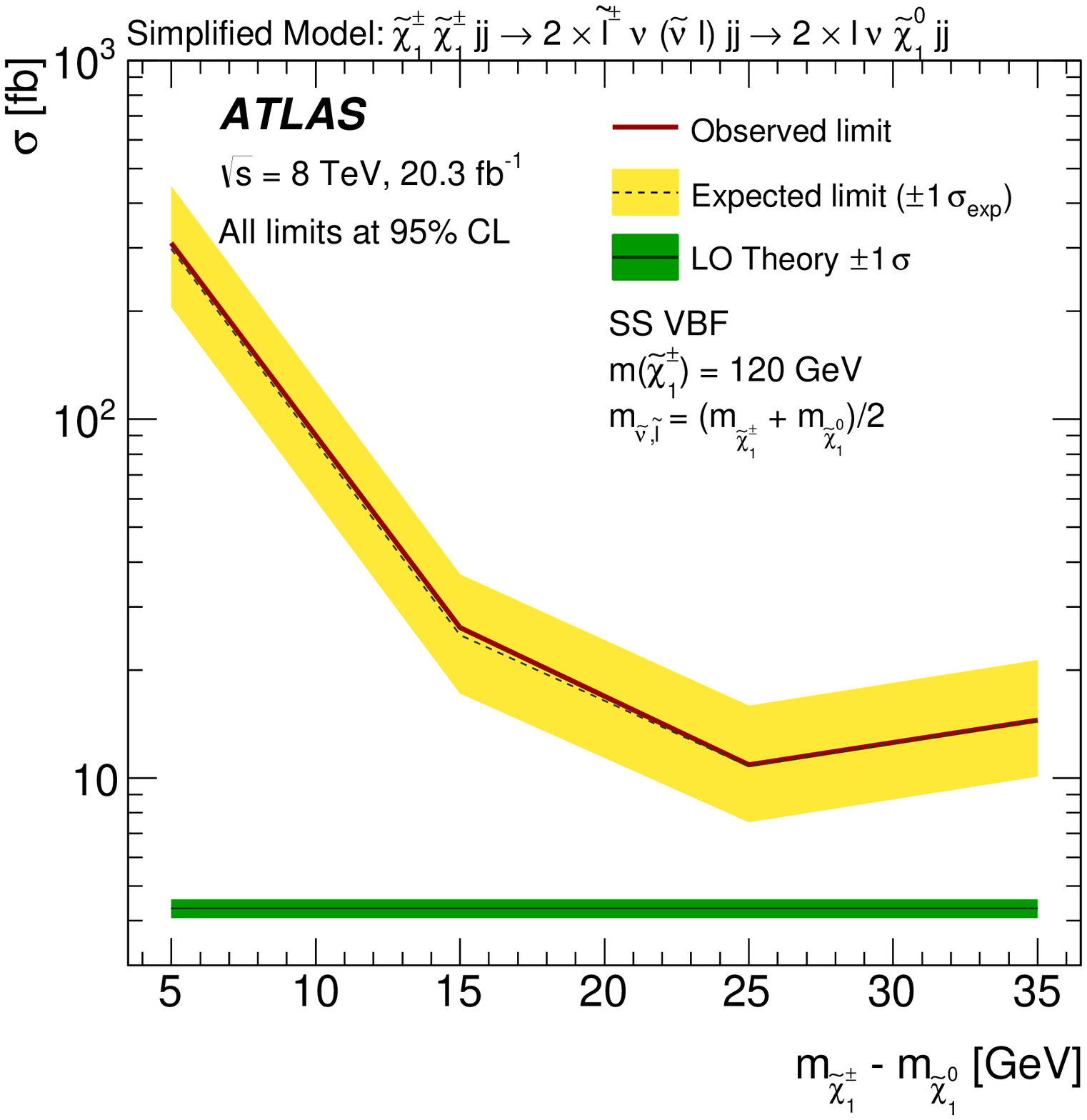}}
  \caption{ The 95\% CL upper limit on the signal cross-section for VBF $\chinoonepm\chinoonepm$ production for (a) $m(\chinoonepm)$ = 110$\GeV$ and (b) $m(\chinoonepm)$ = 120$\GeV$. The limits are set with respect to the mass difference between the $\chinoonepm$ and $\ninoone$, and use the results from the same-sign, two-lepton VBF analysis. 
\label{fig:InterpC1C1VBF}}
\end{figure}

The same-sign, two-lepton VBF analysis described in Section~\ref{sec:samesignvbf} is used to set limits on VBF $\chinoonepm\chinoonepm$ production, where the $\chinoonepm$ decays through sleptons. 
Figures~\ref{fig:InterpC1C1VBF}(a) and \ref{fig:InterpC1C1VBF}(b) show the 95\% CL upper limits on the cross-section for $m$($\chinoonepm$)\,=\,110$\GeV$ and $m$($\chinoonepm$)\,=\,120$\GeV$, as a function of the mass splitting between the chargino and the neutralino. 
The best observed upper limit on the VBF $\chinoonepm\chinoonepm$ production cross-section is found for a $\chinoonepm$ mass of 120$\GeV$ and $m$($\chinoonepm$) $-$ $m$($\ninoone$) = 25$\GeV$, where the theoretical cross-section at LO is 4.33 fb and the excluded cross-section is 10.9 fb.
The sensitivity is slightly stronger for higher $\chinoonepm$ masses, since these scenarios were used for optimizing the signal selection.

\subsection{Direct neutralino production~\label{sec:interp_n2n3}}

\begin{figure}[h]
 \centering
  \subfigure[]{ \includegraphics[trim=0 0.7cm 0.4cm 0, clip=true, width=0.49\textwidth]{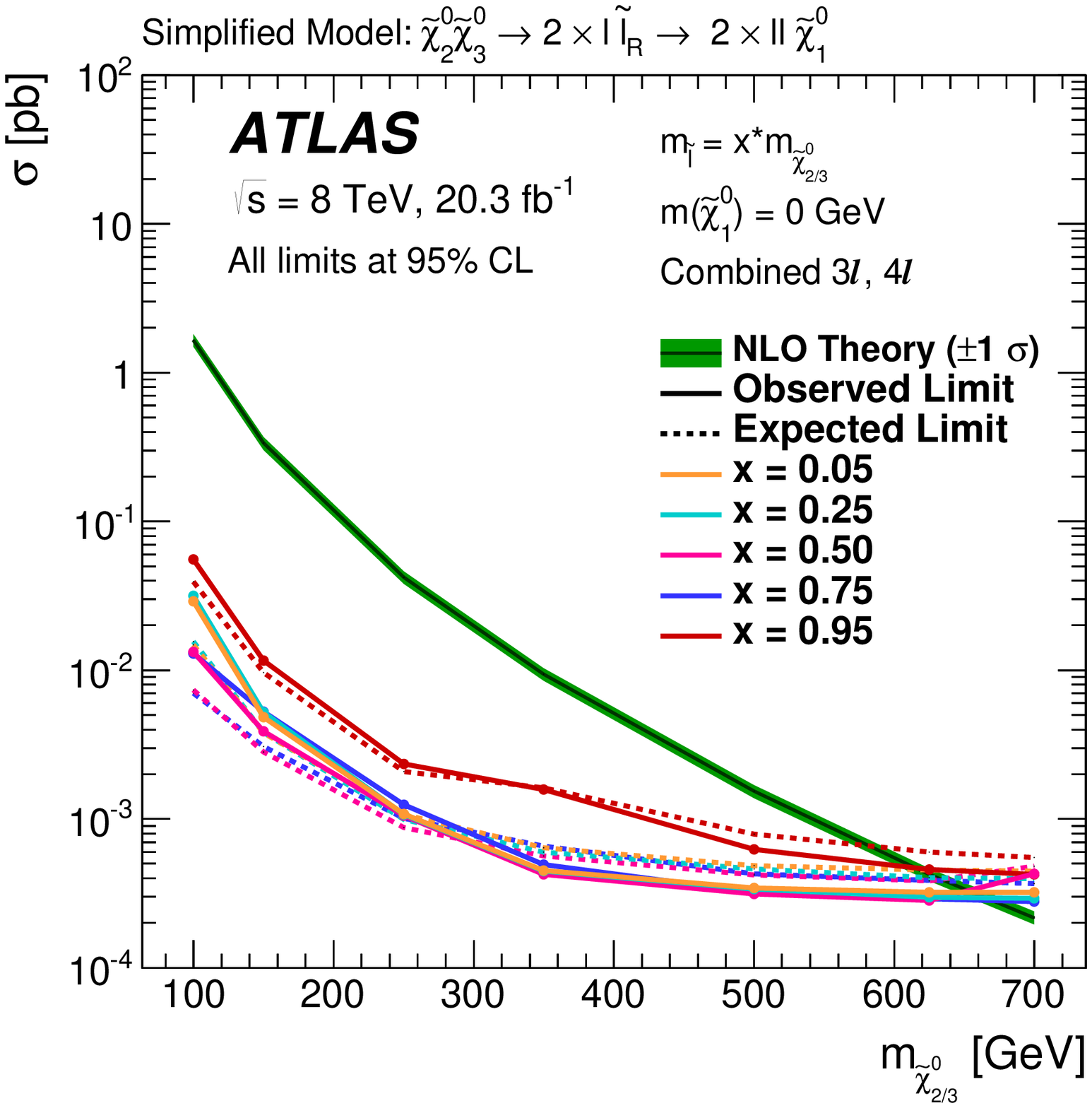}}
  \subfigure[]{ \includegraphics[width=0.49\textwidth]{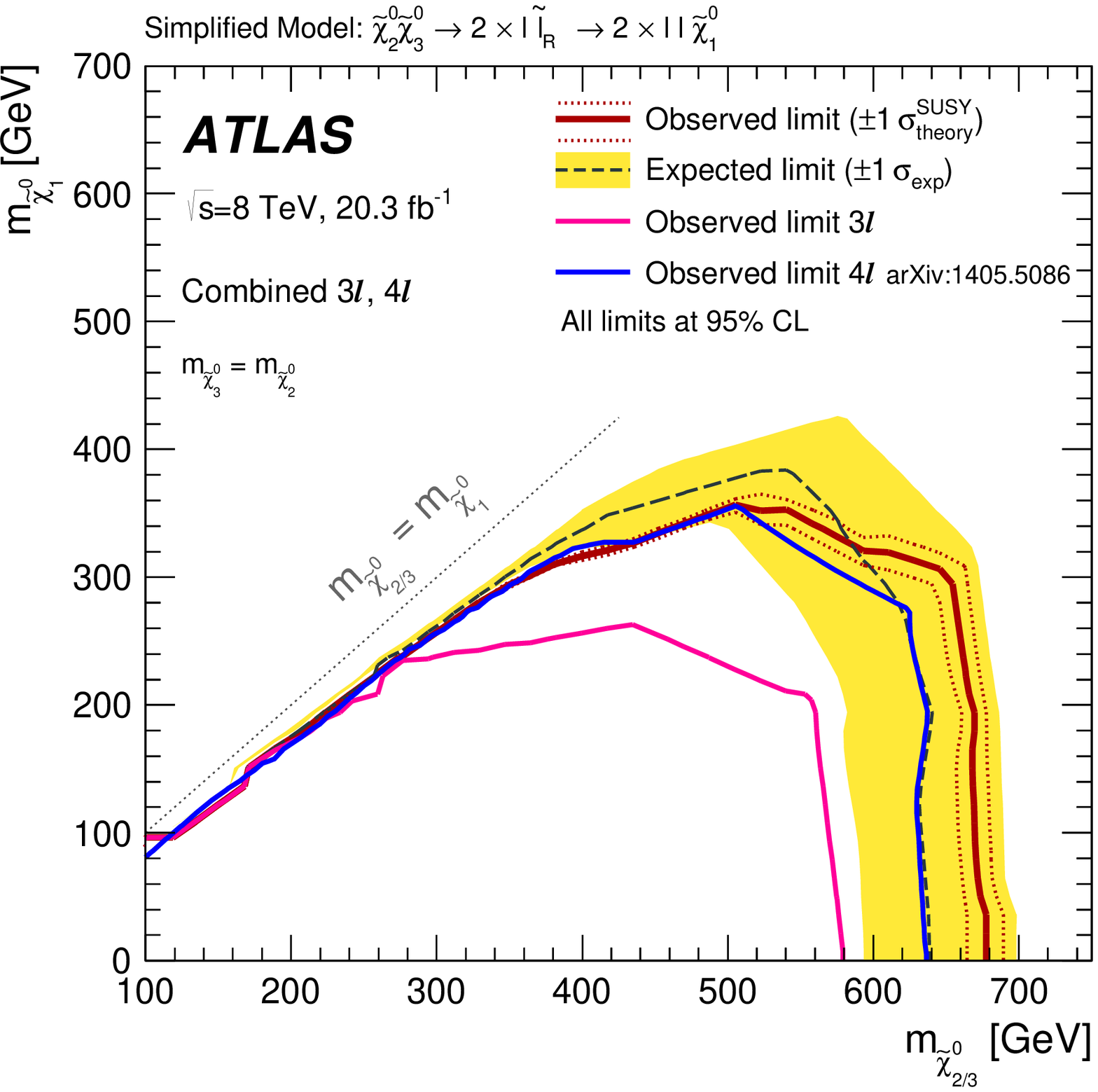}}
  \caption{ 
  The 95$\%$ CL exclusion limits on $\ninotwo\ninothree$ production with $\slepR$-mediated decays, (a) where the $\ninoone$ is massless and the intermediate slepton mass is set to 5$\%$, 25$\%$, 50$\%$, 75$\%$ and 95$\%$ of the $\ninotwo$ mass, and (b) as a function of the $\ninotwo$ and $\ninoone$ masses, where the slepton mass is halfway between the $\ninotwo$ and $\ninoone$ masses.  The limits in (a) and (b) are set using a combination of the 3$\ell$ analysis from Ref.~\cite{Aad:2014nua} and the 4$\ell$ analysis from Ref.~\cite{Aad:2014iza}. 
\label{fig:InterpN2N3variableSlep}}
\end{figure}

The combination of the three-lepton analysis in Ref.~\cite{Aad:2014nua} and four-lepton analysis in Ref.~\cite{Aad:2014iza} is used to set limits on $\ninotwo\ninothree$ production with $\slepR$-mediated decays, where the slepton mass is varied between the $\ninotwo$ and $\ninoone$ masses. 
Scenarios where the slepton mass is 5$\%$, 25$\%$, 50$\%$, 75$\%$ and 95$\%$ of the $\ninotwo$ mass are studied for a massless $\ninoone$, and the limits are shown in Figure~\ref{fig:InterpN2N3variableSlep}(a). 
For the majority of $\ninotwo$ masses considered, the slepton mass does not have a significant effect on the sensitivity and $\ninotwo$ masses are excluded up to $\sim$600$\GeV$.
The sensitivity is reduced for a very small mass splitting between the $\ninotwo$ and slepton ($x$ = 0.95) as the lepton produced in the $\ninotwo\rightarrow\ell\slepR$ decay has low-momentum. 
The reduced sensitivity is not seen for a very small mass splitting between the slepton and the LSP ($x$ = 0.05) as the lepton produced in the $\slepR\rightarrow\ell\ninoone$ decay can carry some of the momentum of the slepton.

Limits are also set in the $\ninotwo\ninothree$ scenario with $\slepR$-mediated decays, with slepton masses set halfway between the $\ninotwo$ and the $\ninoone$ masses, where both the $\ninotwo$ and the $\ninoone$ masses are varied. 
The combination of the three- and four-lepton analysis is again used here and limits are shown in Figure~\ref{fig:InterpN2N3variableSlep}(b), where $\ninotwo$, $\ninothree$ masses up to 670$\GeV$ are excluded, improving the previous limits by 30$\GeV$ for $\ninoone$ masses below 200$\GeV$.

\subsection{Direct neutralino--chargino production~\label{sec:Interp_c1n2}}

\begin{figure}[h]
 \centering
     \includegraphics[width=0.5\textwidth]{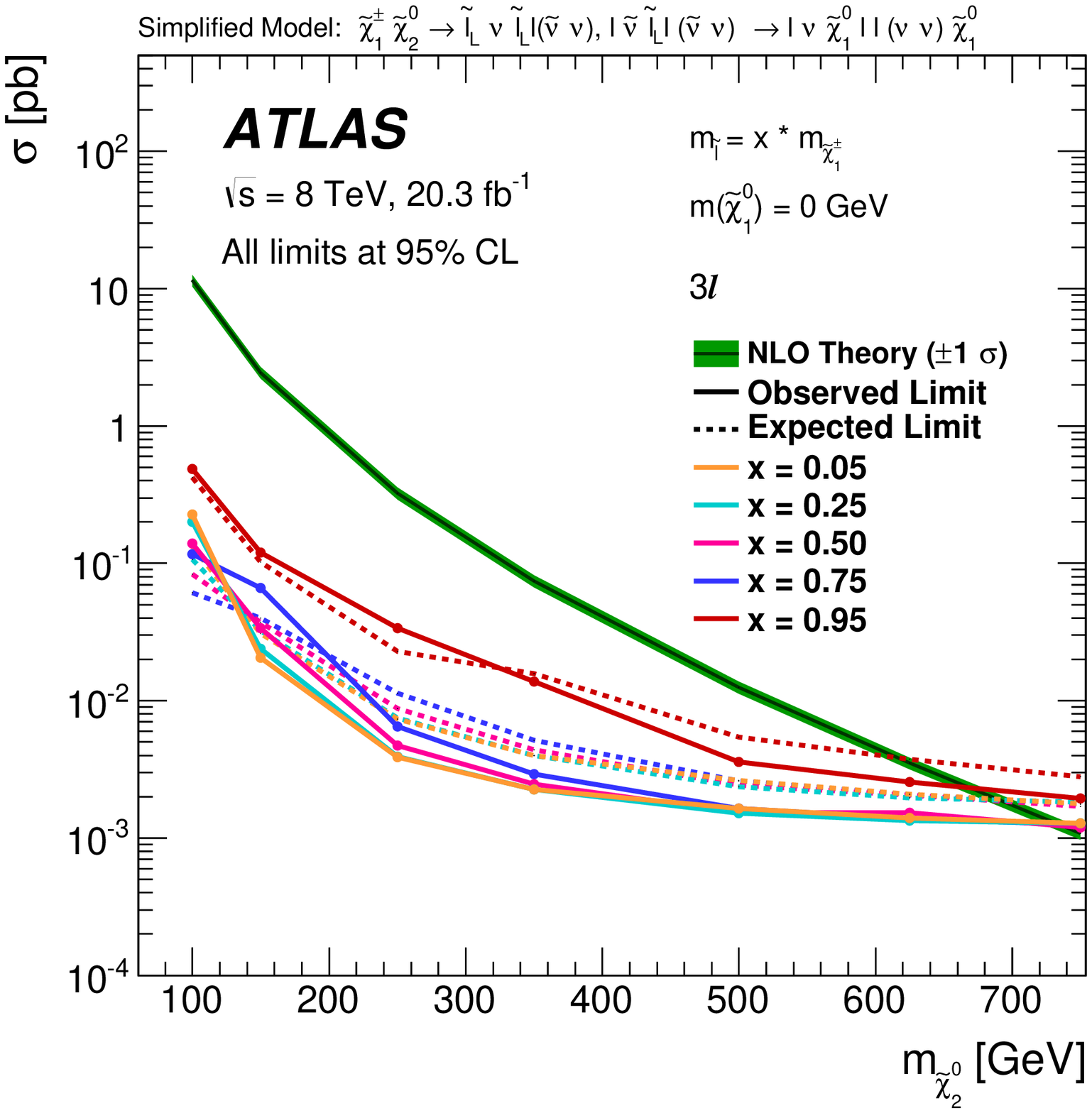}
  \caption{
  The 95$\%$ CL upper cross-section limits on $\chinoonepm\ninotwo$ production with $\slepL$-mediated decays, where the $\ninoone$ is massless and the intermediate slepton mass is set to 5$\%$, 25$\%$, 50$\%$, 75$\%$, and 95$\%$ of the $\chinoonepm$ mass.  The limits are set using the 3$\ell$ analysis from Ref.~\cite{Aad:2014nua}. 
\label{fig:InterpC1N2variableSlep} }
\end{figure}

The three-lepton analysis in Ref.~\cite{Aad:2014nua} is used to reinterpret the limits on $\chinoonepm\ninotwo$ production decaying through sleptons. 
Scenarios where the slepton mass is 5$\%$, 25$\%$, 50$\%$, 75$\%$ and 95$\%$ of the $\chinoonepm$ mass are studied for a massless $\ninoone$. 
The limits on these variable slepton mass scenarios are shown in Figure~\ref{fig:InterpC1N2variableSlep}. 
For the majority of $\chinoonepm$ masses considered, the slepton mass does not have a significant effect on the sensitivity and $\chinoonepm$ masses are excluded up to $\sim$700$\GeV$. 
The same reduction in sensitivity is seen for a small mass splitting between the $\ninotwo$ and slepton ($x$ = 0.95) as in the $\ninotwo\ninothree$ interpretation in Section~\ref{sec:interp_n2n3}. 
For $\chinoonepm\ninotwo$ production scenarios decaying through SM $W$, $Z$ or Higgs bosons~\cite{Aad:2014nua}, the results in Figure~\ref{fig:InterpC1N2variableSlep} would be degraded due to lower branching fractions into leptonic final states. 
The pMSSM scenario in Section~\ref{sec:interp_pMSSM} shows the sensitivity to SUSY scenarios without sleptons in the $\chinoonepm\ninotwo$ decay chain.

\begin{figure}[h]
 \centering
   \subfigure[]{\includegraphics[width=0.49\textwidth]{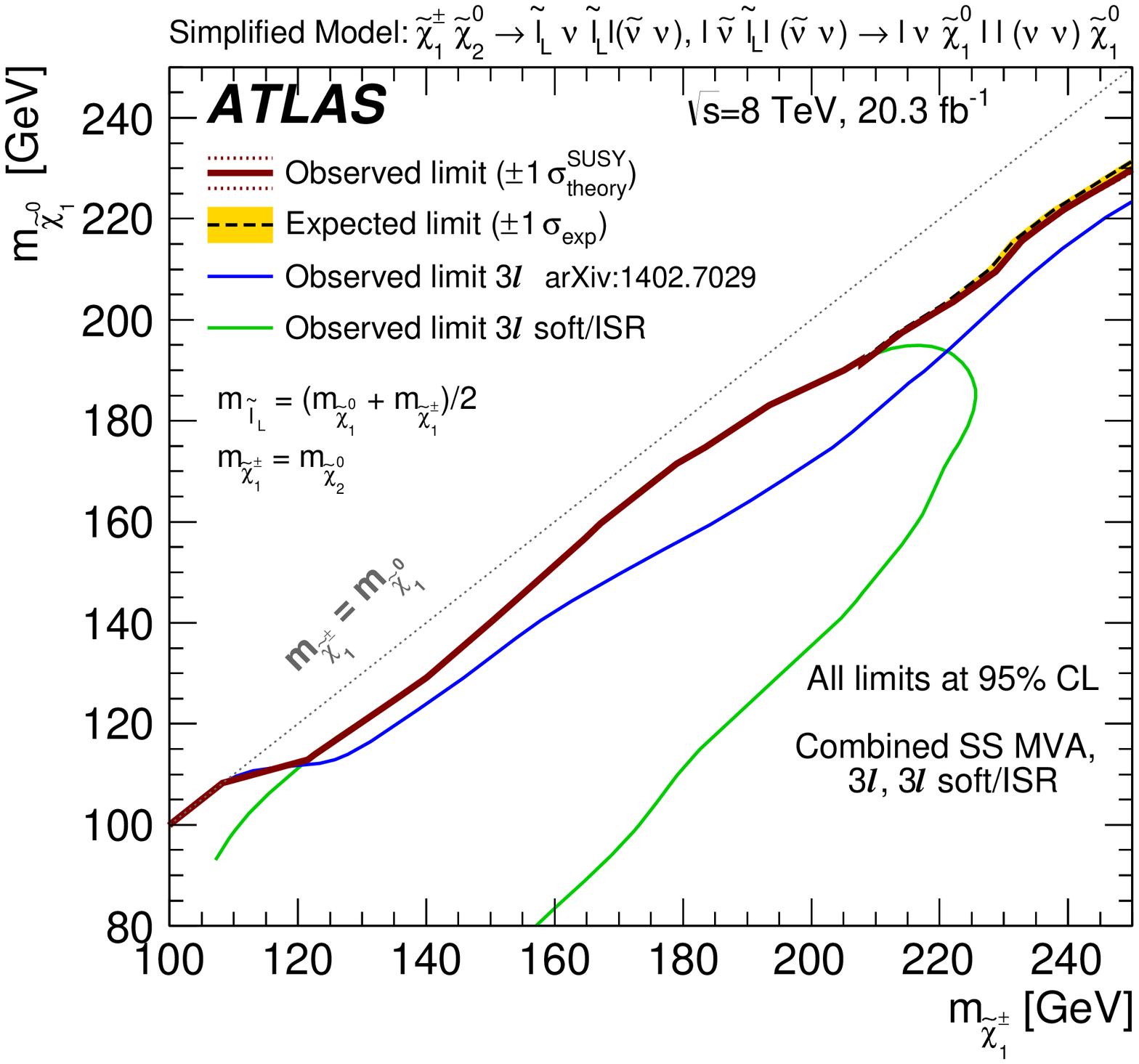}}
   \subfigure[]{\includegraphics[width=0.49\textwidth]{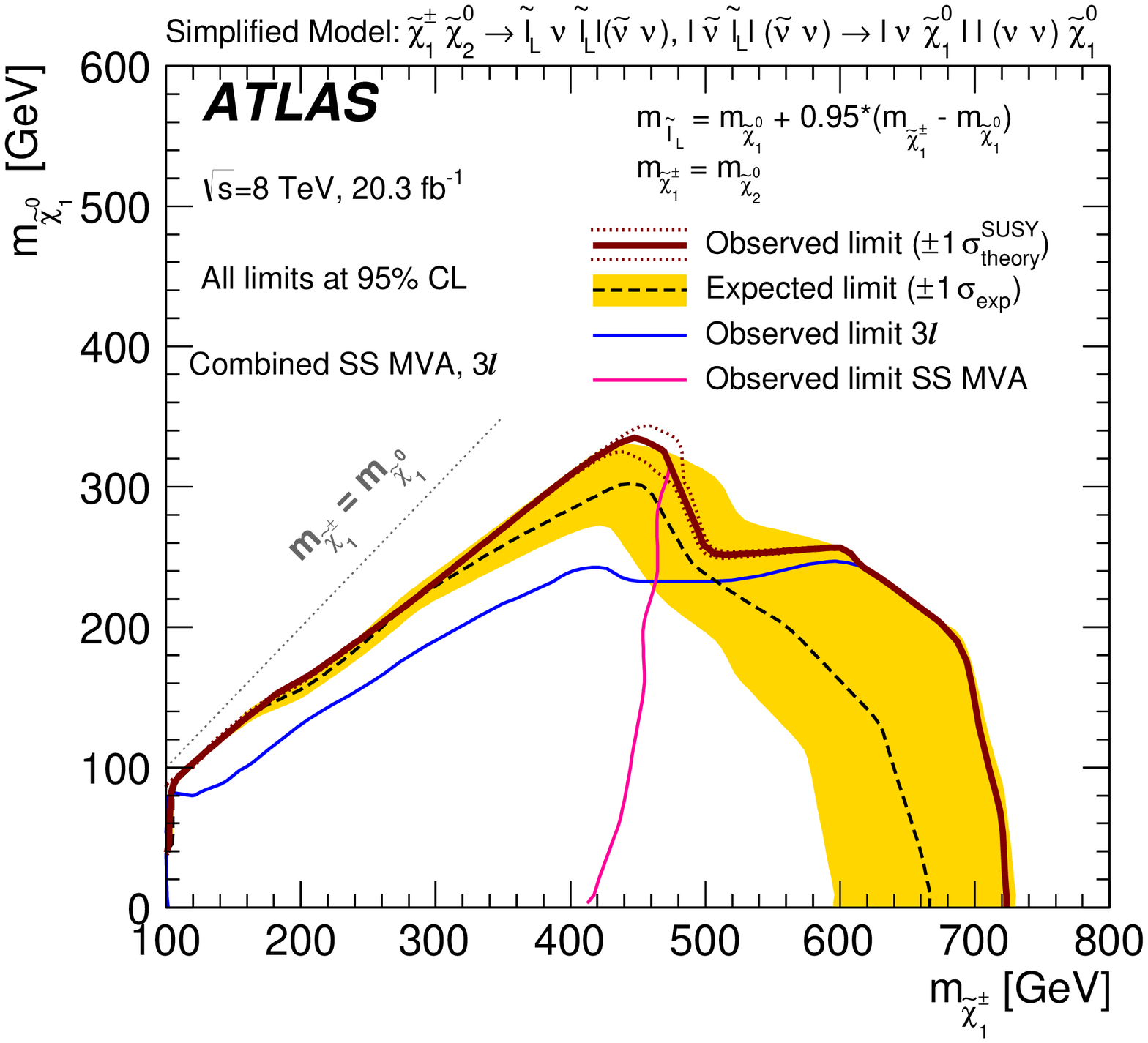}}
  \caption{
  The 95$\%$ CL exclusion limits on $\chinoonepm\ninotwo$ production with $\slepL$-mediated decays, as a function of the $\chinoonepm$ and $\ninoone$ masses, where the intermediate slepton mass is set to the $\ninoone$ mass plus (a) 50\% or (b) 95\% of the difference between the $\chinoonepm$ and the $\ninoone$ masses. The limits in (a) are set using a combination of the 3$\ell$ analysis from Ref.~\cite{Aad:2014nua} and the same-sign, two-lepton anaysis from this article, while the limits in (b) use the combination of the three-lepton and same-sign, two-lepton anayses from this article.
\label{fig:InterpC1N250Slep} }
\end{figure}

Limits are also set in the $\chinoonepm\ninotwo$ scenarios with $\slepL$-mediated decays, with slepton masses set halfway and at 95\% between the $\chinoonepm$ and the $\ninoone$ masses, where both the $\chinoonepm$ and the $\ninoone$ masses are varied. 
Figures~\ref{fig:InterpC1N250Slep}(a) and \ref{fig:InterpC1N250Slep}(b) show that the combination of the published and new analyses gives an improved sensitivity to compressed scenarios up to $\chinoonepm$ masses of $\sim$250$\GeV$. 
In scenarios with large mass splittings, $\chinoonepm$ masses are excluded up to $\sim$700$\GeV$ for slepton masses set to the $\ninoone$ mass plus 50\% or 95\% of the difference between the $\chinoonepm$ and the $\ninoone$ masses. 
In the compressed areas of the $\chinoonepm\ninotwo$ scenario with $\slepL$-mediated decays, and slepton masses set halfway (95\%) between the $\chinoonepm$ and the $\ninoone$ masses, the three-lepton (same-sign, two-lepton) analysis has the strongest sensitivity.

\begin{figure}[h]
 \centering
 \includegraphics[width=0.5\textwidth]{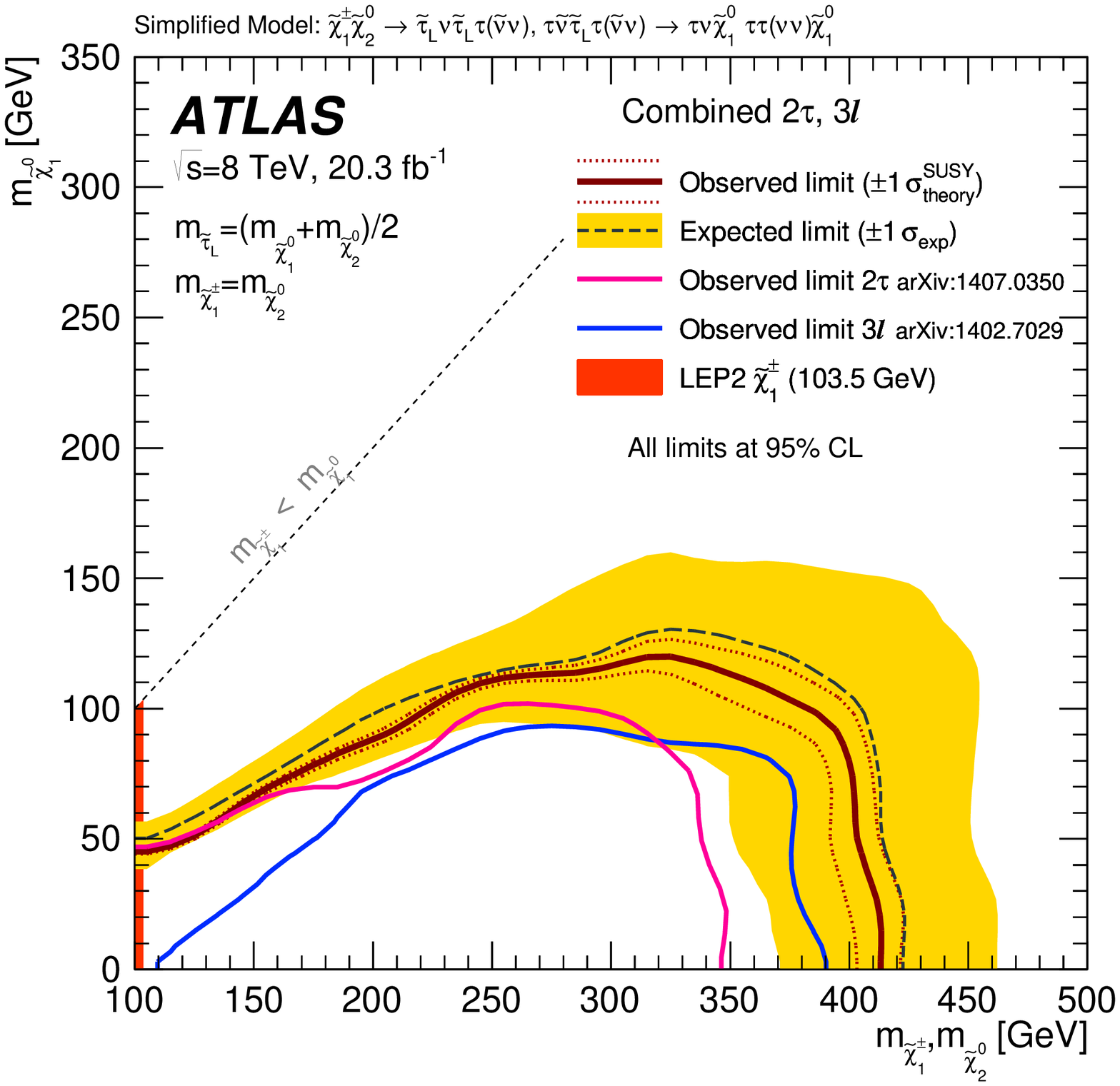}
  \caption{
  The 95$\%$ CL exclusion limits on $\chinoonepm\ninotwo$ production with $\stau$-mediated decays, as a function of the $\chinoonepm$ and $\ninoone$ masses. The limits are set using a combination of the 3$\ell$ analysis from Ref.~\cite{Aad:2014nua} and the 2$\tau$ analysis from Ref.~\cite{Aad:2014yka}. 
\label{fig:InterpC1CN2Stau} }
\end{figure}

Finally, limits are set in the $\chinoonepm\ninotwo$ scenario with $\stau$-mediated decays, using combined results from the two-tau analysis in Ref.~\cite{Aad:2014yka} and the three-lepton analysis in Ref.~\cite{Aad:2014nua}. 
Figure~\ref{fig:InterpC1CN2Stau} shows that the sensitivity to large $\chinoonepm$ masses is improved by 20$\GeV$ with the new combination, where $\chinoonepm$ masses are excluded up to $\sim$400$\GeV$ for massless $\ninoone$.

\FloatBarrier

\subsection{Summary of simplified electroweakino production \label{sec:summaryplots}}

\begin{figure}[h]
 \centering
   \subfigure[]{\includegraphics[width=0.65\textwidth]{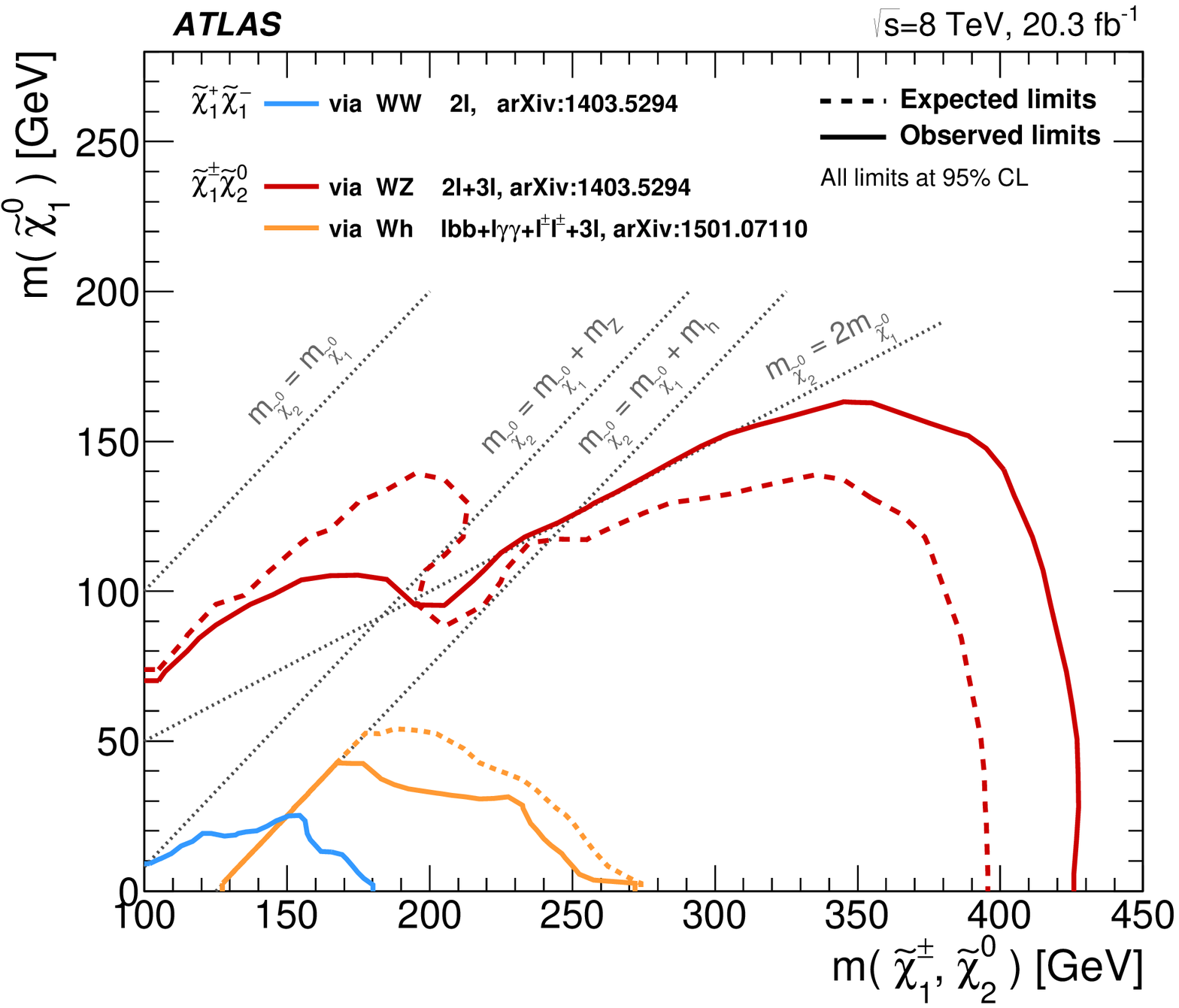}}
   \subfigure[]{\includegraphics[width=0.65\textwidth]{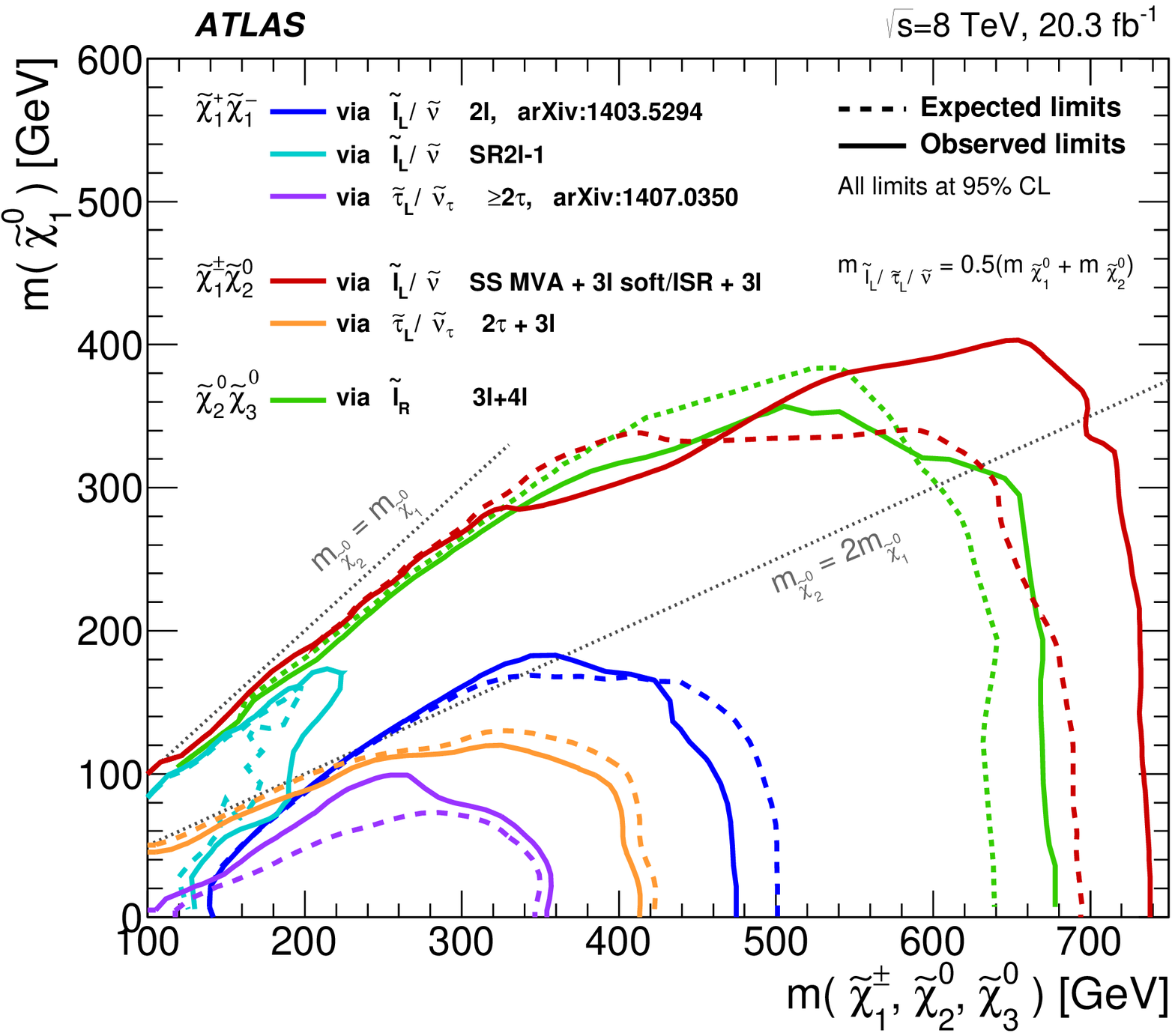}}
  \caption{
  The 95$\%$ CL exclusion limits on $\chinoonep\chinoonem$, $\chinoonepm\ninotwo$ and $\ninotwo\ninothree$ production with (a) SM-boson-mediated decays and (b) $\slepton$-mediated decays, as a function of the $\chinoonepm$, $\ninotwo$ and $\ninoone$ masses. 
The production cross-section is for pure wino $\chinoonep\chinoonem$ and $\chinoonepm\ninotwo$, and pure higgsino $\ninotwo\ninothree$. 
\label{fig:InterpSummary} }
\end{figure}

The ATLAS results for electroweakino searches at 8$\TeV$ in the framework of simplified models are summarized in Figures~\ref{fig:InterpSummary}(a) and \ref{fig:InterpSummary}(b) in the $m$($\chinoonepm$,$\ninotwo$)--$m$($\ninoone$) plane. 
As explained in Section~\ref{sec:susysignals}, each of the $\chinoonepm/\ninotwo/\ninothree$ decays considered in the plot is assumed to have 100\% branching fraction, and the production cross-section is for pure wino $\chinoonep\chinoonem$ and $\chinoonepm\ninotwo$, and pure higgsino $\ninotwo\ninothree$. 
The limits for $\chinoonep\chinoonem$ and $\chinoonepm\ninotwo$ production with decays mediated by SM bosons are summarized in Figure~\ref{fig:InterpSummary}(a). 
All of the limits are from the two-lepton, three-lepton, and $Wh$ analyses from Refs.~\cite{Aad:2015jqa,Aad:2014vma,Aad:2014nua}. 
The new analyses targeting compressed spectra presented in this article only have a small sensitivity to these scenarios and did not significantly improve upon published limits. 
The limits for $\chinoonep\chinoonem$, $\chinoonepm\ninotwo$ and $\ninotwo\ninothree$ production with $\slepton$-mediated decays are summarized in Figure~\ref{fig:InterpSummary}(b). 
The limits are from the new analyses in Sections~\ref{sec:Interp_c1c1}--\ref{sec:Interp_c1n2} and the previously published analyses. 

\FloatBarrier
\subsection{pMSSM \label{sec:interp_pMSSM}}

\begin{figure}[!h]
  \centering
\includegraphics[width=0.5\textwidth]{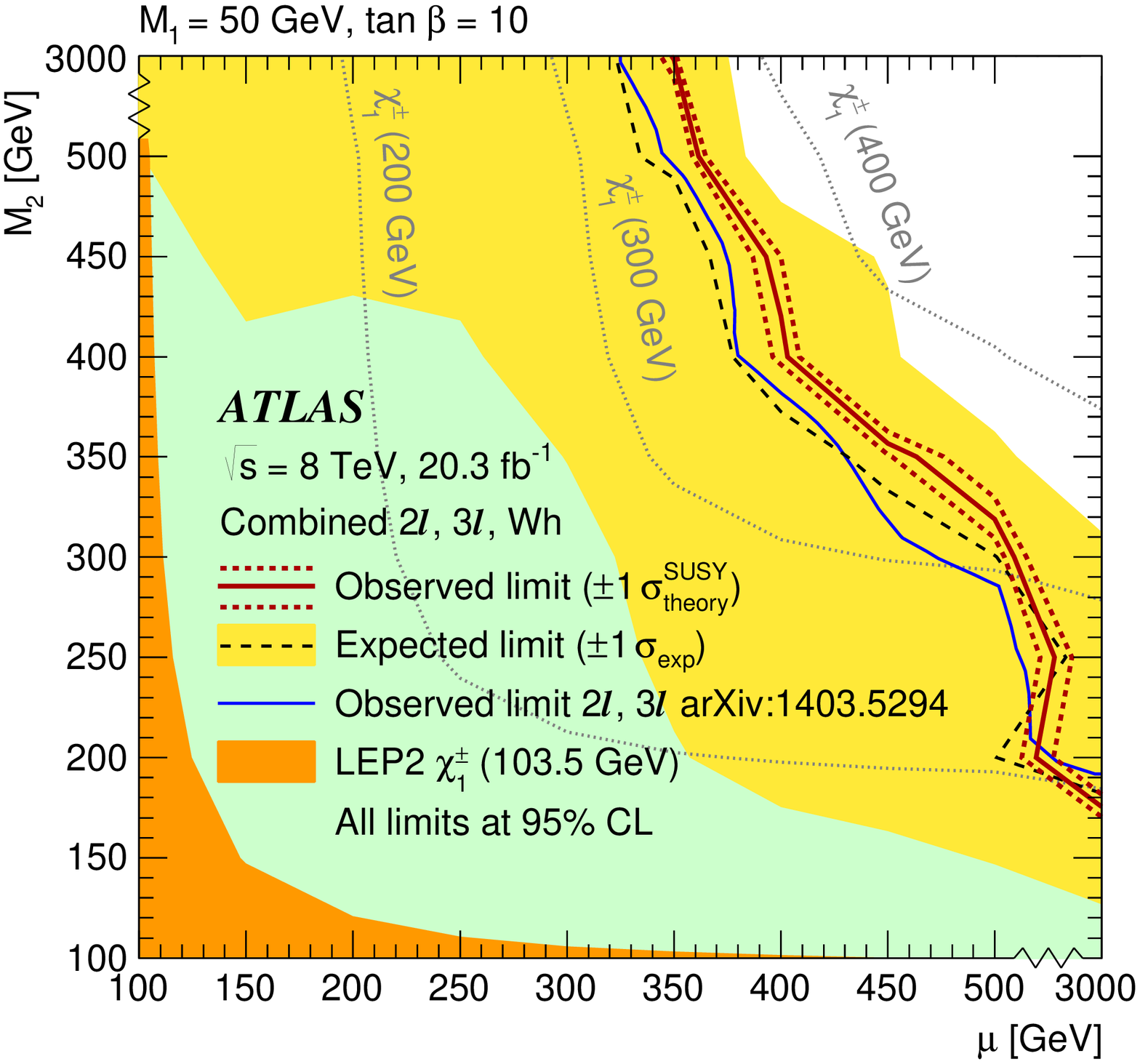}
  \caption{
The 95$\%$~CL exclusion limit in the pMSSM scenario, using a combination of the 2$\ell$ and 3$\ell$ analyses from Ref.~\cite{Aad:2014vma} and the $Wh$ analysis from Ref.~\cite{Aad:2015jqa}. 
The areas excluded by the $-1\sigma$ expected limit are shown in green. 
The blue contour corresponds to the limits from the combination of the 2$\ell$ and 3$\ell$ analyses from Ref.~\cite{Aad:2014vma}. 
The grey dotted contours show the chargino mass isolines.
\label{fig:InterpModelsPMSSM} }
\end{figure}

The two-lepton, three-lepton, and $Wh$ analyses from Refs.~\cite{Aad:2015jqa,Aad:2014vma,Aad:2014nua} are combined to improve the sensitivity in the considered pMSSM scenario where the EW SUSY production and the decays through $W$, $Z$, or $h$ bosons are dominant. 
The 95\% CL exclusion in the pMSSM $\mu$--$M_2$ plane for the scenario of heavy sleptons, $\tan \beta =10$, and $M_1\,$=$\,$50$\GeV$ is shown in Figure~\ref{fig:InterpModelsPMSSM}. 
Including the $Wh$ analysis in the new combination results in a stronger limit at high values of $M_2$, in particular in the intermediate $\mu$ region.

\FloatBarrier

\subsection{NUHM2 \label{sec:Interp_nuhm}}
\begin{figure}[!h]
  \centering
\includegraphics[width=0.5\textwidth]{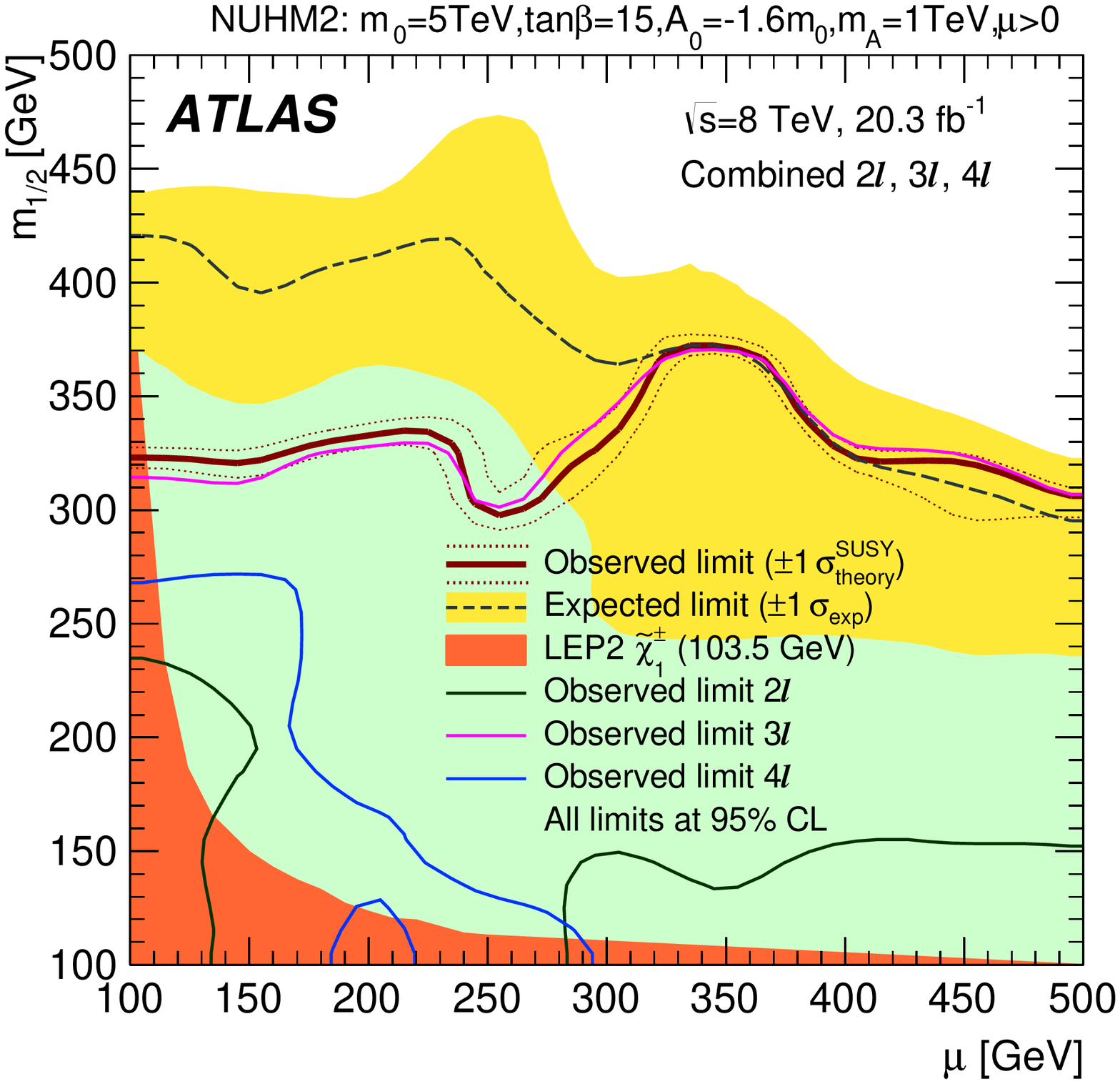}
  \caption{
The 95$\%$~CL exclusion limit in the NUHM2 scenario, using a combination of the 2$\ell$, 3$\ell$ and 4$\ell$ analyses from from Refs.~\cite{Aad:2014vma,Aad:2014nua,Aad:2014iza}. 
The areas excluded by the $-1\sigma$ expected limit are shown in green.      
The black, pink and blue contours correspond to the limits from the 2$\ell$, 3$\ell$ and 4$\ell$ analyses respectively. 
\label{fig:InterpModelsNUHM2} }
\end{figure}

The two-, three- and four-lepton analyses from Refs.~\cite{Aad:2014vma,Aad:2014nua,Aad:2014iza} are combined to set limits in a new interpretation for the NUHM2 model. 
The 95\% CL exclusion in the NUHM2 $m_{1/2}$--$\mu$ plane is shown in Figure~\ref{fig:InterpModelsNUHM2}, where the three-lepton analysis offers the best sensitivity and  drives the combined limit. 
The results in the three-lepton signal regions lead to a weaker observed exclusion than expected for the compressed scenarios in the high-$m_{1/2}$, low-$\mu$ region. 
In general, $m_{1/2}$ values up to 300$\GeV$ are excluded in the NUHM2 model.

\FloatBarrier

\subsection{GMSB \label{sec:Interp_gmsb}}

\begin{figure}[!h]
  \centering
\includegraphics[width=0.5\textwidth]{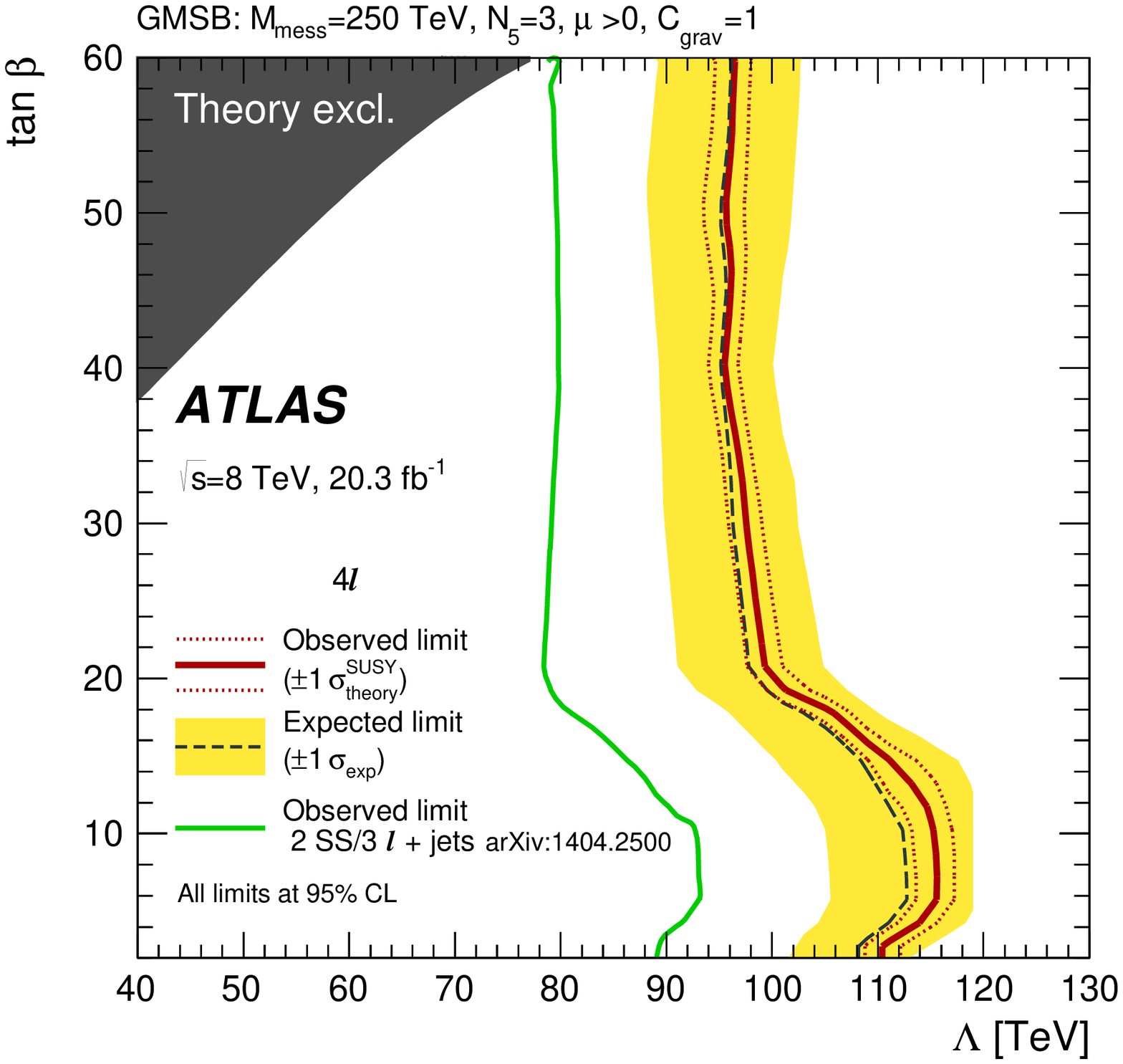}
  \caption{
The 95$\%$~CL exclusion limit in the GMSB scenario, using the 4$\ell$ analysis from Ref.~\cite{Aad:2014iza}. 
The green contour corresponds to the limit from the 2SS/3$\ell$+jets analysis from Ref.~\cite{Aad:2014pda}.
\label{fig:InterpModelsGMSB} }
\end{figure}

The four-lepton analysis from Ref.~\cite{Aad:2014iza} is reinterpreted in the GMSB model described in Section~\ref{sec:susysignals}. 
The 95\% CL exclusion in the GMSB $\Lambda$--$\tan\beta$ plane is shown in Figure~\ref{fig:InterpModelsGMSB}, where $\Lambda$ values up to 94$\TeV$ are excluded for all values of $\tan\beta$. 
For $\tan\beta=10$, $\Lambda$ values below 113$\TeV$ are excluded. 
These results improve upon the previous limit in Ref.~\cite{Aad:2014pda} by 20$\TeV$ (15$\TeV$) in the low (high) $\tan\beta$ region.

\FloatBarrier

\section{Conclusion \label{sec:conclusion}}
This article summarizes and extends the search for the production of electroweak SUSY particles using 20 fb$^{-1}$ of $\sqrt{s}$ = 8$\TeV$ $pp$ collision data collected with the ATLAS detector at the LHC. 
New analyses targeting scenarios with compressed mass spectra, VBF production of charginos and neutralinos, and the direct production of stau pairs provide sensitivity to EW SUSY scenarios not optimally covered in previous publications. 
The new and previous results are combined to set exclusion limits in a wide range of simplified and phenomenological SUSY models. 
For $\chinoonep\chinoonem$ production with $\slepL$-mediated decays, $\chinoonepm$ with masses up to $\sim$500$\GeV$ are excluded. 
In the $\chinoonepm\ninotwo$ and $\ninotwo\ninothree$ scenarios with $\slepL$-mediated decays, $\chinoonepm$ and $\ninotwo$ masses are excluded up to 700$\GeV$ and 670$\GeV$ respectively.
For all three $\slepL$-mediated decay scenarios, the value of the slepton mass is not seen to have a significant effect on the sensitivity. 
Exclusions are also set in pMSSM, NUHM2, and GMSB models, improving upon previous limits.
\FloatBarrier

\noindent
{\bf Acknowledgments}

We thank CERN for the very successful operation of the LHC, as well as the
support staff from our institutions without whom ATLAS could not be
operated efficiently.
We thank Sabine Kraml for her advice while preparing the model files of the NUHM2 grid.
We acknowledge the support of ANPCyT, Argentina; YerPhI, Armenia; ARC,
Australia; BMWFW and FWF, Austria; ANAS, Azerbaijan; SSTC, Belarus; CNPq and FAPESP,
Brazil; NSERC, NRC and CFI, Canada; CERN; CONICYT, Chile; CAS, MOST and NSFC,
China; COLCIENCIAS, Colombia; MSMT CR, MPO CR and VSC CR, Czech Republic;
DNRF, DNSRC and Lundbeck Foundation, Denmark; EPLANET, ERC and NSRF, European Union;
IN2P3-CNRS, CEA-DSM/IRFU, France; GNSF, Georgia; BMBF, DFG, HGF, MPG and AvH
Foundation, Germany; GSRT and NSRF, Greece; RGC, Hong Kong SAR, China; ISF, MINERVA, GIF, I-CORE and Benoziyo Center, Israel; INFN, Italy; MEXT and JSPS, Japan; CNRST, Morocco; FOM and NWO, Netherlands; BRF and RCN, Norway; MNiSW and NCN, Poland; GRICES and FCT, Portugal; MNE/IFA, Romania; MES of Russia and NRC KI, Russian Federation; JINR; MSTD,
Serbia; MSSR, Slovakia; ARRS and MIZ\v{S}, Slovenia; DST/NRF, South Africa;
MINECO, Spain; SRC and Wallenberg Foundation, Sweden; SER, SNSF and Cantons of
Bern and Geneva, Switzerland; NSC, Taiwan; TAEK, Turkey; STFC, the Royal
Society and Leverhulme Trust, United Kingdom; DOE and NSF, United States of
America.

The crucial computing support from all WLCG partners is acknowledged
gratefully, in particular from CERN and the ATLAS Tier-1 facilities at
TRIUMF (Canada), NDGF (Denmark, Norway, Sweden), CC-IN2P3 (France),
KIT/GridKA (Germany), INFN-CNAF (Italy), NL-T1 (Netherlands), PIC (Spain),
ASGC (Taiwan), RAL (UK) and BNL (USA) and in the Tier-2 facilities
worldwide.

\clearpage

\appendix
\part*{Appendix}
\addcontentsline{toc}{part}{Appendix}
\section{Cross-section calculation for the same-sign chargino-pair production via vector-boson fusion \label{sec:app_vbfxsec}}

The cross-sections for same-sign chargino-pair production via vector-boson fusion (including radiative processes) 
are calculated to leading order (LO) in the strong coupling constant using MadGraph 5-1.3.33~\cite{Alwall:2007st}. 
The default value of 99 is used for the maximum number of QCD and QED couplings. 
Same-sign chargino-pairs are generated in association with two additional partons with $|\eta|\,<\,5$ and no \pT\ requirement.
No jet-parton matching is performed. 
All SUSY particles, except for the \chinoonepm, \ninoone, \ninotwo, \slepton, and \snu, are decoupled by setting their physical masses to $\sim$100$\TeV$. 
The \chinoonepm\ and \ninotwo\ are assumed to be mass degenerate. 
The sleptons are assumed to be mass degenerate with sneutrinos, and have masses set halfway between \chinoonepm\ and \ninoone\ masses. 
Cross-sections are also calculated using MadGraph 5-2.2.3 and are in agreement with those calculated using MadGraph 5-1.3.33.
Details from the ``\texttt{proc\_card.dat}'' are provided below.

{\scriptsize
\begin{verbatim}
import model mssm
define p = g u c d s u~ c~ d~ s~
define j = g u c d s u~ c~ d~ s~
define l+ = e+ mu+
define l- = e- mu-
define vl = ve vm vt
define vl~ = ve~ vm~ vt~
generate     p p > x1+ x1+ j j @1
add process  p p > x1- x1- j j @2
output -f
\end{verbatim}
}

\clearpage

\bibliographystyle{bibtex/bst/atlasBibStyleWoTitle}
\raggedright
\bibliography{paper}

\newpage
\begin{flushleft}
{\Large The ATLAS Collaboration}

\bigskip

G.~Aad$^{\rm 85}$,
B.~Abbott$^{\rm 113}$,
J.~Abdallah$^{\rm 151}$,
O.~Abdinov$^{\rm 11}$,
R.~Aben$^{\rm 107}$,
M.~Abolins$^{\rm 90}$,
O.S.~AbouZeid$^{\rm 158}$,
H.~Abramowicz$^{\rm 153}$,
H.~Abreu$^{\rm 152}$,
R.~Abreu$^{\rm 116}$
% ...
\bigskip
\\
% ...
$^{11}$ Institute of Physics, Azerbaijan Academy of Sciences, Baku, Azerbaijan\\
% ...
$^{85}$ CPPM, Aix-Marseille Universit{\'e} and CNRS/IN2P3, Marseille, France\\
% ...
$^{90}$ Department of Physics and Astronomy, Michigan State University, East Lansing MI, United States of America\\
% ...
$^{107}$ Nikhef National Institute for Subatomic Physics and University of Amsterdam, Amsterdam, Netherlands\\
% ...
$^{113}$ Homer L. Dodge Department of Physics and Astronomy, University of Oklahoma, Norman OK, United States of America\\
% ...
$^{116}$ Center for High Energy Physics, University of Oregon, Eugene OR, United States of America\\
% ...
$^{151}$ Institute of Physics, Academia Sinica, Taipei, Taiwan\\
$^{152}$ Department of Physics, Technion: Israel Institute of Technology, Haifa, Israel\\
$^{153}$ Raymond and Beverly Sackler School of Physics and Astronomy, Tel Aviv University, Tel Aviv, Israel\\
% ...
$^{158}$ Department of Physics, University of Toronto, Toronto ON, Canada\\
% ...
\end{flushleft}


\end{document}